# Exploits, advances and challenges benefiting beyond Li-ion battery technologies

A. El Kharbachi,[a,*] O. Zavorotynska,[b] M. Latroche,[c] F. Cuevas,[c] V. Yartys,[d] M. Fichtner [a,e,*]

[(a)] Helmholtz Institute Ulm for Electrochemical Energy Storage (HIU), Helmholtzstr. 11, 89081 Ulm, Germany
[(b)] Department of Mathematics and Physics, University of Stavanger, P.O. Box 8600 Forus, NO-4036 Stavanger, Norway
[(c)] Université Paris Est, Institut de Chimie et des Matériaux Paris Est, ICMPE, CNRS-UPEC, F- 94320 Thiais, France
[(d)] Institute for Energy Technology, P.O. Box 40, NO-2027 Kjeller, Norway
[(e)] Institute of Nanotechnology, Karlsruhe Institute of Technology (KIT), P.O. Box 3640, 76021 Karlsruhe, Germany

Corresponding Authors: kharbachi@kit.edu and m.fichtner@kit.edu

## Abstract

The battery market is undergoing quick expansion owing to the urgent demand for mobile devices, electric vehicles and energy storage systems, convoying the current energy transition. Beyond Li-ion batteries are of high importance to follow these multiple-speed changes and adapt to the specificity of each application. This review-study will address some of the relevant post-Li ion issues and battery technologies, including Na-ion batteries, Mg batteries, Ca-ion batteries, Zn-ion batteries, Al-ion batteries and anionic (F- and Cl-) shuttle batteries. MH-based batteries are also presented with emphasize on Ni*M*H batteries, and novel MH-accommodated Li-ion batteries. Finally, to facilitate further research and development some future research trends and directions are discussed based on comparison of the different battery systems with respect to Li-ion battery assumptions. Remarkably, aqueous systems are most likely to be given reconsideration for intensive, cost-effective and safer production of batteries; for instance to be utilized in (quasi)-stationary energy storage applications.

**Keywords:** post-lithium batteries; (mono)multi-valent systems; MH-based batteries, rechargeable battery assessment

# 1. Introduction

The Li-ion Batteries (LIBs) are the most advanced technologies for electrochemical storage and conversion and undergoing a market expansion with respect to the increase of the electrical vehicles sales and appearance of a panoply of mobile applications. Although extensive studies have been undertaken in order to increase the energy density and power in LIBs, however, the achieved energy storage capability so far is still not adequate to meet the continuous demand from the growing markets, and keep up with challenges for building "sustainable" batteries in terms of performance/energy density as well as cost-efficiency and safety.

In actual fact, LIBs suffer from the rare abundance of Li metal, and the apparent decrease of the price of LIBs in general, owing to mass production, does not justify the diminution of the overall resources involved in LIB components and processes. For more than a quarter century of commercialization, LIBs have been embraced as high energy density and long-cycle-life technology, and consequently dominated portable electronics and rechargeable battery systems for the emerging electric/hybrid vehicles.

Even though this technology is considered as a possible choice for future electric vehicles and grid-scale energy storage systems; one must admit, insufficiency on a global scale of lithium resources and safety factors will strongly limit its further use in large-scale applications [1]. It is predicted, indeed, that the possibility of lithium supply will run out on long term basis (Fig. 1a), depending on the forthcoming political decisions for large-scale energy storage. Although there are opportunities of cost-effective recycling and exploring new sources, however, the gap between offer and demand could result in price significant fluctuations [2]. In the near future, market forecast of rechargeable batteries predict large-scale battery markets with electric vehicles (xEVs) and energy storage systems (ESSs) for smart grids with the matching of the volumes of the produced renewable energies (Fig. 1b). One can expect, if not already felt, that the market is moving from the small-scale to large-capacity industry sector. Fig. 1b shows the rapid growth phase and market expansion which is governed by the emerging applications. The mobile device sector of the market is also expected to continue expanding at constant rates [3].

On the other hand, LIBs struggle to satisfy the current EVs and electricity-grid needs regarding high energy density and low cost. For instance, the addition of more battery stacks in electric cars does not solve really the issue of long range, neither the excessive costs. The challenge for grid storage is the existence, at certain conditions, of inexpensive easy production and output modulation, using power plants that provide electricity costing five time less than that could be supplied by currently available batteries [4]. To overcome the lack of reliable energy storage and conversion, and revolutionize the transport and electricity-grid, novel electrochemical storage technologies beyond Li-ion batteries are highly required.

Electrochemical energy storage systems and technologies are in continuous development owing to the worldwide demand to overcome the current energy issues and satisfy the daily needs in which rechargeable batteries play a key role [5]. Sodium-ion batteries (SIBs) and potassium-ion batteries (KIBs) are the most evident alternatives to LIBs since these technologies are using relatively abundant and cheap sodium (potassium) elements and they have similar chemical properties to lithium, though they have been pointed out regarding their low energy density, and use of highly toxic and flammable electrolytes, as well as having rather high operating costs at their early stage of development [6]. The SIB is a complex cell when in operation compared to LIBs. Such batteries need to be explored and studied in the aim to establish alternative battery chemistries with low-cost, high safety and long cycle life. Here, we will review the recent battery developments beyond classical LIBs, taking in consideration electrode materials and electrolytes for cationic shuttles (Na, Mg, Ca, Zn and Al), as well as anionic shuttles such as halides. An overview of the state-of-the-art of MH-based batteries will be also presented, including Ni*M*H batteries, and metal hydrides accommodated LIBs. Furthermore, we will discuss the scientific challenges of the most relevant battery technologies, and how this will affect our perception of future batteries according to the specificity of the application. Finally, we will summarize the outcome of this review work in the conclusion part and provide new perspectives for possible battery research directions.

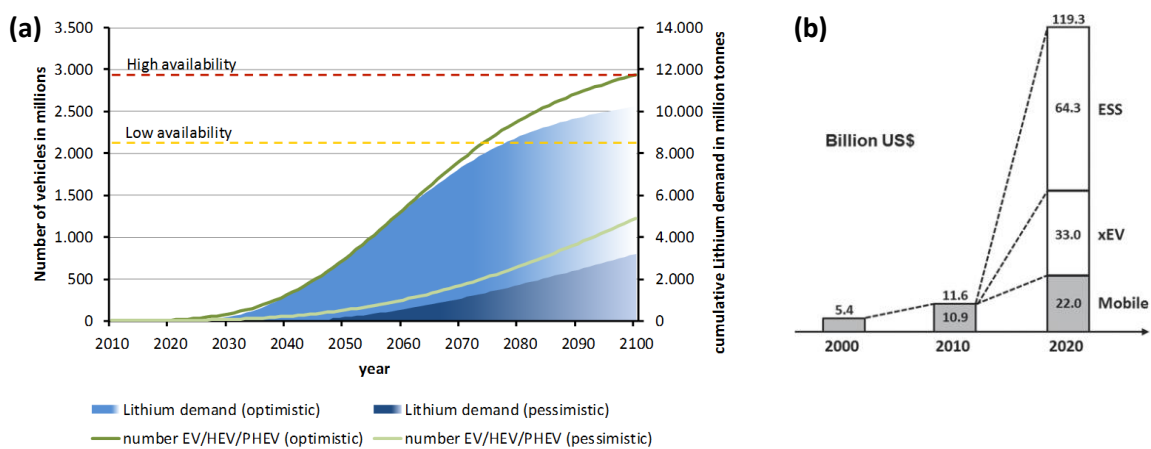

Figure 1.
(a) Long-term assessment of lithium availability and demand, and number of electric vehicles (EV, HEV and PHEV) over time. Lithium run out could be expected for low availability and optimistic electric vehicles production [7]; (b) Market expectation of rechargeable batteries. xEV: all electric vehicles such as full (EV), hybrid (HEV) and plug-in hybrid (PHEV) types. ESS: other Energy storage systems as a part of smart grids and renewable energies [3].

# 2. Cationic shuttles

## 2.1. Monovalent systems: Na-ion batteries

### 2.1.1. Motivation and current development

On the first plan, Na-ion batteries are presented as alternative to Li-ion technology owing to cost-efficiency, safety and long-term sustainability. The abundance of Na compared to Li makes the cost factor decisive for the choose between the two technologies [2]. In addition, Al current collector can be used with Na instead of Cu with higher costs, and no alloying between Al-Na takes place. Na-ion batteries are considered safer with less thermal runaway [8]. By comparing to LIB, sodium-ion battery has similar chemistry during the (de)insertion; it is believed that concept, manufacturing and end-products commercialization processes can be adapted to the existing ones for Li-ion technology.

However, the Na-ion battery system counts a few drawbacks which hinder its quick development as alternative to Li-ion battery [1]. Due to the higher atomic size and larger specific weight of Na, the theoretical capacities of the metal and of the electrode materials are lower, as well as the corresponding energy densities [9]. The anode consists usually of hard carbon, as the graphite can not allow the intercalation of $Na^+$ ions between the carbon layers. The most common electrolytes allowing transport of $Na^+$ ions are based on either carbonate-based solvents or ionic liquids. A large number of potential cathode materials have been explored in the last years, consisting of structurally stable polyanionic materials and layered transition metal oxides such as $NaTMO_2$ providing high energy density and high operating voltages [10]. The cathode material $Na(Ni_{0.5}Mn_{0.5})O_2$ has the specific capacity of 125 mAh.g$^{-1}$ (2.2-3.8 V) and high rate capability [11]. Nevertheless, the long term cycling stability is still a challenge for the layered oxides owing to the large structural changes caused by the volume expansion/contraction during $Na^+$ (de)intercalation. For quick development, the Sodium-ion batteries face the challenge to improve the specific capacity and reaching higher working voltages. This technology seems to follow the same trend as for Li-ion one regarding major challenges, i.e. most research efforts are first put on the study of the cathode structure, as well as cycle life, capacity fading, degradation aspects, interfaces and electrolyte composition (with or without additives).

On the other hand, analysis of the costs normalized to energy density has demonstrated that Na-ion battery is equally expensive compared to a Li-ion battery [12]. At present, Na-ion battery are not competitive as compared to the high energy lithium-ion systems, such as based on lithium cobalt oxide or lithium iron phosphate cathodes, however it is thought that sodium-ion batteries will become a complementary electrochemical storage solution depending on a particular application concerned and solicitation in connection to other electrical devices and power diverters. For instance, Na-ion batteries at their initial development are more likely to fit in the stationary storage of energy endeavoring to become a commercial product at larger scale.

## 2.1.2. State-of-the art of Na-ion batteries
### 2.1.2.1. Cathode materials

Metal oxide cathode materials are the most developed and promising cathodes employed in SIBs [13-35]. Similar to LIBs, the $Na_xCoO_2$ has been studied already in the early 1980s [32]. This cathode demonstrates reversible intercalation of $Na^+$ ions in the phase $Na_xCoO_2$ (0.5 < x < 1), accompanied by a phase transition of the layered structure involving a change from octahedral or trigonal prismatic coordination to the monoclinically distorted phase packing [2,9,32,36]. Similarly to $Na_xCoO_2$, $Na_xMnO_2$ polymorphs are widely investigated as cathode materials for SIBs [13,37-46]. The α-$Na_xMnO_2$ phase is structurally more stable than its homologue high-$T$ orthorombic β-$Na_xMnO_2$ phase (Fig. 2a), and shows a layered structure with monoclinic distortion. Based on ab initio studies, it has been found that the structure shown in Fig. 2b for $Na_{0.44}MnO_2$ have the lowest energy in the S-shaped tunnel [39]. The intercalation of $Na^+$ in the α-phase allows 185 mAh.g$^{-1}$ at C/10 rate with 71% capacity retention over 20 cycles, meanwhile 70% after 100 cycles is delivered when comparing to β-phase [46]. The charge/discharge profiles shown in Fig. 2c indicate a multi-step processes in relation to the presence of intermediate phases' transformations [38]. It seems that not all these transformations and reactional pathways are well understood and consequently not yet determined in details [36].

Significant improvement of the long-term cyclability of the α-$NaMnO_2$ phase has been achieved when the electrolyte 1M $NaBF_4$/tetraethylene glycol dimethyl ether (TEGDME) is used instead of 1M $NaClO_4$/EC:DEC [47]. Though in the presence of EC:DEC-based electrolyte the cell shows lower bulk and interfacial resistances. The electrolyte substitution with TEGDME-based one allows the stabilization of the interface resistance hence leading to better long-term cyclability. Other materials have been studied showing a weaker electrochemical performance compared to $NaMnO_2$. These include $NaCrO_2$ and $NaFeO_2$ phases, as well as multi-cations oxides such as $Na_x(Ni_{2/9}Co_{1/9}Mn_{2/3})O_2$ [48-55].

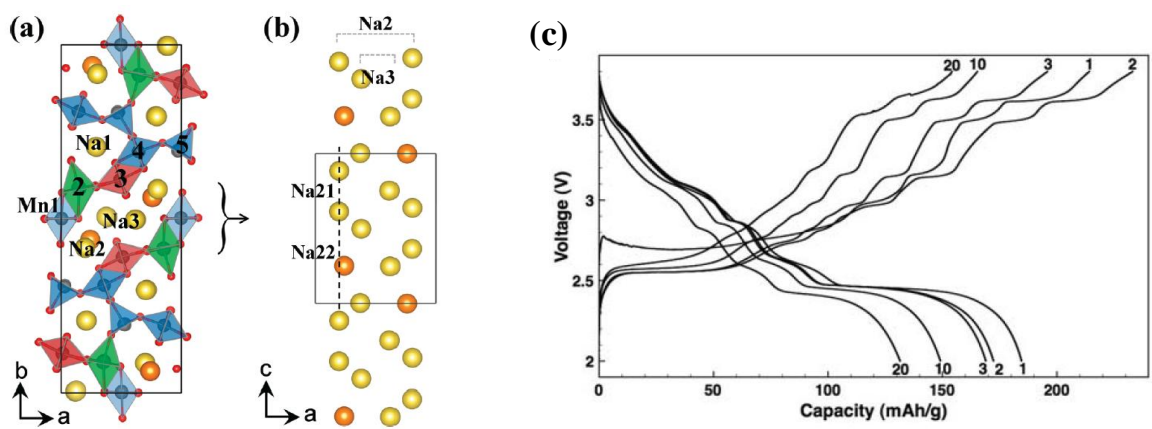

Figure 2.
(a) Crystal structure of $Na_{0.44}MnO_2$ with five crystallographic sites for manganese and three sites for sodium ions [39], (b) most probable sodium configuration is in the *S*-shaped tunnel along the *c*-axis where the Na2 site has two different sodium positions, Na21 and Na22 [39], (c) voltage profile of $NaMnO_2$ after multiple cycles at C/10. The cell is galvanostatically cycled between 2.0 V and 3.8 V [38].

Multiple cation transition metal oxides can be synthesized using co-precipitation in aqueous solution and extensive rinsing with distilled water [56]. The materials show high reversibility and good capacity retention with a specific capacity of 135 mAh g$^{-1}$ and a Coulombic efficiency 99.7% over 250 cycles in ionic liquid medium [56]. In fact, the solubility of Mn has been pointed out in many studies. Then, substitution of carbonate-based electrolyte with an ionic liquid demonstrates uniform SEI layer at low and high voltage operation. A specific capacity of 200 mAh.g$^{-1}$ has been reached with a capacity retention of about 80% after 100 cycles in the presence of 10 mol.% NaTFSI/*N*-butyl-*N*-methylpyrrolidinium bis(fluorosulfonyl)imide electrolyte [56]. The structural study of the phase O3-$NaNi_{0.5}Ti_{0.5}O_2$ has been reported to be suitable as cathode material for SIBs [57]. A schematic illustration of the structural model of O3-$NaNi_{0.5}Ti_{0.5}O_2$ is presented in Fig. 3a. In this model, nickel and titanium ions are positioned at the octahedral sites of the $MeO_2$ layer (3a sites, Me = Ni and Ti), while sodium ions are located at the octahedral sites of the $NaO_2$ layer (3b sites). The model shows no cation intermixing between sodium and nickel ions due to their large difference in ionic diameter [57]. The material exhibits reversible structural behavior during (de)sodiation with an average voltage of 3.1 V *vs*. Na$^+$/Na redox couple and a capacity of 121 mAh g$^{-1}$ at C/5. At high rate (5C), 60% of the initial discharge capacity is obtained. Fig. 3b/d shows the good cyclability and stability of the electrode lifespan over 100 cycles at two different cycling rates (C/5 and 1C). Rate capability tests of the Na/$NaNi_{0.5}Ti_{0.5}O_2$ cells at different rates are shown in Fig. 3c. The cell delivers a reversible capacity of about 90 mAh g$^{-1}$ with small polarization even at 1C rate. The charge–discharge of the $NaNi_{0.5}Ti_{0.5}O_2$ electrode material at 10C rate can still deliver 27 mAh g$^{-1}$ [57].

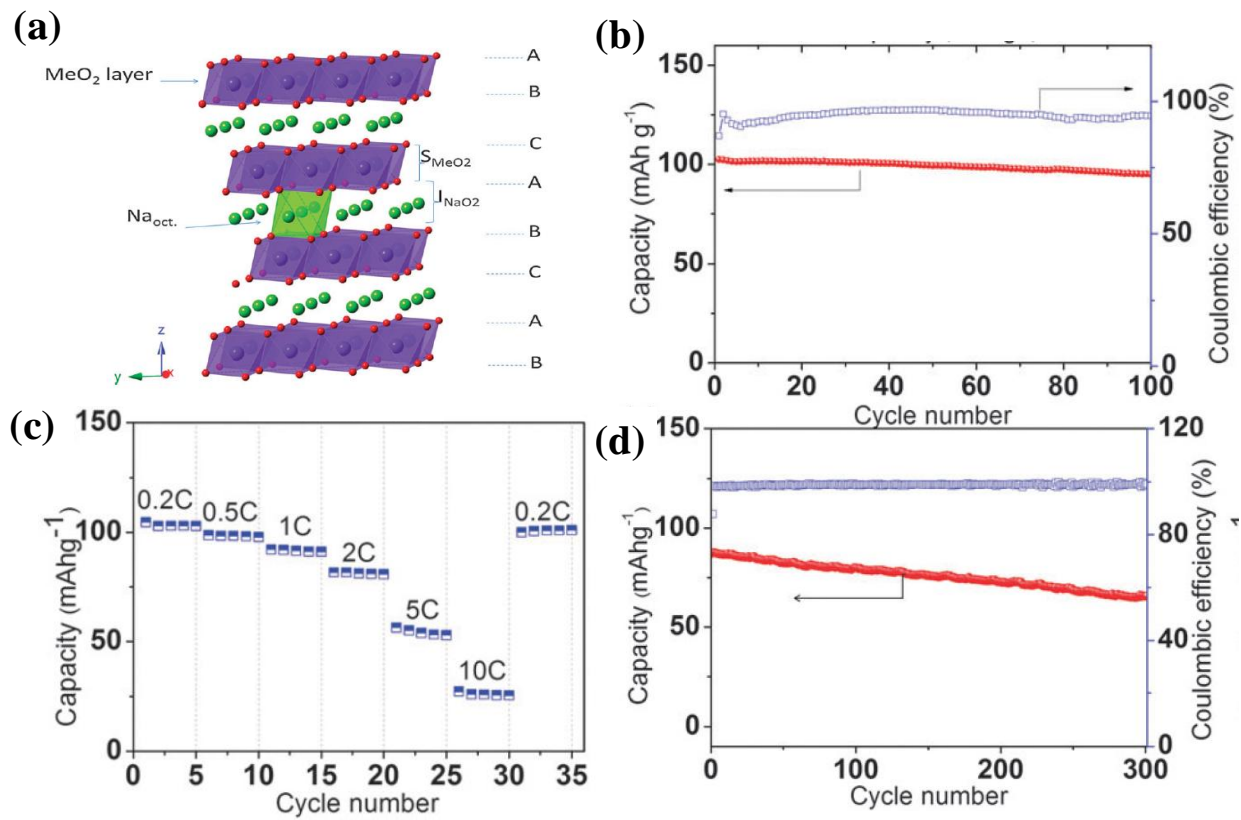

Figure 3.
Structural and electrochemical performance of O3-NaNi$_{0.5}$Ti$_{0.5}$O$_2$ cathode material, (a) schematic illustration of the crystal structure consisting of Me octahedra (blue) and Na octahedra (green), (b) and (d) cycle performance (2-4 V *vs.* Na$^+$/Na) as function of cycle number and relative coulombic efficiency of the Na/cathode material cells at C/5 and 1C rates, respectively, (c) rate capability performance of the cell [57].

Further to the oxides, a series of sulfate, phosphate and fluoride materials has been studied as cathodes for SIBs. Na-S batteries undergo the same challenges as Li-S batteries regarding polysulfide dissolution and dendrite formation, which will not be approached in this review work [36,58]. In this category of material cathodes, Na$_2$Fe$_2$(SO$_4$)$_3$ showed the most interesting electrochemical features with a voltage of 3.8 V and delivering a capacity of 100 mAh.g$^{-1}$ and 50% capacity retention at high rate 20C. The crystallography of this system is under study, where Na seems to occupy three different specific sites. The (de)intercalation of Na in this material is enhanced by the fast Na transfer, thanks to the 3D alluaudite framework with large tunnels along the *c*-axis [59].

The NASICON Na$_3$V$_2$(PO$_4$)$_3$ has been synthesized in nanograins and reached 98.6% of the theoretical capacity (117.6 mAh g$^{-1}$) with high capacity retention at high C-rate [60]. In this structure, corner shared VO$_6$ and PO$_4$ polyhedra form a framework with large diffusion channels for Na-ions [60,61].

From the fluorophosphates family, Na$_{1.5}$VPO$_{4.8}$F$_{0.7}$ can be synthesized and it crystallizes in a pseudolayered structure (space group *P*4$_2$/*mnm*).

During (de)intercalation, 1.2 e$^-$ can be exchanged/f.u. Na$_{1.5}$VPO$_{4.8}$F$_{0.7}$. According to the working potential of 3.8 V (vs. Na$^+$/Na) of vanadium redox couple, this leads to an energy density of 600 Wh kg$^{-1}$ with 95% capacity retention for 100 cycles and ~84% for 500 cycles respectively [62].

### 2.1.2.2. Anode materials

Graphite anodes commonly used in LIBs, are not suitable for the intercalation of Na$^+$ ions with larger ionic radii. Hard carbon was demonstrated as a host to accommodate inserted Na$^+$ ions [63,64]. Furthermore, similar to lithium, pure metals, alloys, hydrides and oxides have been studied as different alternative known mechanisms operating in addition to intercalation, such as alloying and conversion reactions [65-75].

The first tests with hard carbon anodes led to an initial capacity of 220 mAh g$^{-1}$ in NaClO$_4$ / EC:DMC electrolyte, which decreases during cycling [64]. Hard carbon C1600 was reported as anode of Na ion battery. These electrodes were tested in different electrolyte media. A capacity retention was 90% over 50 cycles where an initial capacity 413 mAh g$^{-1}$ is obtained in the presence of 1M NaClO$_4$ / EC:DMC [76]. Previously, Ponrouch *et al.* [77] have demonstrated a half-cell battery with hard carbon having 200 mAh g$^{-1}$ capacity, with a decent rate capability and cyclability over 180 cycles when using the same electrolyte. At present, hard carbon is selected to be the most suitable anode for SIBs, although a wide series of carbonaceous materials with different shapes and nanostructures are under study as well [78,79].

### 2.1.2.3. Electrolytes for SIB

Interfacial reactions are even more crucial for SIBs than for LIBs, because of the slow diffusion of Na$^+$.

Interfaces, SEI layer formation and charge transfer resistances are the factors that can be dependent on the electrolyte composition; hence this plays a determining role in a better optimization of the battery operation. Studies aiming at selection of the suitable electrolytes showed that the performance is electrode dependent. The electrolyte 1M NaClO$_4$ / EC:PC offered more stable electrochemical performance of the Na$_4$Fe$_3$(PO$_4$)$_2$(P$_2$O$_7$) electrode with hard carbon anode for use in SIBs. The substitution of NaClO$_4$ with NaPF$_6$ offered better SEI thermal stability [77,80].

Improvements have been made in electrolyte composition by adding a small amount of DMC, with low viscosity and dielectric constant compared to EC/PC solvents. The solvation shell of Na$^+$ cations is mainly composed of EC with negligible combination from other solvents or anions [81]. As there is no significant modification of the solvation by DMC addition, the increased ionic conductivity was attributed to the decrease of the viscosity of the mixed-solvents used for the preparation of the electrolytes. For instance, $EC_{0.45} \cdot PC_{0.45} \cdot DMC_{0.1}$ was selected for use in testing hard carbon anode and $Na_3V_2(PO_4)_2F_3$ cathode *vs.* $Na^+/Na$ redox couple. The assembled Na-ion full cells demonstrated a working voltage of 3.65 V, low polarization and good capacity retention with a reversible capacity of ~97 mAh g$^{-1}$ over 120 cycles with a coulombic efficiency > 98.5% [81].

The use of ionic liquids allowed to work with high voltage cathodes (> 4.2 V) such as $Na_{0.45}Ni_{0.22}Co_{0.11}Mn_{0.66}O_2$, where also the dissolution of Mn can be avoided at low voltages [56], in addition to the improved safety features (low flammability and volatility), wide electrochemical/thermal stability, low vapor pressure and high ionic conductivity [82,83]. More safe electrolyte for SIBs is an aqueous-based one [84,85]. However, the low operating voltage 0-0.9 V *vs.* SCE (1M $Na_2SO_4$), may not be suitable for the current urgent demand for high power and high energy density applications [86].

## 2.2. Multivalent systems

### 2.2.1. Mg batteries

#### 2.2.1.1. Motivation, principle and historical development

The dynamic interest in high energy density electrochemical storage systems such as "rechargeable magnesium batteries" (RMBs) has been fueled by the high capacity of Mg metal anode, dendrite-free Mg plating and stripping, and the promise of economic efficiency and sustainability. Indeed, magnesium metal anode has the theoretical capacities of 2200 mAh g$^{-1}$ and 3835 mAh cm$^{-3}$, the latter being almost double of that of Li. Another advantage of Mg over Li is the abundancy of Mg in the Earth crust being at least four orders of magnitude higher than that of Li [87,88]. Moreover, Mg is safer to handle than Li; it does not usually form toxic and/or dangerous compounds promising cost-effective and eco-friendly industrial processes.

Early work of Brenner in the 1970$^{th}$ on the electrodeposition of magnesium from a solution containing decaborane, $MgCl_2$ and anhydrous THF at room temperature (R*T*) can be regarded as the beginning of magnesium battery research [89]. In the following years, sporadic reports appeared until the 1990$^{th}$ when Gregory *et al*. published an extensive study on non-aqueous electrochemistry of magnesium [90]. They have examined a range of various intercalation cathodes (sulfides, oxides, and borides of transition metals), electrolytes (solutions of organomagnesium compounds), solvent-solute-intercalation cathode combinations, and the strategies improving the electrochemical properties of RMB. The cells systems with the best performance were prepared of Mg sheet anode, magnesium dibutyldiphenylborate solution in THF-DME as electrolyte and $Co_3O_4$ cathode. The authors reported poor stability of the electrolytes towards the transition metal oxide or sulfide cathodes with the largest found reversible capacities. It was also estimated that for a battery with an operating voltage of 1.5 V, minimum acceptable specific capacity of the cathode material should be around 230 mAh g$^{-1}$. For a decade after this work, only scarce reports on magnesium battery components appeared emphasizing the difficulties to find suitable intercalation cathodes, electrolytes and chemically stable cell systems (Fig. 4). In the 2000$^{th}$, Aurbach *et al.* reported a reversible Mg-battery composed of Mg organohaloaluminate-based electrolytes and Chevrel phase intercalation cathodes, mainly Mo$_6$S$_8$. These electrolytes exhibited higher anodic stability of 2.2 V compared to that of 1.5 V of Gregory *et al.* For the batteries with the best performance, based on the THF/Mg(AlCl$_2$BuEt)$_2$ electrolyte and Mg$_x$Mo$_3$S$_4$ cathode, more than 2,000 charge–discharge cycles at 100% depth of discharge of the cathodes (rates 0.1–1 mA cm$^{-2}$) with less than 15% capacity deterioration was demonstrated. The initial capacity of the systems was 60-90 mAh g$^{-1}$. As noted in these early studies, despite of the obvious advantages of Mg anodes, the difficulties in finding the suitable combination of electrolyte-cathode-anode chemistries constituted (and still do) significant technical challenges and determine the direction of the research efforts [90,91]. Fig. 4 demonstrates in fact that cathodes and electrolytes have been the subject of the main research activity in the field [92].

The "cathode challenges" have been caused by difficulties of intercalating divalent high charge density magnesium cations in most of the known electrode hosts. A variety of structures with different geometries and chemical compositions has thus been explored so far. Among them, Chevrel-type cathodes have demonstrated the best performance in terms of specific capacity, Mg$^{2+}$ intercalation kinetics, Coulombic efficiency, reversibility, and operational voltage [91,93]. On the other hand, these cathodes still possess rather low specific capacity of ~ 230 mAh g$^{-1}$. The novel structures such as functionalized 2D sheets and fullerenes have been recently suggested as promising novel intercalation cathodes [94-96]. As an alternative, conversion cathodes such as sulfur or iodine were explored [97,98]. The conversion cathodes can offer high specific capacities of ca. 820 mAh g$^{-1}$ but are not stable towards high operational voltages (> 1.5 V) [93]. In order to achieve the compromise between operational voltage and specific capacity, breakthrough solutions are needed.

The field of electrolyte development has been driven by search for compositions stable towards electrodes, forming favorable electrode-electrolyte interface, delivering high Mg$^{2+}$ diffusion rates and not corroding the cell components. Several different types of electrolyte have been developed through the years, many targeted at a particular cathode type. First prototype batteries included electrolytes based on ether solutions with Mg organo-borate or organo-aluminate salts [90]. These compounds allowed for reversible plating and stripping of magnesium but were unstable towards high voltages (>1.5 V) and high capacity reversible electrophilic cathodes. The next-generation electrolytes based on ethereal solutions of magnesium halo-alkyl aluminate complex have improved the reversibility of Mg deposition and demonstrated high-voltage (2.5 V) stability [91]. These dichloro-complex electrolytes (DCC) demonstrated excellent reversibility of several thousand cycles, in particular in systems with Chevrel type cathodes and faster Mg$^{2+}$ intercalation kinetics [99]. All phenyl complex (APC) electrolytes were the next step on the way to high voltage electrolytes approaching 3 V [100]. Both DCC and APC electrolytes have shown a large dependence on a particular chemical composition, solvents, additives, working temperature, the electrode composition, etc., and have possessed a challenging task even for a skilled organic chemist. Thus some attempts to develop all inorganic electrolytes [101], including solid-state electrolytes for all-solid-state RMB [102], and halide-free non-corrosive electrolytes [103,104] have been undertaken. The demonstration of the first proof-of-concept Mg-S battery required the development of non-nucleophilic electrolyte to exclude the chemical reaction with sulfur. For the conversion cathodes, electrolytes based on hexamethyldisilazide magnesium chloride (HMDSMgCl) have been developed [97]. At the same time, modifications of the anode material have been undertaken in order to increase its stability towards the electrolyte [105].

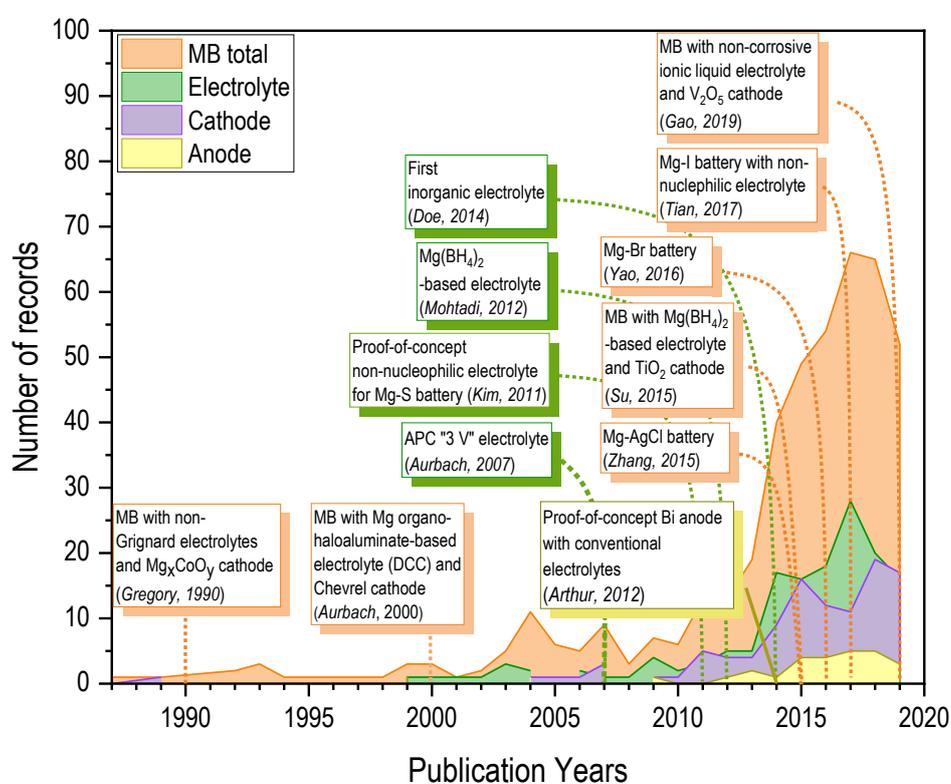

Figure 4.
Search results from ISI Web of Science database with "magnesium batteries" (MB) in the *Topic* field and "electrolyte", "cathode", or "anode" in *Title* search field. Significant achievements are indicated on the graph. The search was performed over all publication years. The data were obtained in October 2019. References [90,91,97,98,100,101,105-110].

## 2.2.1.2. State-of-the art of Mg batteries

### 2.2.1.2.1. Cathode materials

Cathodes for RMB have been one of the largest hurdles on the way of rechargeable batteries. For an efficient battery, the chemically stable cathodes with high electrochemical potential vs. Mg, and high capacities stable over many cycles are required. In addition, composition of non-toxic, abundant elements is highly desirable. In order to achieve the abovementioned requirements, several classes of cathodes have been explored as described below.

**Intercalation cathodes.** Intercalation-type cathodes are a commercialized technology for Li-ion batteries, and have been considered as a benchmarking technology for magnesium batteries. By a sharp contrast to lithium, however, electrochemical insertion of the divalent $Mg^{2+}$ into a solid host is significantly hampered by the increased charge density on the cations. This leads to strong interactions with the host, and thus slow kinetics and unfavorable thermodynamics of the insertion and diffusion processes. A multivalent cation diffusion depends significantly also on structure of the host that determines the diffusion pathway [111]. A variety of compounds has been proposed as candidates for intercalation magnesium cathodes. These structures can be classified as 3D diffusion channels (Chevrel phase, spinel), 2D layered structures, and 1D polyanion structures [112].

Chevrel phases are ternary molybdenum chalcogenides $M_xMo_6X_8$ (X = chalcogen) with structures spanning from 3D lattices where the third element M can be inserted, up to a condensation of clusters giving rise to a 1D material [113]. Chevrel-type 3D cathodes have shown an excellent reversible intercalation kinetics for $Mg^{2+}$ with capacities of ~120 mAh g$^{-1}$ at 1.2 V [93,112]. The fast insertion kinetics for bivalent $Mg^{2+}$ ions in the 3D $Mo_6S_8$ is attributed to the unusual structure of the Chevrel that allows for neutralizing the extra positive charge brought in by the guest $Mg^{2+}$ and offers a large number of closely located sites for diffusion [114]. Furthermore, the special surface structure of the phase facilitates the desolvation of complex cations of $Mg^{2+}$ from electrolyte at the electrolyte/cathode interface [115]. A significant disadvantage of the $Mo_6S_8$ cathode resides in the strong temperature dependence of the intercalation kinetics and partial (20-25%) irreversibility at R*T* due to the cation trapping [112]. The $Mo_6Se_8$ phase provides more open and more polarizable structure with faster intercalation kinetics and ionic mobility for $Mg^{2+}$, however, at the cost of capacity. Mixed phases of $Mo_6Se_{8-x}S_x$ have been synthesized to compromise between the kinetics and the capacity. The $Mg_xMo_6S_6Se_2$ cathode allowed the storage capacity of 110-100 mAh g$^{-1}$ at the voltage 1.1-1.3 V after 100 cycles in a cell with DCC electrolyte and Mg anode [100]. In search for higher voltages and capacities, a range of other structures has been explored for Mg intercalation.

Spinel compounds with general formula $MgT_2X_4$, where T is a transition metal, and X stands for O, S, or Se, offer 3D channels where $Mg^{2+}$ can diffuse along tetrahedral (tetra) → octahedral (octa) → tetra or octa → tetra → octa pathways [112]. The calculated Mg diffusion energy barriers in some spinel oxides are rather high for applications at practical temperatures [116]. Some work has been done with $Mn_2O_4$ owing to the high specific energy density, thermodynamic stability of both the charged and discharged phases, and acceptable volume change of the electrode. Apparently, the crystal structure of the cathode, possibly particle size and morphology, and electrolyte, are critical for intercalation of $Mg^{2+}$ into $Mn_2O_4$ so that this cathode can achieve up to 250, 120 mAh g$^{-1}$ or no appreciable intercalation at all depending on these factors [112,117,118]. Lower diffusion barriers were calculated for sulfide spinels, distinguishing $Cr_2S_4$, $Ti_2S_4$, and $Mn_2S_4$ structures out of 21 3d transition-metal sulfur-spinel compounds [119]. However, the improved mobility of magnesium cations comes at the expense of lower voltage and thereby lower theoretical specific energy. Experiments show that cycling $Ti_2S_4$ cathode in APC electrolyte at 60 °C demonstrated 230 mAh g$^{-1}$ capacity with an average potential of 1.2 V at low rates [120]. At the same time, various attempts to remove Mg from $MgCr_2S_4$ spinel lattice appeared to be unsuccessful [121]. Achieving intercalation of $Mg^{2+}$ at R$T$ in structures with sufficient voltage and specific energy remains the largest obstacle for spinels to be the suitable cathodes for Mg batteries. Further studies in this direction are encouraged as well as the detailed characterization of the intercalation process [112].

In addition to spinel O, S, Se chalcogenides, much attention has been also devoted to layered compounds. In the layered compounds, week van der Waals forces between layers and the structural flexibility could presumably facilitate cation diffusion along the 2D channels. The layered $TiS_2$/$TiS_3$/$TiSe_2$ [122-124], $V_2O_5$ [110,125-127] and $MoO_3$ [126] have been explored demonstrating low to moderate capacities and reversibility. In search for higher capacities, faster intercalation kinetics, and lower migration barriers for $Mg^{2+}$, nanosizing, doping (for example, with $H_2O$, F) and creating defects, and forming solid solutions have been undertaken with variable outcomes [128-131]. Pre-intercalating Na or Li ions in the crystal structure of intercalation cathodes can improve the layered structure stability and electrochemical performance of the materials [132,133]. Fast intercalation kinetics in the layered molybdenum disulfide structures was demonstrated by using solvated magnesium-ions ($[Mg(DME)_x]^{2+}$). The authors suggested the concept of using solvation effect as a general strategy to tackle the sluggish intercalation kinetics of magnesium-ions [134]. One of the strategies employed to increase the capacity of $Mg^{2+}$ insertion is regulation of the interlayer spacing. Thus, in polyanion compounds, such as layered $VOPO_4$, consisting of corner-sharing $VO_6$ octahedra linking to $PO_4$ tetrahedra, the interlayer spacing provides enough diffusion space for fast kinetics of $MgCl^+$ ion flux with low polarization [135]. The Mg battery with the 2D $VOPO_4$ nanosheets cathode, demonstrated the highest capacity of 310 mAh g$^{-1}$ at 50 mA g$^{-1}$, and the highest reversible capacity of 192 mAh g$^{-1}$ at 100 mA g$^{-1}$ retained after 500 cycles. A reversible magnesium-ion storage capability of layered MXenes was theoretically predicted [95], and recently experimentally demonstrated for 3D porous MXene films [94]. MXenes represent a family of transition metal carbides and nitrides with the formula $M_{n+1}X_nT_x$, where M is an early transition metal (Ti, Nb, V, Ta, Cr, Mo), X is carbon and/or nitrogen, $n$ = 1, 2, or 3, and $T_x$ are surface groups such as OH, O, and/or F [136]. The reversible rate-dependent capacities in the range of 55-210 mAh g$^{-1}$ have been demonstrated very recently and call for additional research [94]. In summary, despite a considerable research effort and in some cases higher voltage stabilities (e.g. 2.56 V in $\delta$-$V_2O_5$) [137], none of the layered materials, however, currently satisfies all requirements for a functional cathode calling for further improvements.

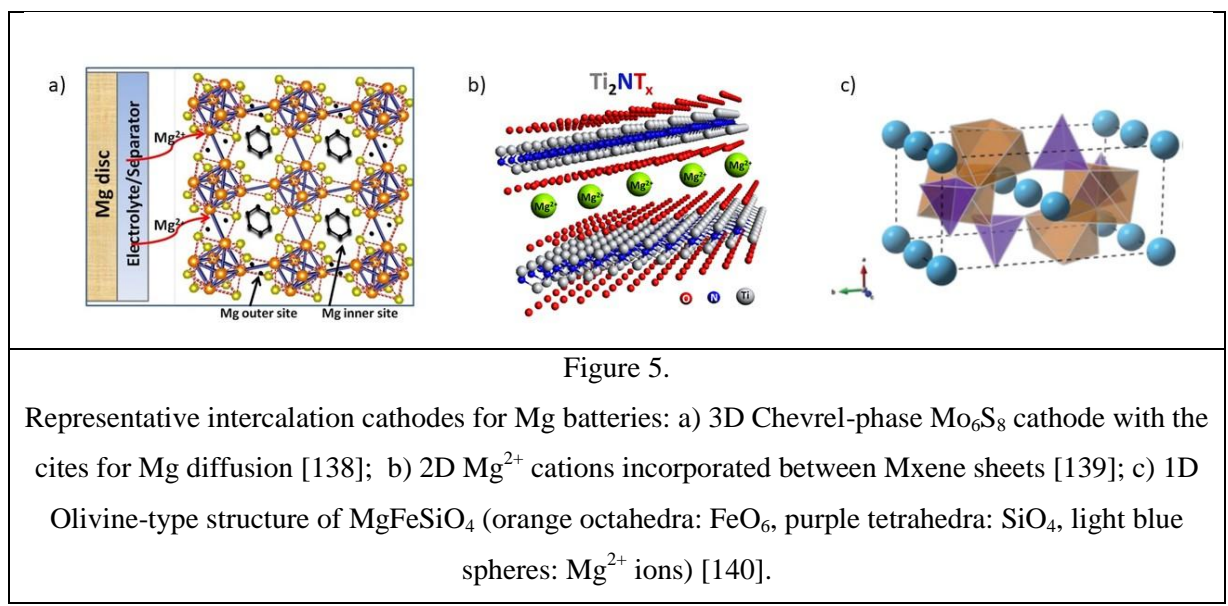

Figure 5.

Representative intercalation cathodes for Mg batteries: a) 3D Chevrel-phase $Mo_6S_8$ cathode with the cites for Mg diffusion [138]; b) 2D $Mg^{2+}$ cations incorporated between Mxene sheets [139]; c) 1D Olivine-type structure of $MgFeSiO_4$ (orange octahedra: $FeO_6$, purple tetrahedra: $SiO_4$, light blue spheres: $Mg^{2+}$ ions) [140].

Polyanion compounds with 1D diffusion channels, such as phosphate and silicate olivine compounds or Prussian blue frameworks, can potentially intercalate $Mg^{2+}$ cations with promisingly high cell voltage ranging from 2.3 V vs. $Mg/Mg^{2+}$ to 2.8–3.0 V [140]. The phosphate compounds, however, have demonstrated a very poor performance. Thus, the olivine $FePO_4$ was shown to deliver ~13 mAh $g^{-1}$ in a non-aqueous electrolyte. The intercalation promoted amorphization of the cathode and thus annihilation of the diffusion/intercalation reaction [141]. On the contrary, ion-exchanged $MgFeSiO_4$ demonstrated a significantly better performance with high reversible capacity exceeding 300 mAh $g^{-1}$ at a voltage of approximately 2.4 V *vs*. Mg for 5 cycles [92,142]. Prussian blue has exhibited very moderate intercalation properties [112].

**Conversion cathodes.** Thermodynamically favorable redox reactions at the conversion electrodes may offer a solution to the slow kinetics of Mg intercalation. These cathodes can be classified into type A and B depending on whether an exchange or a recombination reaction occurs at the electrode [143] :

Type A (exchange reaction): $MX_a + \frac{a}{2} Mg_2 + ae^- \rightleftarrows M + \frac{a}{2} MgX_2$ (1)

Type B (combination reaction): $Mg_2 + X_a + 2e^- \rightleftarrows MgX_a$ (2)

The type A cathodes typically include transition metal halides, oxides, chalcogenides, nitrides or phosphides as $MX_a$ compounds. For example, an AgCl/Mg battery was demonstrated to achieve 95.2% of the theoretical capacity (178-104 mAh g$^{-1}$) at 0.12-10C rates with a flat plateau of ca. 2.0 V. Unfortunately, this system suffered from poor cyclability [109]. In the type A reaction, an intermediate insertion phase is formed with an efficiency depending on ion mobility. Unfortunately, the latter is not a strong side of Mg$^{2+}$ ions as have already been mentioned. Thus, the A-type electrodes, also in case of Li batteries, typically suffer from a poor electronic conductivity, large voltage hysteresis, large volume change and low conversion efficiency [143].

The type B cathodes can be composed of a single element chalcogene (S, Se, Te) or a halogen (Br, I$_2$), often dispersed in a high surface area matrix (e.g. activated carbon, graphite, etc.). Mg-air batteries can also be classified as those with type B cathode (oxygen). Mg-S batteries can demonstrate a theoretical cell voltage of 1.77V and energy density of 1,722 Wh kg$^{-1}$ and 3,200 Wh l$^{-1}$ [143]. Sulfur is usually dispersed in a high surface area matrix, and the matrix itself seems to have a large impact on the performance of the cathode through regulating sulfur loading and formation of soluble polysulfides that deteriorate the cyclability of the cathode. For instance, using ZIF-67 highly porous metal-organic framework (MOFs), a Co- and N-doped carbon support for the sulfur cathode was obtained [144]. This strategy resulted in first discharge capacity of ≈700-600 mAh g$^{-1}$ (at 0.1 and 1 C), and unprecedentedly high cyclic stability, where the ≈300-400 mAh g$^{-1}$ capacity after 150–250 cycles was still maintained when cycling at the up to 5 C rate. The MOF-derivative carbon support doped with N and Co was suggested to trap soluble polysulfides, which in turn allowed for the higher S loading (47%). The addition of Li$^+$ and Cl$^-$ aided in the dissolution of low-order polysulfides, which allowed for the excellent performance. A magnesium/iodine battery have recently been demonstrated [98]. The system showed 180 mAh g$^{-1}$ - 140 mAh g$^{-1}$ at 0.5-1 C and higher energy density by ca. 400 Wh kg$^{-1}$ than the systems with intercalation cathodes. Twenty cycles with about 96% Coulombic efficiency and 3.0 V potential were also shown for Mg-Br$_2$ battery [108]. In these systems, the stability of electrolyte towards cathodes also seems to be a considerable issue. Relatively little research has been made on Mg-air systems although theory predicts promising energy densities [112]. The most notable challenge of this technology seems to be in passivating the surface of the Mg anode, which is very sensitive even to O-containing impurities as discussed below.

### 2.2.1.2.2. Anodes for Mg batteries

The volumetric capacity of magnesium anode is almost twice than that of Li, and the electrochemical deposition-dissolution process is dendrite free at most experimental conditions. On the other hand, upon contact with oxidizing media, magnesium forms surface layers such as MgO and/or Mg(OH)$_2$ that inhibit plating and striping. Electrolyte solutions conventional for Li batteries usually promote the formation of the passivating layers as well as O$_2$ or H$_2$O impurities at levels of 3 ppm [145]. While the main research efforts have been focused on developing electrolytes, the anode modifications have also been proposed with the aim to tune the anodic high reduction capacity.

Nanostructured Mg anode has demonstrated a discharge capacity of 170 mAh g$^{-1}$ and high reversibility. This performance was explained by a decrease in the thickness of the passivation surface layer [146]. We note that theory predicts 1042 mAh g$^{-1}$ Mg capacity for defective graphene [147].

Alloying Mg may enable fast and efficient magnesium plating and striping, adequate redox potential over the whole range of magnesiation, low toxicity and reasonable cost [93]. Bismuth and Bi-based compounds appear to be most interesting candidates [148]. The rhombohedric crystalline structure of Bi facilitates the formation of high capacity Mg$_x$Bi$_y$ alloys, assuming the reaction:

$$2Bi + 3Mg^{2+} + 6e^- \rightarrow Mg_3Bi_2$$

The theoretical capacity 385 mAh g$^{-1}$ can be achieved which is comparable to that of Li-graphite technology (372 mAh g$^{-1}$). Arthur *et al.* [105] presented a study of electrochemical magnesiation / demagnesiation in one cycle of Bi in a Mg(N(SO$_2$CF$_3$)$_2$)$_2$ / acetonitrile solution as a proof-of-concept for compatibility of a Bi anode with conventional battery electrolytes. They have obtained specific anode capacities of 257-222 mAh g$^{-1}$ over 100 cycles in an electrolyte composed of ethylmagnesium chloride, diethylaluminum chloride in THF. The compatibility of Mg$_3$Bi thin films electrodes with the acetonitrile and glyme-based solutions was also recently demonstrated [149]. Approximately half of the Bi-Mg theoretical capacity, but with significant deterioration in the subsequent cycles, was demonstrated in the first cycle in a Bi-carbon nanotubes composite electrodes in acetonitrile-0.5M(Mg(ClO$_4$)$_2$)-ether electrolyte [150]. Application of Bi nanotubes in a Mg(BH$_4$)$_2$-LiBH$_4$-diglyme electrolyte yielded in a specific capacity of 350 mAh g$^{-1}$ with 95-100 % Coulombic efficiency for 200 cycles. The cell composed of Mg$_3$Bi$_2$ anode, Mo$_6$S$_8$ cathode, and a conventional Mg(TFSI)$_2$–diglyme-electrolyte [151] showed a similar performance. Alloying Bi with Sb in Bi$_{0.88}$Sb$_{0.12}$ ratio yielded in 298 mAh g$^{-1}$, which decreased to 215 mAh g$^{-1}$ over 100 cycles at 1C rate; whereas pure antimony anodes demonstrated very poor capacity [105]. Mg$_3$Sb$_2$, which has similar crystal structure and chemical properties to Mg$_3$Bi$_2$, was found electrochemically inactive in acetonitrile and glyme-based solutions [149].

Tin-based anodes have also attracted much attention due to higher theoretical specific capacities and higher availability than that of Bi. The electron-exchange reaction between Sn and Mg yields four electrons per Sn atom:

$$Sn + 2Mg^{2+} + 4e^- \rightarrow Mg_2Sn$$

Other intermetallic anodes, such as Mg$_3$B, Mg$_2$Sn, Mg$_3$Bi$_x$Sb$_{1-x}$, Mg-Sb have also been investigated.

An exotic anode composed of layered Na$_2$Ti$_3$O$_7$/MgNaTi$_3$O$_7$/ Mg$_{0.5}$NaTi$_3$O$_7$ nanoribbons exhibited a reversible Mg$^{2+}$ insertion−extraction multi-step reaction with a practical capacity of 78 mAh g$^{-1}$. The MgNaTi$_3$O$_7$ anode was used in full Mg-ion batteries with Mg(ClO$_4$)$_2$–diglyme electrolyte and V$_2$O$_5$ cathode and demonstrated a reversible capacity of 75 mAh g$^{-1}$ corresponding to an energy density of 53 Wh kg$^{-1}$ [152]. In all, if the alloy anode can offer high stability towards conventional electrolyte solutions, this compatibility comes at the cost of high equilibrium potential and reduced specific capacity [149]. Various Mg-based alloys have been also studied for primary (non-rechargeable) seawater and air batteries [153]. Recently, arsenene (single-layer arsenic nanosheet) has been predicted to be a potential anode candidate for Li/Na-ion and Mg batteries. In the latter case, Arsenene can store Mg via adsorption with theoretical capacity 1430 mAh g$^{-1}$ and low voltage [154].

### 2.2.1.2.3. Electrolytes for Mg batteries

Electrolyte, along with the cathode, has been another hurdle on the way to an efficient rechargeable magnesium battery. The main requirements to the electrolyte are favorable electrochemical properties, fast magnesium conductivity, chemical stability towards electrodes, non-corrosive, non-toxic and potentially inexpensive composition. Reversible Mg deposition and dissolution do not occur in most polar organic electrolytes used in LIBs [90]. The reversible reaction can occur in Grignard electrolytes solutions in ethereal solvents (R-Mg-X, where R is an alkyl or aryl group, and X is Cl or Br). However, these are highly reducing and are unstable towards high capacity electrophilic cathodes [155]. A large effort has therefore been devoted to developing compatible electrolyte-electrode chemistries.

First prototype batteries were based on electrolytes composed of ether solutions with Mg organo-borate or organo-aluminate complexes [90] and ethereal solutions of magnesium halo-alkyl aluminate complex [91]. The oxidative stability of magnesium organohaloaluminate electrolytes and the Coulombic efficiency have been gradually improved. Firstly, by tuning the ratio of organomagnesium to the Lewis acid, the DCC (dichloro complex) electrolyte was developed with higher oxidative stability of 2.2V *vs.* Mg and 100% Coulombic efficiency [91,100,156], though questioned at some point [157]. These electrolytes have demonstrated superior reversibility of Mg$^{2+}$ intercalation in particular with Chevrel-type cathodes and improved conductivity. However, the contradictions in the reported properties of DCC aroused the concerns that its synthesis was too complicated for practical use [100], and the electrochemistry was temperamental and dependent on strict conditions of synthesis and quality of the starting materials [155]. Moreover, higher oxidation stabilities were desirable. Substitution of the alkyl groups in DCC with aromatics led to the synthesis of all phenyl complex (APC) electrolyte allowing for increase in the oxidative stability to 3.0 V - 5 V vs. Mg [100,158]. Higher potentials are reached when AlCl$_3$ is substituted with aluminium triphenoxide [155,158], or fluoro-compounds added to the electrolyte [159,160]. Both solution and crystallized form of APC (Mg$_2$(μ-Cl)$_3$·6THF)(Ph$_n$AlCl$_{4-n}$), n=1, 2, 3, 4), appear to be electrochemically active [97,100,155,161]. Similar oxidative stabilities with 90-99% Coulombic efficiencies are attainable in all-inorganic electrolytes (in THF or glyme solutions) where MgCl$_2$ is used instead of organomagnesium [101,161]. Utilizing inorganic $MgCl_2$ instead of organomagnesium simplifies the synthesis and decreases overall costs. These are so-called MAAC (magnesium aluminum chloride complex) electrolytes. The electrolyte systems composed of $MgCl_2$–$AlCl_3$, $MgCl_2$–$AlPh_3$, and $MgCl_2$–$AlEtCl_2$, also demonstrated high oxidation stability (up to 3.4 V *vs.* Mg), improved electrophile compatibility and electrochemical reversibility (up to 100% Coulombic efficiency), and clean and dendrite-free Mg bulk plating [161]. The largest oxidative stability to date of 3.7 V for the electrolyte containing magnesium dimer was reported for the crystallized magnesium organoborate $(Mg_2(\mu\text{-}Cl)_3 \cdot 6THF)(B(C_6F_5)_3Ph)$ [162]. $MgCl_2$-ionic liquid electrolytes ($\delta$-$MgCl_2$ in 1-ethyl-3-methylimidazolium tetrafluoroborate ($EMImBF_4$) ionic liquid) have also shown promising electrolytic performance [163]. Non-nucleophilic electrolytes $(Mg_2(\mu\text{-}Cl)_3 \cdot 6THF)(HMDSAlCl_3)$ compatible with sulfur reduction cathodes have been also developed [97,164].

In all the electrolyte systems mentioned above, a halide plays a significant role enabling and/or facilitating Mg diffusion and/or intercalation. However, the corrosive nature of halides, in particular, chloride, has been pushing for alternative solutions. Starting from $1990^{th}$, halide-free boron-based electrolytes have been developed and investigated [155]. Magnesium organoborate $Mg(BBu_4)_2$ electrolyte demonstrates the oxidative stability of 1.9 V and a low overpotential. Trispentafluorophenylborane $(B(C_6F_5)_3)$ demonstrated the stability of 3.7 V *vs.* Mg [162]. Reversible Mg deposition and dissolution was demonstrated for magnesium(II) bis(-trifluoromethanesulfonyl)imide $(Mg(TFSI)_2)$ in glyme, but with high overpotential and a low Coulombic efficiency [165]. Mohtadi *et al.* [107] first demonstrated that the reversible Mg deposition/stripping with the cycling capability (4 cycles) of 128.8 mAh $g^{-1}$ and 94% coulombic efficiency was possible from electrolyte containing $Mg(BH_4)_2$ in dimethoxyethane (DME or glyme) with $LiBH_4$ additive, $Mo_6S_8$ anode and Mg metal cathode. Those results have also shown that for $Mg(BH_4)_2$, the electrochemical performance in DME is higher than that in THF by contrast to organomagnesium electrolytes [166], and that $LiBH_4$ additive significantly improves the electrochemical properties of the electrolyte. The oxidative stability of this electrolyte is close to 1.5 V *vs.* Mg. Other reports on $Mg(BH_4)_2$-based electrolytes emphasized the crucial effect of solvents and dopants on the electrochemistry [167-169]. Watkins *et al.* demonstrated the possibility to substitute the volatile and flammable solvents with ionic liquids [170]. They reported a fully inorganic and halide-free Mg electrolytes based on $Mg(BH_4)_2$ to show reversible Mg deposition and stripping with 90% Coulombic efficiency [170]. Zhao-Karger *et al.* [104] have developed chemically stable non-corrosive electrolytes based on $Mg(BH_4)_2$ and fluorinated alkoxyborate. The electrolyte demonstrated a high anodic stability, ionic conductivity and Coulombic efficiency. Incorporating larger boron cluster, such as carboranes ($CB_{11}H_{12}^-$), in the electrolyte with the final composition $Mg(CB_{11}H_{12})_2$/tetraglyme (MMC/$G_4$) demonstrated ionic conductivities around 1.8 mS $cm^{-1}$, stabilities of ~ 3.8 V, and a Coulombic efficiency of 94% in the first cycle [103]. Combining this electrolyte with a high-voltage cathode, such as $\alpha$-$MnO_2$, allowed for cell charging up to 3.5 V, thus marking the first time coin cells employing highly performing electrolytes to examine high voltage Mg-based cathodes. The cell demonstrated a reduction in the discharge capacity from 180 to ca. 90 mAh g$^{−1}$ after 10 cycles, which was a sound improvement over the APC electrolyte deactivating after the 1$^{st}$ cycle at this high voltage. The improved mobility of Mg$^{2+}$ ions was achieved by adding (NH$_4$)$^+$ ions to Mg(BH$_4$)$_2$ solutions [171]. Other hydride-based compounds for use as Mg-battery electrolytes have been recently reviewed [172]. Using ab initio calculations, nuclear magnetic resonance, and impedance spectroscopy measurements, Canepa *et al.* [102] argued a substantial (~ 0.01–0.1 mS cm$^{−1}$ at 298 K) magnesium ion mobility in close-packed frameworks, specifically in the magnesium scandium selenide spinel. They suggested that high magnesium ion mobility is possible in other chalcogenide spinels as well, enabling a realization of magnesium solid ionic conductors for all solid-state magnesium battery. Table 1 summarizes the performance of several most researched and/or promising magnesium battery configurations.

Table 1. Properties of selected rechargeable magnesium cell prototypes reported in literature

| Year | Composition anode / electrolyte / cathode (I or C) $^§$ | Properties | | | | Comments | Ref. |
|---|---|---|---|---|---|---|---|
| | | Operating $T$, $^o$C | Cathode capacity (1$^{st}$ – last cycle)/ mAh g$^{-1}$ | Operating voltage / V | Stability (cycles and/or CE$^*$) | | |
| 1990 | Mg / Mg(BBu$_2$Ph$_2$) in THF-DME / Co$_3$O$_4$ (I) | R$T$ | ca. 185 | 1.5 | 4 | low potential, high polarisation, low oxidative stability of the electrolyte | [90] |
| 2000 | Mg / THF/Mg(AlCl$_2$BuEt)$_2$ /Mo$_6$S$_8$ (I) | -20 to 80 °C | ca.90 – ca. 75 | 1-1.3 | 580 | low capacity, but long durability (up to 2000 cycles) | [91] |
| 2015 | Mg / Mg(BH$_4$)$_2$+LiBH$_4$ in tetraglyme / TiO$_2$ (I) | R$T$ | 168-148 | 0.9-1.1 | 100 | good stability and rate capability | [173] |
| 2016 | Mg / Mg(TFSI)$_2$ in DME/ Diglyme(1:1vol)+ Mg(TFSI)$_2$ -PYR$_{14}$TFSI(IL)$^#$-MgBr$_2$ / Br (C) | R$T$ | ca. 275 | 2.4-3.2 | 20, 95% | dual-electrolyte, few cycles only demonstrated | [174] |
| 2017 | Mg / Mg-HMDS / I$_2$ (C) | R$T$ | 180 | 2.2 | 120 | Absence of solid-state diffusion, suitable for semi-flow batteries | [98] |
| 2018 | Mg / Mg-Li dual-salt / Na$_2$C$_6$O$_6$ (co-I) | R$T$ | 450-125 (at various rates) | 1.1 | 600 | Multi-process intercalation, dominated by Li-ions | [175] |
| 2019 | Mg / [Mg(BH$_4$)$_2$]$_{0.3}$[N$_{07}$TFSI]$_{0.7}$-PYR$_{14}$TFSI(IL)$^#$/ V$_2$O$_5$ aerogel (I) | R$T$ | 100 - 80 | 1.4 – 1.8 | 40 | halide-free non-corrosive electrolyte; large capacity loss with cycling | [110] |

$^*$ CE: Coulombic efficiency (%)
$^§$ I: Intercalation cathode; C: Conversion cathode
$^#$ IL: Ionic liquid

## 2.2.2.  Ca-ion batteries: State-of-the-art

Calcium anode has a similar volumetric capacity of that of Li (2072 mAh cm$^{-3}$), and a similar potential of 0.17 V *vs.* Li [176]. Calcium abundance (exceeding that of magnesium) [87], lower charge density of Ca$^{2+}$ ions and superior safety over LIBs [177], have been fueling the efforts for a calcium battery (CAB). The polarizing power of Ca$^{2+}$ cations is in between those of Mg$^{2+}$ and Li$^+$ promising moderate interaction with solvent and intercalation host. Moreover, similar to Li/Li$^+$ the value of the standard electrode potential gives the prospect of high-voltage batteries, by contrast to RMB. The problems encountered with CAB are similar to those of RMB and other multivalent batteries, i.e., low diffusion rates of Ca$^{2+}$ and high reduction potential towards electrolyte, formation of passivation layers at anodic surface, and thus a challenge of finding the suitable cathodes, electrolytes, and steadily efficient compatible battery chemistries.

Metallic Ca anodes, similar to those of Mg, would offer superior volumetric and gravimetric capacities with respect to the graphitic anodes in Li-ion battery technology (2072 mAh cm$^{-3}$ and 1337 mAh g$^{-1}$ *vs.* 300–430 mAh cm$^{-3}$ and 372 mAh g$^{-1}$, respectively) [178]. However, the non-conducting surface layers rapidly formed on the electrolyte/anode interface upon discharge, appear to be detrimental for the calcium deposition. Depending on the electrolyte, these films can consist, for example, of Ca(OH)$_2$, CaCO$_3$, calcium alkoxides [179], CaF$_2$ [180]. The reversible deposition process is possible at elevated temperatures at ~100 °C [180].

It has been noted recently (2017) that "Ca-ion batteries currently remain a curiosity" [181]- despite the fact that early research started already in the 1980$^{th}$. The field is indeed at its initial stage with around fifty scientific reports appearing over the last three decades, most of them after 2010. The early reports can be traced back to the 1980$^{th}$ when the first studies suggested a calcium-thionyl chloride battery as a safer alternative to high-power lithium batteries [177]. The systems contained a calcium foil anode, a 7% Ba(AlCl$_4$)$_2$ or Sr(AlCl$_4$)$_2$ electrolyte solution in thionyl chloride, and a carbon intercalation cathode. The formation of passivation layers on calcium anode that deteriorated battery performance was noted. These anode surface layers were found to be formed in various organic electrolytes as well. The reversible plating and stripping of Ca from the anode was thought to be impossible in the 1990$^{th}$ [179], and the research onwards focused on the primary Ca battery [179,182]. Thus, See *et al.* reported a primary conversion-reaction Ca-S cell based on Ca anode, S-infiltrated mesoporous carbon cathode, and 0.5 M Ca(ClO$_4$)$_{(2)}$ in CH$_3$CN electrolyte. Discharge capacities of 600 mAh g$^{-1}$ at a discharge rate of C/3.5 were demonstrated [182]. The first reversible calcium electrodeposition was demonstrated only in 2016 by Ponrouch *et al.* [172]. The system contained the electrolyte composed of 0.45 M Ca(BF$_4$)$_2$ in a mixture of conventional polar aprotic solvents and could be cycled for 30 cycles at 50-100 °C. No electrochemical activity was observed at room temperature. By suggesting an electrolyte for the reversible Ca electrodeposition, this work has opened the way for optimizing electrolyte chemistries and testing various Ca cathodes for the rechargeable CAB. The reversible plating and stripping of $Ca^{2+}$ at room temperature was achieved in the system based on $Ca(BH_4)_2$ in tetrahydrofuran (THF) [168]. The capacities of 1 mAh cm$^{-2}$ at a rate of 1 mA cm$^{-2}$, with low polarization (close to 100 mV) and more than 50 cycles were demonstrated. In this system, a small amount of $CaH_2$ was found to form by reaction between the deposited calcium and the electrolyte, protecting the calcium metal anode at open circuit. Wang *et al.* [183] reported the reversible CAB in a new cell configuration with graphite as the cathode, tin foils as the anode as well as the current collector, and the $(PF_6)^-$ anions counterpart based electrolyte operating at R*T*. This system demonstrated a working voltage of up to 4.45 V with capacity retention of 95% after 350 cycles. The hexafluorophosphate (de)intercalation at the cathode and the Ca-involved (de)alloying reaction at the anode were suggested to cause the outstanding battery performance. At the same time, Wu *et al.* demonstrated a system with a reversible discharge capacity of 66 mAh g$^{-1}$ at a current rate of 2 C with a high working voltage of 4.6 V. The stability over 300 cycles with a high capacity retention of 94% and the final discharge capacity of 62 mAh g$^{-1}$ was demonstrated [184]. In this Ca-ion full battery both electrodes were carbon-based intercalation type layered structures, where $Ca^{2+}$ was plated and stripped reversibly to/from the $Ca(PF_6)_2$ electrolyte in carbonate solution (a mixture of ethylene carbonate (EC), dimethyl carbonate (DMC) and ethyl methyl carbonate (EMC)). Similar to the system of Wang *et al.*, $(PF_6)^-$ was intercalated in the cathode in parallel to $Ca^{2+}$, realizing the "dual graphitic carbon intercalation chemistry".

The insertion of $Ca^{2+}$ into the crystalline $V_2O_5$ cathode with a discharge capacity of ~ 450 mAh g$^{-1}$ was demonstrated but only at very low current densities of 50 μA cm$^{-2}$ at R*T*. The intercalation causes the formation of a new phase coexisting with the pristine $V_2O_5$ phase [185]. Bervas *et al.* have achieved the specific capacity of 465 mAh g$^{-1}$ for $Ca^{2+}$ intercalation in the vanadate nanocomposite with propylene carbonate (PC). Two to ten times better performance over the conventional $V_2O_5$ structure owing to the presence of PC was demonstrated [186]. $CaMO_3$ perovskites (M = Mo, Cr, Mn, Fe, Co, Ni) were shown to be unsuitable as Ca-intercalation cathodes [187].

Lipson *et al.* first demonstrated the feasibility of $Ca^{2+}$ intercalation into layered fluorinated sodium iron phosphate in 0.2 M $Ca(PF_6)_2$ / EC-PC (3:7) electrolyte [188]. They obtained an initial capacity of 60 mAh g$^{-1}$ which actually increased to 80 mAh g$^{-1}$ after the 50$^{th}$ cycle at the average voltage of 2.6 V vs. $Ca/Ca^{2+}$. The authors suggest that diffusion limits the intercalation process of calcium into $Na_2FePO_4F$. A recent report has shown that reversible Ca plating and stripping in conventional alkyl carbonate electrolytes at moderate temperature is feasible, prompting moderate optimism [180]. Yet promoting diffusion of large divalent ions in intercalation hosts implies great challenges. Overall, and despite the outstanding fundamental studies, the development of Ca-based rechargeable battery in the twenty-first century seems unlikely, unless the challenges are addressed through new creative approaches by battery chemists [189].

We *et al.* found that the a mixture of ethylene carbonate (EC), dimethyl carbonate (DMC) and ethyl methyl carbonate (EMC) with $Ca(PF_6)_2$ exhibits superior performance as electrolyte, where higher amount of EC is beneficial for improved solubility of $Ca(PF_6)_2$ [184].

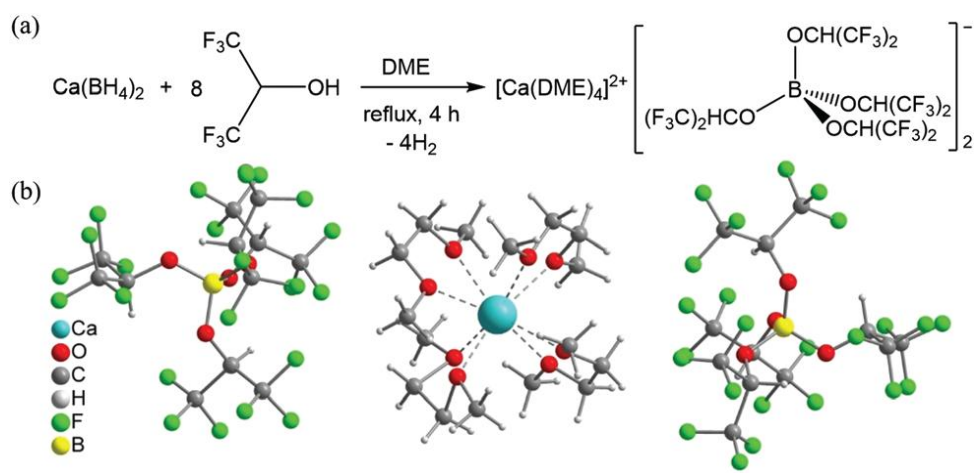

Figure 6.
Synthesis procedure (a) and single-crystal structure (b) of $Ca[B(hfip)_4]2·4DME$ [190].

Similar to the RMB system, an effort has been made to overcome the extreme sensitivity of the anode surface. Theoretical investigations have shown that hydrogenation of graphite can enhance intercalation of Ca generating an electrical capacity of 591 mAh $g^{-1}$ [191]. Yao *et al.* [192] investigated the electrochemical calcium deposition using density functional theory (DFT) calculations. Their work suggests that many metalloids (Si, Sb, Ge) and (post)transition metals (Al, Pb, Cu, Cd, $CdCu_2$) can be promising anode candidates for CAB.

Recently, a breakthrough has been achieved by using a new electrolyte which can reversibly strip and plate Ca at room temperature. The system is easy to synthesize in various solvents, it has a high oxidative stability up to 4.5 V and a high ionic conductivity of >8 mS $cm^{-1}$ [190]. The salt is constituted of calcium tetrakis(hexafluoroisopropyloxy)borate $Ca[B(hfip)_4]_2$, which can be directly synthesized in a one-step reaction from $Ca(BH_4)_2$ and the isopropylate, with $H_2$ as the only by-product (Fig. 6a). The single crystals have been isolated from the DME solution and analyzed using X-ray crystallography (Fig. 6b). The crystal unit consists of the counter anion $[B(hfip)4]^-$ bonded with four hexafluoroisopropyloxy groups with a tetrahedral geometry. The $Ca^{2+}$ ion is solvated with four DME molecules. Due to the larger size of $Ca^{2+}$ ion compared to $Mg^{2+}$ ion, and the weaker O-interaction, it is assumed that the desolvation energy for $Ca^{2+}$ ion can be lower than that of $Mg^{2+}$ ion, which is beneficial for the intercalation mechanism [190]. The high ionic conductivity is possible due to the weak interaction of the large fluorinated anion with the calcium, thereby enhancing the mobility of the cation. The new electrolyte seems to be versatile with a number of host materials and can enable further progress in the field.

### 2.2.3. Overview of Zn batteries

Rechargeable aqueous Zinc-ion batteries (ZIBs) are considered as an easy to realize alternative battery chemistry. The overall drawbacks and the high cost of batteries requiring use of inert atmosphere, such as LIBs/SIBs with toxic and flammable electrolytes based on carbonate solvents, motivate elaboration of novel research ideas regarding alternative chemistries and simplified large-scale manufacturing, where cost, safety, long-cycle life and recycling has to be reconsidered [5,6,193].

Furthermore, aqueous electrolytes have a higher ionic conductivity (~1 S cm$^{-1}$) as compared to non-aqueous electrolytes (~0.01 S cm$^{-1}$). The ZIBs operating in aqueous media offer potential applicability in grid-scale related energy storage [194,195]. Different battery configurations have been tested and reported in literature [196,197]. Efforts have been made for further improvements and detailed mechanistic studies have been reported for rechargeable alkaline Zn-MnO$_2$ batteries, which suffered from the formation of Zn dendrites and irreversible discharge capacities [198-200].

Indeed, many research works have been published dealing with aqueous ZIBs, which include use of Zn anode and various cathode materials, such as manganese and vanadium-based oxides, Prussian blue analogs as well as polyanion compounds etc [194,201-205].

The mechanism of the energy storage in aqueous ZIBs is not straightforward as the system acts in slightly acidic aqueous medium. For instance, many compounds with tunnel-type and layered structure enable the insertion/extraction of Zn$^{2+}$ ions [194], according to the following reactions in the presence of diluted ZnSO$_4$ or Zn(NO$_3$)$_2$:

Cathode:   $Zn^{2+} + 2e^- + 2\alpha MnO_2 \leftrightarrow ZnMn_2O_4$   (1)

Anode:   $Zn \leftrightarrow Zn^{2+} + 2e^-$   (2)

Other polymorphs of MnO$_2$ may show more complicated electrochemistry, with multi-step phase transitions (e.g. from tunnel-type to layered structures owing to the expansion of the structure) [206,207].

Later studies have included the work on Zn/vanadium oxides batteries [208].

Indeed, V-based cathodes for ZIBs offer more stability and various V oxidation states compared to Mn-based cathodes. Furthermore, V-O coordination can adopt different polyhedral units, including tetrahedron, trigonal bipyramid, square pyramid, distorted octahedron and regular octahedron, which can change based on the V oxidation state [209,210]. Different vanadium oxide frameworks can be constructed by corner and/or edge sharing of these polyhedra, for an eventual reversible $Zn^{2+}$ (de)intercalation. Fig. 7 enumerates the V-based cathodes used for ZIBs and compared to $V_2O_5$, upon the measured electrochemical performance. The number of inserted Zn-ions seems to depend on the electrolyte system (i.e. solvent and salt conc.). Compared to Mn, several V-based cathodes approach the specific capacity 400 mAh g$^{-1}$. Particularly, the cathode $Zn_{0.25}V_2O_5 \cdot nH_2O$ has been studied in more details, regarding the electrochemical properties and structure, which can be modified by intercalation of water molecules. The presence of water seems to facilitate the $Zn^{2+}$ intercalation owing to the expanded structure. The electrochemical behavior is highly reversible where 1.1 $Zn^{2+}$ are exchanged to form $Zn_{1.35}V_2O_5 \cdot nH_2O$. The mechanism is rather complex which may include multi-step reaction pathways [208,211].

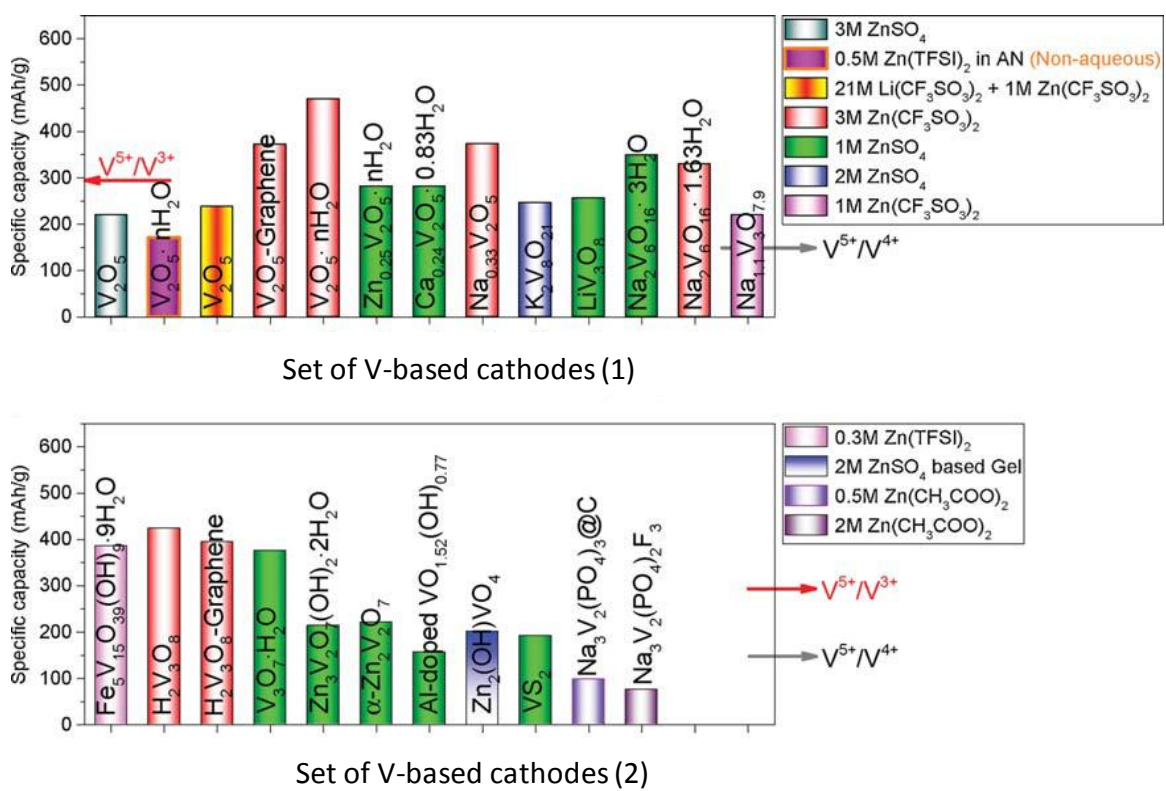

Figure 7.

Specific capacities of a series of V-based cathode materials studied for application in ZIBs. The utilized aqueous electrolyte (except indication) is given for comparison. Adapted from Ref. [210].

More recently, aqueous batteries made of $Zn_{0.3}V_2O_5 \cdot 1.5H_2O$ hierarchically porous cathode, 3M $Zn(CF_3SO_3)_2$ aq. electrolyte and Zn anode are designed and fabricated. Upon cycling, the cathodes undergo phase transition to form hierarchical cathode nanoflowers morphology, providing abundant surface contact between electrode and electrolyte as well as active Zn storage sites [212]. Likewise, the presence of crystal water has been highlighted to contribute to the stabilization of the host structure, and the $Zn^{2+}/H^+$ ion insertion preserves the $V_2O_5$ interlayer spacing, benefiting long-term cycling stability. This aqueous battery demonstrated high specific capacity of 426 mAh $g^{-1}$ at 0.2 A $g^{-1}$, a specific energy of ~268 Wh $kg^{-1}$ at 1400 W $kg^{-1}$ (cathode only), and a long-term cycling stability with 96% capacity retention after 20,000 cycles at 10 A $g^{-1}$. Taking into consideration the electrochemical performance, safety of aqueous electrolyte, low-cost electrode material, the battery could be promising for grid-scale energy storage applications.

### 2.2.4. Al batteries: Electrolyte challenges

Available Al batteries are consisting of Al and pyrolytic graphite/carbon. These components are relatively cheap , except the battery makes use of room temperature ionic liquids which are non-flammable [213]. High power has been demonstrated in a such configuration with 7,500 cycle lifespan. However, the battery is still at the early stage of development as it shows low energy density, due to the difficulties related to the intercalation of large cations such $Al^{3+}$ into the host cathode material. As a consequence, the application domain is restrained and the initial cost is rather high. In the following, we will present the most relevant results for future applications.

The first Al-ion rechargeable battery was demonstrated in 2011 [214]. The battery used $V_2O_5$ nanowires as the cathode and Al metal as anode in ionic liquid-based electrolyte. The Al-ion battery demonstrates a well-defined $Al^{3+}$ intercalation plateau at 0.55 V (*vs*. $Al^{3+}$/Al). In the first cycle, it exhibits a capacity of 305 mAh $g^{-1}$ against 273 mAh $g^{-1}$ at the end of 20 cycles. Another Al-ion battery consisting of a fluorinated natural graphite nanosheet as cathode was obtained with a charge capacity of approximately 300 mAh $g^{-1}$, but the columbic efficiency of the cell (75%), though stable during 40 cycles, is abnormal [215]. Recently, an Al ion battery with 3D graphitic-foam as the positive electrode was presented [213]. The cell exhibits well-defined discharge voltage plateaus near 2V (*vs*. $Al^{3+}$/Al) and long cycle life up to 7500 cycles at ultrahigh current densities. Although the beneficial properties of ionic liquid-based electrolytes (chloroaluminate) of the above Al-ion batteries, the cost is still high.
An aqueous rechargeable aluminum battery was assembled using these graphite nanosheets as the positive electrode and zinc as the negative electrode in $Al_2(SO_4)_3$/$Zn(CHCOO)_2$ aqueous electrolyte [216]. This battery could be rapidly charged and discharged at a high current density. The average charge and discharge voltages are 1.35 and 1.0 V, respectively. The graphite nanosheets show a discharge capacity of 60 mAh $g^{-1}$ even at 2 A $g^{-1}$. In comparison to rechargeable Al-ion batteries in ionic liquid electrolytes, one must highlight the high working voltage and stable cycling behavior (over 200 charging/discharging cycles) at low cost using aqueous electrolytes. One of the current issues of this kind of aqueous rechargeable aluminum battery is that it needs highly concentrated Al salt electrolyte to obtain high energy density; which would lead to high acidity in the electrolyte (hydrolysis of $Al^{3+}$ ions), being aggressive/corrosive to the negative electrode (Zn).

## 3. Anionic shuttles

### 3.1. F-ion battery: a novel solid-state battery

The fluoride ion has a wide electrochemical stability window of 6 V and it is very stable for charge transfer in a battery owing to its high electronegativity. The transfer of the fluoride ion between two electrodes enables reversible electrochemical storage. In a simple concept, the ion shuttles between two metal electrodes and the difference in the free energy of formation of the respective halide results in an electromotive force. Due to the theoretical voltages in the range of 1-3 V, due to the change of several oxidation states of the respective metal and due to the high densities of metal fluoride, high theoretical energy densities are possible. The reactions at the electrodes during discharge are as follows:

$$\text{Cathode: } xe^- + MF_x \rightarrow M + xF^-$$
$$\text{Anode: } xF^- + M' \rightarrow M'F_x + xe^-$$

The first reversibly working F ion battery based on $BiF_3$ cathode and Ce anode was demonstrated by Anji Reddy *et al.*, using a solid state electrolyte based on a tysonite structure [217,218]. However, although tysonites are "superionic conductors" for fluoride, the ionic conductivity is still low at room temperature ($10^{-7}$ S cm$^{-1}$) so that the battery was operated at 160 °C, one reason was the electrolyte layer which was thick (600 μm) for mechanical stability reasons. Follow-up work focused on understanding the role of interfaces in the ionic transfer in bulk materials and on improvement of the overall ionic conductance, e.g. by using thin film electrolytes and sintered materials [219]. Recently, progress was made by the development of a first battery working at room temperature [220,221]. This was possible due to a novel electrolyte based on ternary fluorides such as $BaSnF_4$ and $BiSnF_4$ which offer higher ionic conductivities, in the order of $10^{-1}$ S cm$^{-1}$. Fig. 8a shows the cyclic voltammograms (CVs) of the Sn/$BaSnF_4$/$BiF_3$ cell in the potential range of 0.05 to 1.2 V at 25 °C. A cathodic peak appeared at 0.1 V which corresponds to the reduction of $BiF_3$ to Bi metal, while in the forward scan, a broad anodic peak at 0.58 V was observed that corresponds to the oxidation of Bi metal to $BiF_3$. Notable cycling performance (Fig. 8b-8f) has been demonstrated at moderate temperatures (R$T$-150°C). The cell cycled at R$T$ delivered a first discharge capacity of 120 mAh g$^{-1}$ at a current density of 10 μA cm$^{-2}$. The low capacities compared to the theoretical specific capacity of $BiF_3$ (302 mAh g$^{-1}$) were attributed to incomplete conversion of $BiF_3$ to Bi. Attempts to build secondary batteries based on liquid electrolytes have not been convincing, so far. Major problems are the shielding of the F-ion which attracts protons from electrolyte components and forms HF, or the reversibility in general, as the cells are fading rapidly after a few cycles.

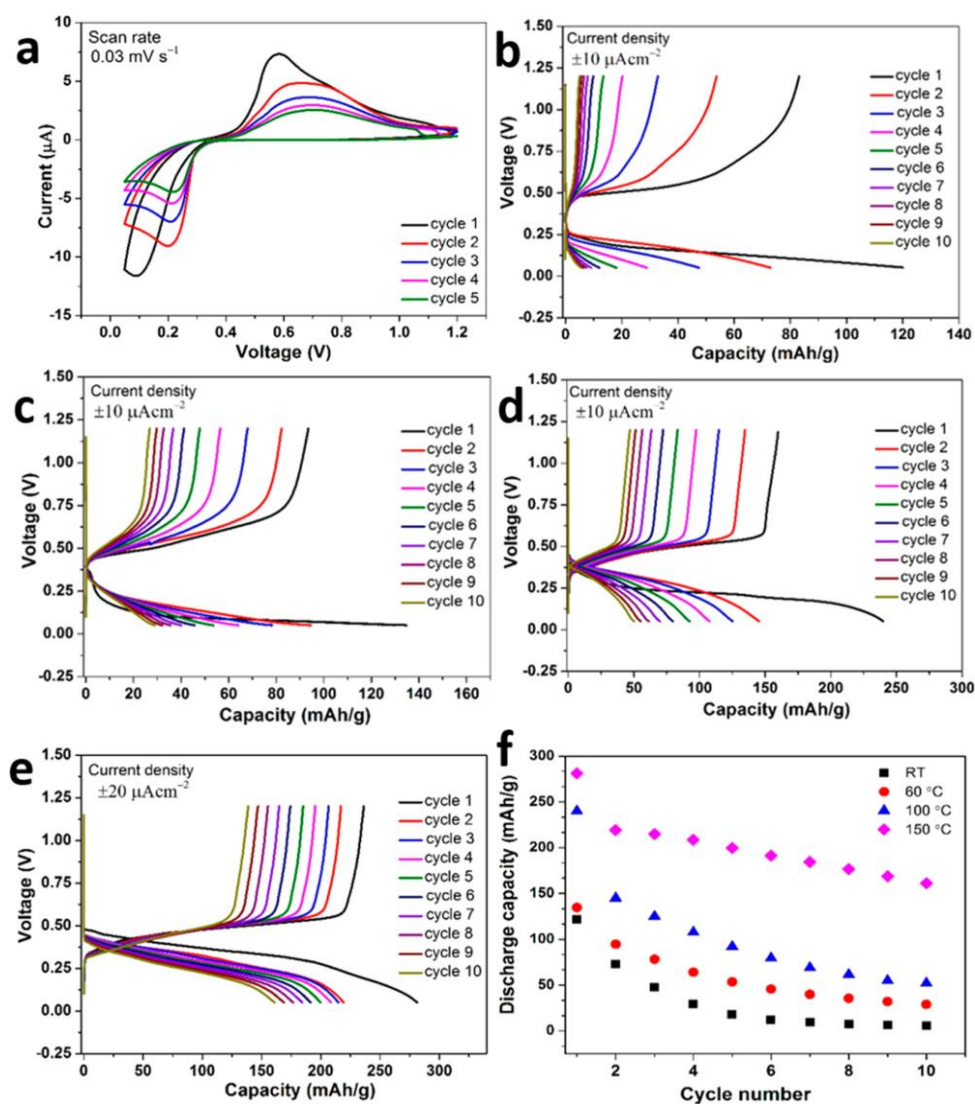

Figure 8.

Electrochemical properties of the F-ion cell configuration Sn/BaSnF$_4$/BiF$_3$ as function of the operating temperature: (a) cyclic voltammetry at 25 °C, galvanostatic cycling at (b) 25 °C, (c) 60 °C, (d) 100 °C, and (e) 150 °C. (f) Cycling performance of the corresponding cells [221].

## 3.2. Cl-ion battery: Early stage of development

The Chloride Ion Battery is conceptually similar to the Fluoride Ion Battery, just that the Cl⁻ replaces the fluoride ion. A series of metal chloride/metal combinations demonstrate theoretical energy densities above those of the current LIBs which makes them attractive [222]. The evident advantage of this battery system lies in the fact that active material electrodes can be built from abundant material resources and it is possible to use various metals such as Li, Na, Mg, Ca, and Ce, as well as their corresponding chlorides. Rechargeable batteries are comparably easy to realize because a liquid electrolyte transporting chloride ions can be made from a mixture of suitable ionic liquids with chloride ion and large cation and other solvents. Such systems have been demonstrated to work at room temperature and tested at comparably high rates, up to 2C [223].

A key challenge in this concept is the solubility of many chlorides in the electrolyte which limits the application of, e.g. $FeCl_3$ as electrode material. Here, the development goes into the direction of using chloride hosts with low solubility, oxychlorides for example. Another approach would be the development of solid electrolytes with high Cl-ion conductivity. Solid inorganic compounds such as $PbCl_2$, $SnCl_2$, and LaOCl show fast chloride mobility only at high temperatures, even higher than the melting points of some metal chlorides [222,224]. More promising ionic conductivity (1 mS cm$^{-1}$, 100 °C) has been reported for the cubic phase $CsSnCl_3$. However, the electrochemical properties for suitability in the battery system needs to be studied further [225,226].

## 4. MH-based Batteries

### 4.1. Ni$M$H batteries

Metal hydrides are extensively studied for their properties to store reversibly hydrogen gas making them suitable for solid state storage tanks. Beside the solid-gas route, these materials can also serve as suitable anodes for rechargeable alkaline Ni$M$H batteries [227-231]. Indeed, they have led to practical applications to power light electronic devices or emergency light units. Recently, they found a new market as auxiliary storage units in hybrid cars (either ICE or FCEV). Despite lower efficiency compared to lithium batteries, Ni$M$H are cheaper, they offer much more safe operation in case of overheating, and they support high rate (dis-)charge currents. Current batteries developed by Panasonic store about 1.3 kWh for c.a. thirty to forty kg of a battery cell with a lifetime of 8 years. A large scale Battery Power System for railways has been recently developed by Kawasaki Heavy Industries which provides a stable discharge operation at a current density of 3$C$ and a peak current density of 20$C$ and instant regenerative breaking charge performance for a system built from 5.6 kWh stacks. Such a system has been used for the electricity driven trains [232] both in Japan and internationally. The main benefits of the system include energy saving, peak shaving, line voltage stabilization together with safety of operation and absence of thermal runaway.

The Ni$M$H batteries work in alkaline medium using concentrated KOH as electrolyte. The following reactions take place during discharge for the cathode (1) and the anode (2):

Cathode ($E° = +0,49$ V): $x$Ni(OH)$_2$ + $x$OH$^-$ ↔ $x$NiOOH + $x$H$_2$O + $x$e$^-$ (1)

Anode ($E° = -0,83$ V): $M$ + $x$H$_2$O + $x$e$^-$ ↔ $M$H$x$ + $x$OH$^-$ (2)

First generation of anodes were mainly based on LaNi$_5$-type intermetallics that have been finely modified regarding their chemical compositions to address thermodynamic stability and resistance to corrosion in highly concentrated KOH [228]. This was achieved by substitution of nickel by other elements like manganese, aluminum, cobalt or iron, giving materials ability to sustain thousands of cycles with specific capacities exceeding 300 mAh g$^{-1}$. Next generations of alloys are now based on so-called intergrowth phases. They consist in $RT_y$ ($R$: Rare earth or Mg; $T$: Transition metals; 2<$y$<5) stacking structures made of [$R_2T_4$] and [$RT_5$] sub-units. Khan [233] describes these structures using the general formula: $y=(5n+4)/(n+2)$ (where $n$ is the number of [$RT_5$] units). These phases are usually polymorphic, depending on the repetition of the sequence $n[RT_5]+[R_2T_4]$ along the $c$ axis; either three times for rhombohedral or two times for hexagonal structure. A decisive breakthrough was achieved for these materials by developing ternary Mg-containing materials. Indeed, part of the rare earth can be substituted, exclusively in the [$R_2T_4$] sub-units, by magnesium [234-236]. In this way, amorphization easily induced by repeated hydrogenation of the binary phases can be reduced. In addition, stabilization of multi-plateau pressure features commonly observed in $RT_y$ binary hydrides can be lessened into one single plateau stabilized in the practical pressure range. Consequently, working anode materials with significant reduction of the molar mass are obtained leading to weight capacity enhancement up to about 400 mAh g$^{-1}$ [236-238].

Complex multistep transformations taking place during the heating of La-Mg-Ni alloys proceed between 700 and 1050 °C [239] and show that to synthesize single phase intermetallics of a particular type, a strict control over the temperature and content of easily sublimated Mg is required. Rapid solidification when its conditions are properly selected allows both increase the content of Mg-containing layered structures and also to synthesize nanosized alloys [240]. Hydrogen storage and electrochemical performance (Fig. 9) with capacity reaching 420 mAh g$^{-1}$ can be optimized by (a) selecting stoichiometry – AB$_3$ or A$_2$B$_7$- and type of the alloy's structure; (b) changing the ratio between rare earth metal and magnesium (optimal compositions are close to A$_2$MgNi$_9$ and A$_3$MgNi$_{14}$) and type and content of rare earth metal – La, Pr, Nd, Sm, Gd; (c) applying nanostructuring via rapid solidification; (d) by in situ studying and accounting the mechanism of phase-structural transformations during electrochemical formation and decomposition of the hydrides on charge and discharge of the MH electrode; (e) by modelling the processes of hydrogen exchange in the electrodes on their charge and discharge [241-248].

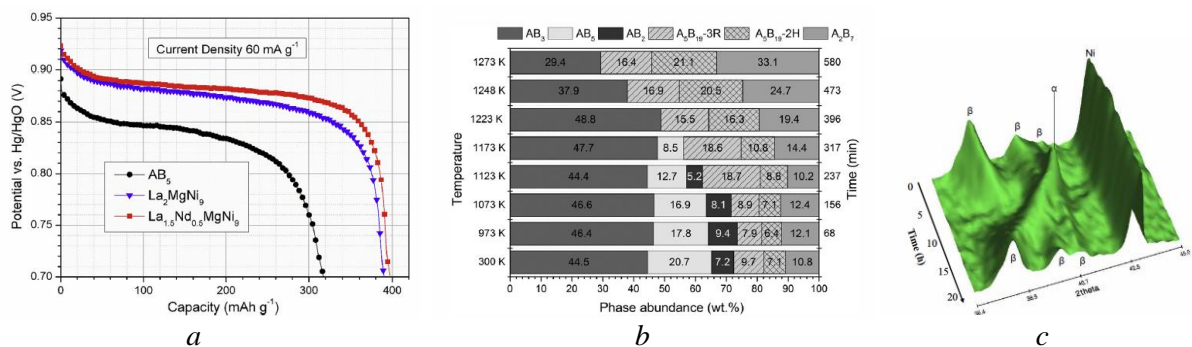

*a*          *b*          *c*

Figure 9.

Discharge capacity of the $La_{1.5}Nd_{0.5}MgNi_9$ and $La_2MgNi_9$ alloys showing their superior performance as compared to the commercial AB5 alloys and improvement of the characteristics on La substitution by Nd because of faster H diffusion in the alloy [245] (a); Complex temperature-dependent phase-structural transformations in the $La_{1.5}Nd_{0.5}MgNi_9$ alloy causing gradual disappearance of the electrochemically inactive AB5 type intermetallic – in situ neutron powder diffraction data [245] (b); Evolution of the NPD pattern of the $La_2MgNi_9$ anode studied by *in-situ* neutron diffraction during its charge and discharge [242] (c).

For a drastic increase of the capacity, the light Mg-$H_2$ system remains a key one but the slow hydrogenation kinetic, the poor resistance of Mg to corrosion and the high thermodynamic stability of $MgH_2$ remains an issue [249]. Corrosion resistance can be improved by suitable surface engineering such as Nafion coating [250]. A step forward is foreseen by developing richer Mg-based materials using nanoscaled structures [251,252]. Long Period Stacking Order (LPSO) phases can be prepared to form a serie of Mg and $(R,Mg)T_y$ layers stacked along the *c*-axis, in a very similar way to the $RT_y$ compounds mentioned above, expecting higher capacities.

Light weight intermetallic compounds, such as $AB_2$, $AB$ and V-based BCC solid solutions likely offer the best though still challenging compromise between high capacity and resistance to corrosion in alkaline media. Progress in $AB_2$ and V-based compounds have been extensively considered in recent reviews [253]. These compounds offer high discharge capacities reaching 400 mAh $g^{-1}$ but sometimes suffer from poor activation while their cost should be optimized.

Recent studies of the Zr- and Ti- containing Laves-type intermetallics (Fig. 10) showed their excellent high rate performance with electrochemical storage capacities exceeding 420 mAh $g^{-1}$ and possibilities to optimize electrochemical behaviors as related to a) type of structure - $C15/C14$; b) ratio between Zr and Ti; c) selection of chemical composition of *B* (Mn, Ni, Fe, V, Sn, Al) and B/A ratio; d) Presence of catalytic additive – small amounts of La providing easy activation of the alloys; e) Metallurgical route of alloy's preparation with benefits of increased H diffusion rates created by rapid solidification processing of the alloys [254-259].

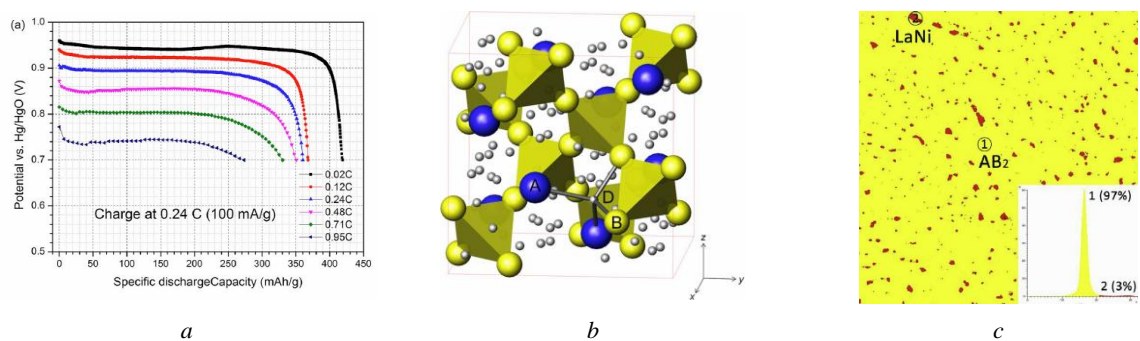

*a*  *b*  *c*

Figure 10.
Discharge performance of the C15 type $Ti_{0.2}Zr_{0.8}La_{0.03}Ni_{1.2}Mn_{0.7}V_{0.12}Fe_{0.12}$ alloy showing maximum discharge capacity of 420 mAh $g^{-1}$ [256] (a); Crystal structure of $AB_2D_{2.9}$ deuteride of the C15 alloy studied by neutron powder diffraction with H/D atoms filling $A_2B_2$ tetrahedral interstices [256] (b); Easy activation of these types of alloys is achieved because of presence of a secondary LaNi intermetallic (3 %) catalyzing hydrogenation of the main Laves type intermetallic (97 %) [257] (c).

As for *AB*-compounds, they are considered as attractive anode materials since early studies in Ni*M*H batteries [260]. Being exempt of scarce and expensive rare earths, they have gained renewed interest in recent years [261-264]. Research works have been focused on $A$ = Ti and $B$ = Fe, Co and Ni elements. Within this family, TiFe might be considered at a first sight as the most interesting alloy since it has the highest hydrogenation capacity (2 hydrogen atoms by formula unit (H/f.u.) at normal thermodynamic conditions. Potentially, it can provide as high as 500 mAh $g^{-1}$ in electrochemical units. Unfortunately, due to surface passivation, TiFe is not electrochemically active in alkaline media, a fact that can be partially solved by partial replacement of Fe by Ni [265]. Binary TiCo has electrocatalyic activity but suffers from severe Co corrosion [266]. Finally, the best choice turns to be TiNi as this compound has excellent electrocatalytic activity as well as good resistance to corrosion in alkaline media, both being attributed to Ni [260,267,268]. However, TiNi has a modest electrochemical capacity of 150 mAh $g^{-1}$ at *C*/10 regime, much lower than its hydrogenation capacity at normal thermodynamic conditions: 1.4 H/f.u. (i.e. 350 mAh $g^{-1}$ in electrochemical units). In fact, most hydrogen stored in TiNi is too tightly bonded for being extracted electrochemically. Chemical substitutions have been chased to tackle this issue.

Emami et *al.* have studied $TiNi_{1-x}B'_x$ alloys with B' = Co [269] and Cu [270] and $0 \leq x \leq 0.5$. Neither Co nor Cu substitutions significantly modify the hydrogenation capacity of TiNi alloy. However, they drastically affect the shape and stability of Pressure-Composition-Temperature (*PCT*) isotherms. For B' = Co, *PCT* isotherms have a multi-plateau behavior which is beneficial to enhance the cycle-life of TiNi-based electrodes at low substitution contents (x ≤ 0.3). At higher substitution ratios, significant capacity decay attributed to Co corrosion is observed. Cu-substitution is even more interesting for eventual applications. It enlarges the cell volume of TiNi but, contrary to geometric expectations, it leads to significant destabilization of the TiNi hydride. As a consequence, hydrogen bonding to TiNi is weakened and the experimental capacity can be significantly raised. By replacing 20% of Ni by Cu, the reversible capacity doubles: 300 mAh $g^{-1}$ at *C*/10 regime. Furthermore, $TiNi_{0.8}Cu_{0.2}$ electrodes have easy activation and good kinetics up to *C*/2 regime.

## 4.2. LIBs accommodating metal hydrides as anodes and electrolytes

Beside aqueous alkaline systems, LIBs are also developed with the aim of increasing the energy density of the electrodes. Most of the current anodes use Li intercalation into carbonaceous materials, mainly graphite. Though efficient in terms of cycle life (with formation of a stable SEI) and kinetics, these materials are intrinsically limited in capacities (typically 370 mAh g$^{-1}$ for graphite). To improve the energy density, other reactions are foreseen like alloying with highly capacitive elements (Si, Sn) or conversion reactions [271]. For the latter, metallic hydrides have shown over the past decade that they offer an interesting alternative to other materials [272-276]. This was first demonstrated by Oumellal *et al.* [277] that established the following conversion reaction:

$$M\text{H}x + x\text{Li}^+ + x\text{e}^- \leftrightarrow M + x\text{LiH} \qquad (3)$$

Using metal hydrides brings several advantages: low potential for anodes, low polarization (the lowest compared to other binary *MYx* compounds such as oxides, phosphides, nitrides…) and very high capacities. This latter parameter depends on both the *M* molar mass and the H content *x*. Again, Mg-based hydrides have been studied due to the light weight of Mg, its abundance and low cost. From the early works, it was observed that the reaction is thermodynamically favorable (as $\Delta G_f(\text{MgH}_2)/2 > \Delta G_f(\text{LiH})$) but kinetically sluggish. This can be overcome by a cautious nanostructuring of MgH$_2$ [277,278] and by tuning the composition of the composite electrode (by using additives of carbon and polymer binder, like carboxymethyl cellulose) [279,280] to handle the poor electronic conductivity of the hydride and large volume changes (84%) upon lithiation. Despite these efforts, even if the lithiation is straightforward, delithiation remains very challenging at room temperature for this hydride. For a better understanding of the limiting factors, the reaction was studied in 2D geometry using a 1 µm thick MgH$_2$ thin film [281]. This model system showed that the lithiation is indeed fully completed at the first cycle with a doubling of the film thickness, while only 25% of delithiation could be achieved. From TEM images it was shown that the cohesion of the film was preserved. Resistivity measurements indicate that the formation of metallic Mg increased the conductivity of the film. Therefore, the poor delithiation rate of the film was assigned to the kinetic limitations related to poor mass transport at room temperature. To overcome this drawback, formulation of composite materials can be valuable. As an example, mixing different amounts of MgH$_2$ and TiH$_2$ brings new insights into the reaction mechanism.

Several groups [282-284] investigated these mixtures of Mg and Ti hydrides as anode materials. Again, lithiation was straightforward but Li extraction was strongly related to the ratio between both hydrides. Depending of the composition, reformation of both hydrides (MgH$_2$-rich mixture), partial reversibility of TiH$_2$ only (equimolar mixture) or full irreversibility (TiH$_2$-rich mixture) of the composite are observed. A cooperative effect between the two phases is clearly observed and, interestingly, good reversibility and lower polarization are obtained for the conversion reaction of the TiH$_2$ phase when cycled in a Mg/LiH matrix. This has been correlated to the modification of the interfaces between the active species (LiH, TiH$_2$ and Mg) and a better volume change adaptation. Other strategies remain opened to improve the efficiency of these Mg-based materials, especially by alloying with transition metals [285-289] as well as by increasing the operation temperature [289-292]. The pure and Li-doped NiTiH systems have been also studied using DFT and MD calculations and predicted to be suitable anode for LIBs [293]. Clear evidence of practical improvements has been highlighted regarding the electrochemical capacity, minor increase in the voltage and volume change for the Li-doped NiTiH system.

Overall, to discard any issue with liquid electrolytes during voltage solicitation, the use of solid-state electrolytes based on $LiBH_4$ have been investigated at high temperatures (R$T$-120°C) with significant improvements of the reversibility and cycling performance of the hydride anodes [290,292,294-296]. A detailed review has been published recently on the contribution of hydrides to solid-state batteries [274]. The study highlights the possibility of next generation LIBs with high capacity and energy density where safer solid-state electrolyte is used instead of a carbonate-based liquid one.

# 5. Discussion and major challenges ahead

## 5.1. Cationic shuttles

**Na-ion batteries.** Significant exploits and rapid progress beyond LIBs have been made in the last decade, as evidenced by the large number of publications reviewed in this work. The intensive research on SIBs can be attributed to the similarity between LIBs and SIBs in their chemistries, mechanistic properties and manufacturing processes. Depending on application area, the developed cathodes and anodes for SIBs can now compete with the performance of a classical LIB. In portable devices, the Li-ion batteries may provide 3.8V and a specific energy of 408 Wh kg$^{-1}$ with 200 cycles or more (electrodes only; e.g. $LiCoO_2$ and graphite anode). This performance is gradually approached by SIBs when combing $Na_{1.5}VPO_{4.8}F_{0.7}$ cathode and nanostructured carbon anode or tin anodes, to make a SIB with 210 cycle life and a specific energy ~300 Wh kg$^{-1}$. If some cathodes could offer the possibility of the insertion of more than one Na ion and a specific capacity higher than 150 mAh g$^{-1}$ [297,298], more research development of these cathodes at the cell level need to be pursued.

Though tin anode can be a choice for substituting to carbon, however, the molar mass will affect significantly the gravimetric energy density. In addition, the alloying-type anodes (Sn, Si, Sb, .. etc) present the issues of low cycle stability due to large volume expansion, resulting in pulverization and loss of the electrical contact. For electrical vehicles, the energy density of SIB with similar electrodes, will remain always lower than that of LIBs. Further improvements in the energy density are needed, for developing electrodes with higher operating voltages and more than one Na ion insertion per formula unit. At the same time, taking in consideration high voltage and wide electrochemical windows, electrolytes with novel properties need to be developed further. Ionic liquids can be an alternative for carbonate-based electrolytes. However, low temperature operating conditions and long synthesis and purification protocols are still a challenge.

**Mg batteries.** There has been a substantial research effort made in the development of magnesium batteries. Several prototype cells have been demonstrated based on various anode-electrolyte-cathode technologies. The key challenge for RMB remains in finding compatible electrode-electrolyte chemistries that are able to intercalate and conduct magnesium ions safely, efficiently and with acceptable cyclability. In particular, current research status for electrolytes can be summarized as follows:

o They possess low electrochemical stability that dictate the voltage charging limits of typically around 3 V but in most cases around 1.5 V.
o Halides and fluorides enabling fast diffusion of $Mg^{2+}$ ions are corrosive towards the cell components and can be explosive. Aluminate ion, also typical for the electrolyte composition, is sensitive to water and air. Therefore, finding alternative electrolyte chemistries should be the focus for future research in high-voltage magnesium batteries.
o The complicated organic chemistries demonstrate dependence on synthetic conditions and history/quality of the chemicals. The complex synthesis reactions can be difficult to reproduce.

The development of magnesium solid ionic conductors [102] for all solid-state magnesium battery seems an interesting and promising direction that could overcome some of the aforementioned challenges.

Furthermore, a significant challenge is within optimizing the electrodes. Intercalation cathodes are more stable towards higher voltages but offer low specific capacities. Conversion cathodes could be especially interesting for high-energy density Mg batteries but they are stable at rather modest cycling. For the anode, the formation of passivation layers promoted by oxygen and moisture present a serious concern. These layers block the diffusion of $Mg^{2+}$ cations and deteriorate the cell's performance. Furthermore, recent findings demonstrate the growth of dendritic magnesium deposits upon the galvanostatic electrodeposition of metallic Mg from Grignard reagents in symmetric Mg–Mg cells [299-303]. The suggested approach is that instead of stating the dendrite-free anode magnesiation and demagnesiation, one should rather clearly define the electrochemical windows where magnesium can be plated/stripped without dendrite formation [304]. Changing the anode material from pure magnesium reduces the theoretical capacity of the RMB and can question this advantage of the divalent battery over LIB. The overall challenge is also in combining all the elements in one working compatible system. As a new perspective towards advanced energy storage, hybrid-ion batteries have been recently reported [305]. In these systems, different metal ions can bring forward the respective advantages.

In addition to the intrinsic problems of the RMB systems, a hurdle lies in comparison of the performance of different systems. The research results are obtained under different conditions, using different types of electrolytes, cathodes, and sometimes anodes. The performance of such systems is defined by the overall cell combination, which makes it difficult to draw a proper conclusion on the performance of an individual components. The non-reliability of the pseudo-reference electrodes as these present both significant potential shifts as well as cause unstable behaviors was noted and an attempt to develop experimental protocols in order to achieve consistent results when using half-cell set-ups was proposed [306].

**Ca-ion batteries.** The diffusion of $Ca^{2+}$, as well as other divalent cations, into the metallic anodes and intercalation hosts is slow, and most reported Ca-ion batteries exhibit low working voltage (<2.0 V), and poor cycling stability (within 100 cycles). Diffusion of polyvalent cations through most inorganic metal oxides or sulphides is slowed down due to the high charge density. The limitations can be overcome by using hydrated compounds where water or hydroxyls shield the strong Coulombic interaction between the high charge density polyvalent guest species and the cation and anions of the host structure [307]. Use of aerogels, for example, have resulted in improvement of the intercalation of polyvalent cations into intercalation compounds to the point where gravimetric capacities of the materials approach that of the Li ion within the same material. Electrolytes would be confined to acetonitrile-based electrolytes for good stability, while corrosion would occur in carbonate-based electrolytes. Ca metal electrode in conventional organic electrolytes is apt to form a surface passivation film, which prevents $Ca^{2+}$ transportation thus leading to irreversible calcium deposition. Although some researchers realized reversible Ca deposition in a molten salt electrolyte, the working temperature is extremely high (550–700 °C) [308], which cannot match the mainstream application conditions. An approach using large, weakly binding anions as demonstrated by Li *et al*. [190], could enable further progress in the field.

Replacing the metallic calcium anode by intercalation-type active material is a feasible strategy to avoid calcium plating and stripping. Nanosized particles, materials engineering, and cell optimization should be considered to achieve technologically sufficient rate capability.

**Zn-ion batteries.** The calculated energy density of ZIBs, assuming Mn-based and V-based cathodes, can reach values of 85 Wh kg$^{-1}$ and 75 Wh kg$^{-1}$, respectively using assumptions similar to those for the LIBs [210]. These values are comparable to those of a Ni-*M*H battery, but lower than the energy densities of LIBs (180–230 Wh kg$^{-1}$). However, taking in consideration safety, cost and environmental aspects, ZIBs could be suitable in some applications while research need to be intensified. Though limited by the achieved operational voltage window, ZIBs in aqueous electrolytes could find their application at extreme temperatures where safe operation can be maintained in comparison to non-aqueous electrolytes. Novel high-rate and stable quasi-solid-state zinc-ion battery – with 2D layered zinc orthovanadate array cathode, Zn array anode supported by a conductive porous graphene foam and a gel electrolyte – has been reported [309]. This was the first application of V-based cathode in flexible ZIBs. Hybrid-flexible ZIBs, using lithium manganese oxide (LMO) and lithium iron phosphate (LFP) as cathodes, were also developed further, in which Pluronic hydrogel electrolytes (PHEs) consisting of a Pluronic polymer and an aqueous solution of mixed salts (0.25M $ZnSO_4$ and 0.25M $Li_2SO_4$) were used [310]. In this configuration, higher voltages of 1.8 V and 1.18 V *vs.* $Zn^{2+}$/Zn can be reached for the hybrid cells Zn/PHE/LMO and Zn/PHE/LFP, respectively. These flexible batteries allowed the demonstration of some ergonomic properties such us a wearable and self-charging system integrated with solar cells; emphasizing the potential of ZIBs in developing parallel market.

**Al-ion batteries.** The early research studies were focused on Al-based primary batteries, with a little attention paid to the rechargeable ones, until advances in nanostructuring of materials have been made. The first Al-ion rechargeable battery is quite recent (2011) compared to other technologies [214], which will require more research work (20 cycles, 273 mAh $g^{-1}$, Eeq = 0.55V *vs.* $Al^{3+}$/Al, ionic liquid medium).

In 2015, an ultrafast Al-ion battery was developed using 3D graphitic foam (cathode), Al anode and ionic liquid-based electrolyte [213]. The cell demonstrated high stability (7500 cycles, 2V *vs.* $Al^{3+}$/Al) at ultrahigh current densities. Although the (de)intercalation of chloroaluminate anions in the graphite has been clarified, some issues still remain to be solved such as high acidity, impurities, cost of ionic liquids and low energy density. Overall, the electrochemical storage performance of the battery needs to be improved further. Successful aqueous electrolyte-based Al-ion batteries have been presented, but this goes on the price of low voltage, resulting in lower energy density. However, this battery could be suitable for stationary energy storage systems. Dual-graphite batteries using graphite as intercalation hosts for both Al ions and anions are under study [311]. In this configuration, the electrolyte acts at the same time as ions carrier/transport and a fuel, which may contribute to the increase of the energy density of the Al battery.

## 5.2. Anionic shuttles

**F-ion and Cl-ion batteries.** Besides rechargeable batteries based on cationic shuttles, novel types of rechargeable batteries with anionic shuttles (Fluoride or chloride) are also under development. The best cycling performances are obtained with solid-state electrolyte operating at high temperatures (150 °C). Efforts are ongoing to improve the energy density and cyclability of the cells, as well as reducing the operating temperature with the appropriate high ion-conducting solid-state electrolyte. The first concept of Cl-ion batteries was proved experimentally by utilizing an ionic liquid electrolyte, a lithium foil as anode and CoCl$_3$, VCl$_3$ or BiCl$_3$ as cathode [222]. Bismuth oxychloride (BiOCl) and iron oxychloride (FeOCl) were also investigated. The Li- FeOCl cell can be cycled for 30 cycles (1 cycle; 158 mAh g$^{-1}$, last cycle; 60 mAh g$^{-1}$). However, the large volume change between charged-discharged states could be a serious factor of limitation, observed in the Li- FeOCl, Mg- FeOCl and Mg- BiOCl electrochemical cells [312]. Further attempts have been followed using a solid polymer electrolyte, demonstrating first all-solid state Cl-ion battery with significant improvements for the working voltage and capacity retention over 20 cycles [313].

## 5.3. MH-based batteries

In the last 20 years, significant progress has been made for Ni-$M$H batteries, particularly as concerns self-discharge issues and enhancement of specific energy density of the negative electrode [314,315]. Indeed, the replacement of LaNi$_5$-type electrodes by superlattice $A_2B_4/AB_5$ alloys brings out higher hydrogen mass capacity (*ca.* 25%) [236]. These breakthroughs added to the intrinsic safety of Ni-$M$H batteries (related to the use of aqueous electrolyte) their low cost and their robustness, have led to high penetration of this technology in hybrid HEVs [316]. However, for applications demanding higher energy densities such as full EVs, Ni-$M$H batteries cannot compete yet with Li-ion ones. This can be noticed in Fig. 11 where performances of Li-ion batteries exceed those of Ni-$M$H in terms of both gravimetric and volumetric energy density.

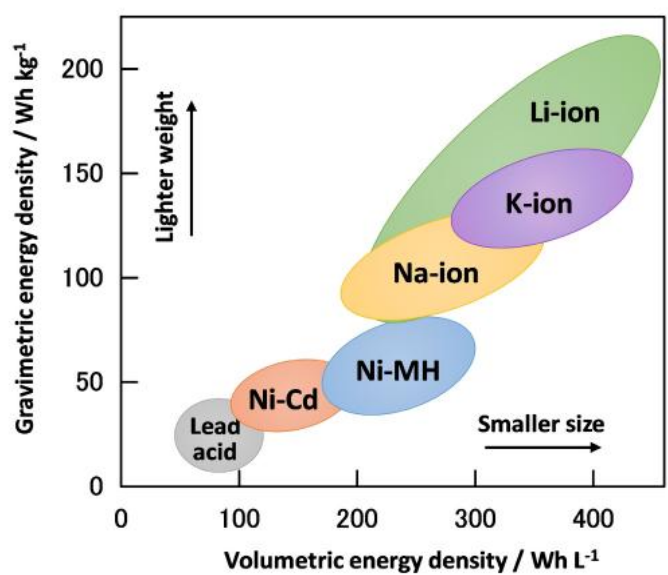

Figure 11.
Energy density of commercial batteries (Lead acid, Ni-Cd, Ni-$M$H and Li-ion) as well as novel monovalent technologies (Na-ion and K-ion) [317].

Nevertheless, there is still a room for the improvements of the specific energy of Ni-*M*H batteries, particularly as concerns the negative electrode by using compounds with higher hydrogen storage capacity. Indeed, reversible mass capacity of $A_2B_4/AB_5$ alloys is limited to *ca.* 1.5 wt.% (400 mAh g$^{-1}$) while alternative metal hydrides with much higher capacities exist such as Zr/Ti Laves type alloys (1.9 wt%, 500 mAh g$^{-1}$), $Mg_2NiH_4$ (3.7 wt%, 1000 mAh g$^{-1}$) or even $MgH_2$ (7.6 wt%, 2000 mAh g$^{-1}$). However, two latter compounds are made of electropositive metals and contain low or no-content of *B*-type late transition metals. They are consequently more prone to corrosion in KOH aqueous media. Two solutions are offered to overcome this issue: either using a suitable coating of the *M*H in the negative electrode or replacing the KOH electrolyte for a lower oxidizing media. For the first one, the coating should be electrochemically active towards water reduction, allowing hydrogen permeation and being mechanically stable upon the cycling of hydrogen charge-discharge. The concept has been successfully applied in Pd-caped Mg-Sc and Mg-Ti thin films providing capacities as high as 1500 mAh g$^{-1}$ [318,319]. However, to find out efficient low-cost coatings and their implementation in current 3D electrode materials remains a challenge. As for the second solution, a low-corroding electrolyte should be able to guarantee high-capacity redox reactions at both electrodes while providing a good ionic conductivity but no electronic conductivity and preserving the intrinsic safety of Ni-*M*H batteries. One attractive option is to replace aqueous KOH by proton-conducting ionic liquids as recently proposed by Meng *et al.* [320]. The other alternative is by dissolving ion-conducting salts in organic solvents. Note that this approach has been used for metal hydrides as negative electrodes in Li-ion batteries [273,274,321], in which Li-ions react reversibly with metal hydrides through a conversion reaction (*cf.* section 4). The advantage of organic electrolytes is that they enable high-voltage positive electrodes to build-up high voltage batteries. This is the key reason of the higher energy densities of Li-ion batteries as compared to Ni-*M*H ones (Fig. 11). However, this solution has a severe drawback because of losing the safety insured in aqueous Ni-*M*H batteries. As evidenced by the vivid interest in aqueous electrolyte batteries [322], safety is a key issue for many applications, even for those in which present Li-ion batteries are currently used.

Table 2. Summary, composition, and applied properties for the different beyond Li-ion battery technologies. Energy density data are referred to commercialized LIB for comparison.

| Battery technology | Cathode | Anode | Electrolyte composition | Assessed performance full-cells | | Major advantages | Drawbacks | Ref. |
|---|---|---|---|---|---|---|---|---|
| | | | | Energy density [a] (Wh kg$^{-1}$) | Durability (cycles) | | | |
| **LIB** | LiFePO$_4$ | Graphitic carbon | 1M LiClO$_4$/ EC-DMC | 90-120 | 2,000 | High energy density | Cost and safety | [323] |
| | LiCoO$_2$ | | 1M LiPF$_6$/ EC-DMC | 150-240 | 500-1000 | High voltage and energy density | | |
| **SIB** | Na$_{1.5}$VPO$_{4.8}$F$_{0.7}$ | Hard or nanostructured carbon | 1M NaClO$_4$/ EC$_{0.45}$·PC$_{0.45}$·DMC$_{0.1}$ | 80-100 [b] | 120-210 | Abundance, low cost at large-scale and better safety, wide temperature range | Moderate energy density | [36,62,64,78,81] |
| | Na$_{0.45}$Ni$_{0.22}$Co$_{0.11}$Mn$_{0.66}$O$_2$ | | 10 mol.% NaTFSI / PYR$_{14}$FSI | 70-114 | 100 | | *idem*, cost of ILs | [56] |
| **RMB** | Mo$_6$S$_8$ | Mg (2200) | Mg(AlCl$_2$BuEt)$_2$/THF | 38-42 | 2,000 | Abundance, low cost | Low capacity | [91] |
| | TiO$_2$ | | Mg(BH$_4$)$_2$-LiBH$_4$ /tetraglyme | 40-45 | 120 | | | [173] |
| **CAB** | Carbon-based layered | Carbon-based layered | 0.7M Ca(PF$_6$)$_2$/EC-DMC-EMC | 40-45 | 300 | Abundance, low cost, low temperature operation | Low energy density | [324] |
| | Na$_2$FePO$_4$F | BP2000 carbon | Ca(PF$_6$)$_2$/EC-PC | 16-20 | 50 | | | [325] |
| **ZIB** | V$_2$O$_5$ | Zn | 3M Zn(CF$_3$SO$_3$)$_2$ aq. electrolyte | 70-75 | 4,000 | Safety, low cost, facile manufacturing, wide temperature | Low voltage | [211] |
| | Zn$_{0.3}$V$_2$O$_5$.1.5H$_2$O | | | 60-85 | 20,000 | | | [195,210] |
| **Al-ion** | 3D graphitic foam | Al | AlCl$_3$/1-ethyl-3-methylimidazolium chloride | 35-40 | 7500 | Cost, safety, high power | High Lewis acidity issues | [213] |
| | Ultrathin graphite nanosheets | Zn | Al$_2$(SO$_4$)$_3$/Zn(CHCOO)$_2$ aq. electrolyte | 15-25 | 200 | Cost, large-scale stationary storage | Low energy density | [216] |
| **Ni*M*H** | Ni(OH)2 | LaNi$_5$-type | Conc. KOH alkaline electrolyte | 40-120 | 300-500 | High specific power | Low voltage (aq. medium) | [245,323] |
| **MH-LIB** | TiS$_2$ | Li | LiBH$_4$ | 110-120 | 300 | High energy density | Operating at 120 °C | [274,326,327] |
| | TiS$_2$ | 75MgH$_2$-25CoO | Li(BH$_4$)$_{0.75}$I$_{0.25}$·(Li$_2$S)$_{0.75}$·(P$_2$S$_5$)$_{0.25}$ | 70-90 [b] | -- | High energy density, dentrite free, safety, low *T* | Low voltage | [274,294,296,326,328] |
| **F-ion** | BiF$_3$ | Ce metal | La$_{0.9}$Ba$_{0.1}$F$_{2.9}$ - BaSnF$_4$ (R*T*-150°C) | < 40 | 10-50 | Safety, energy density | Capacity fading at R*T* | [220,221,329] |

NaTFSI: sodium bis(trifluoromethanesulfonyl)imide; PYR$_{14}$FSI: *N*-butyl-*N*-methylpyrrolidinium bis(fluorosulfonyl)imide
[a] using assumptions similar to those used in LIBs. Values may vary depending on the system and engineering of the cells, e.g. when solid-state electrolyte (polymer or inorganic) is used, higher energy densities are expected.
[b] based on the suggested configuration of full-cell batteries.

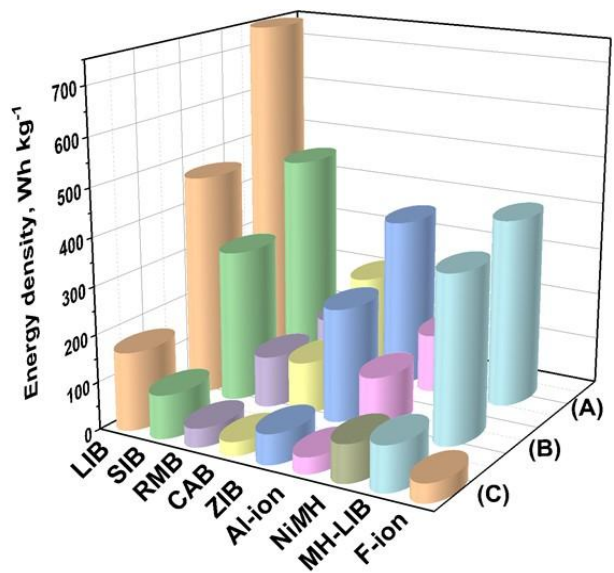

Figure 12.

Energy density comparison for the different battery prototypes reported in table 2. Average values are given based on the available data and taking in consideration (A) cathode only, (B) both electrodes and (C) full-cell units (LIB-normalized). The values may depend on the electrolyte characteristics, as well as current densities. For Ni*M*H and F-ion batteries only (C) case is shown.

# 6. Summary

The present market of mobile devices and electric transportation technologies is undergoing expansion in terms of the need for high energy density and safe rechargeable batteries. The most promising post-Li ion electrochemical storage systems are addressed in this Review. Table 2 summarizes the different available beyond Li-ion batteries, where comparison based on their energy density (referred to LIB) and durability can be made. Fig. 12 gives an overview of the energy density considering the electrodes and full-cell units. Some data are only available at high current rate which may affect the specific energy of the full-cells. The use of sodium-ion batteries is being approaching the performances of the commercial LIBs for energy density issues. Although the technology is under development by some academic and start-up companies, which will result in lower prices compared to LIB, the long-term lifespan requires additional improvements. This could also reach satisfactory levels along the large-scale production, as demonstrated for the recent research discovery of positive electrode materials showing advantages over the electrodes used in LIBs. Additionally, the road map for beyond Li-ion batteries is also being followed worldwide with the development of a variety of multi-valent cation systems such as Mg, Ca, Zn or Al batteries. The design of the suitable post-Li ion batteries will not only depend on electrodes, but also will require their integration in the best cell configuration and assembly, which may involve some engineering work. There is a general agreement that solid-state or gel electrolytes will be closely researched in the field of beyond L-ion batteries, thanks to the safer operation and transportation, higher energy density and long stability of the stacked cells in a wide temperature range.

Exploitable batteries involving aqueous electrolytes with long-term cyclability, for instance in Zn and Al batteries, will undoubtedly attract investors, at least for quasi-stationary energy storage. Ni*M*H batteries are continuing supplying rechargeable battery sectors and much progress has been made to integrate them further in electric vehicles and railways. Moreover, a new category of batteries may see the light, using metal hydride incorporated in a LIB (MH-LIB). This configuration has demonstrated a potential for future applications, mainly as solid-state Li/Na battery with high energy density and safety. For all the technologies, careful electrodes optimization and cell design are essential for beyond Li-ion batteries to take advantage of their materials superiority which may surpass the state-of-the-art LIBs. One of the issues in assessing the performance of post-Li ion batteries is the end-of-life energy density which needs to be specified accurately. Besides energy density, interfaces stability, aging/degradation mechanisms and long-term operation of post LIBs in varied conditions must be studied in detail as it is done for the current LIBs; so that the suitable treatment and creative remedies can be proposed in the aim of better performing batteries and low recycling costs.

**Acknowledgements**

This work contributes to the research performed at CELEST (Center for Electrochemical Energy Storage Ulm-Karlsruhe). VAY is grateful for the support this work has received from the Research Council of Norway (Project 285146 - New IEA Task ENERGY STORAGE AND CONVERSION BASED ON HYDROGEN).

# REFERENCES

[1] J. Deng, W.-B. Luo, S.-L. Chou, H.-K. Liu, S.-X. Dou, Sodium-Ion Batteries: From Academic Research to Practical Commercialization, Advanced Energy Materials, 8 (2018) 1701428.
[2] M.D. Slater, D. Kim, E. Lee, C.S. Johnson, Sodium-Ion Batteries, Advanced Functional Materials, 23 (2013) 947-958.
[3] T.-H. Kim, J.-S. Park, S.K. Chang, S. Choi, J.H. Ryu, H.-K. Song, The Current Move of Lithium Ion Batteries Towards the Next Phase, Advanced Energy Materials, 2 (2012) 860-872.
[4] G. Crabtree, E. Kócs, L. Trahey, The energy-storage frontier: Lithium-ion batteries and beyond, MRS Bulletin, 40 (2015) 1067-1078.
[5] J.M. Tarascon, M. Armand, Issues and challenges facing rechargeable lithium batteries, Nature, 414 (2001) 359-367.
[6] S. Chen, C. Wu, L. Shen, C. Zhu, Y. Huang, K. Xi, J. Maier, Y. Yu, Challenges and Perspectives for NASICON-Type Electrode Materials for Advanced Sodium-Ion Batteries, Advanced Materials, 29 (2017) 1700431.
[7] Research Center for Energy Economics: https://www.ffe.de/en/ (Accessed October 2019).
[8] X. Xia, J.R. Dahn, Study of the Reactivity of Na/Hard Carbon with Different Solvents and Electrolytes, Journal of The Electrochemical Society, 159 (2012) A515-A519.
[9] H. Pan, Y.-S. Hu, L. Chen, Room-temperature stationary sodium-ion batteries for large-scale electric energy storage, Energy & Environmental Science, 6 (2013) 2338-2360.
[10] N. Yabuuchi, M. Kajiyama, J. Iwatate, H. Nishikawa, S. Hitomi, R. Okuyama, R. Usui, Y. Yamada, S. Komaba, P2-type $Na_x[Fe_{1/2}Mn_{1/2}]O_2$ made from earth-abundant elements for rechargeable Na batteries, Nature Materials, 11 (2012) 512.
[11] S. Komaba, N. Yabuuchi, T. Nakayama, A. Ogata, T. Ishikawa, I. Nakai, Study on the Reversible Electrode Reaction of $Na_{1-x}Ni_{0.5}Mn_{0.5}O_2$ for a Rechargeable Sodium-Ion Battery, Inorganic Chemistry, 51 (2012) 6211-6220.
[12] J.W. Choi, D. Aurbach, Promise and reality of post-lithium-ion batteries with high energy densities, Nature Reviews Materials, 1 (2016) 16013.
[13] F. Sauvage, L. Laffont, J.M. Tarascon, E. Baudrin, Study of the insertion/deinsertion mechanism of sodium into $Na_{0.44}MnO_2$, Inorganic Chemistry, 46 (2007) 3289-3294.
[14] M.M. Doeff, M.Y. Peng, Y.P. Ma, L.C. Dejonghe, ORTHORHOMBIC $NAXMNO_2$ AS A CATHODE MATERIAL FOR SECONDARY SODIUM AND LITHIUM POLYMER BATTERIES, Journal of the Electrochemical Society, 141 (1994) L145-L147.
[15] Y. Cao, L. Xiao, W. Wang, D. Choi, Z. Nie, J. Yu, L.V. Saraf, Z. Yang, J. Liu, Reversible Sodium Ion Insertion in Single Crystalline Manganese Oxide Nanowires with Long Cycle Life, Advanced Materials, 23 (2011) 3155-+.
[16] J.M. Tarascon, D.G. Guyomard, B. Wilkens, W.R. McKinnon, P. Barboux, CHEMICAL AND ELECTROCHEMICAL INSERTION OF NA INTO THE SPINEL LAMBDA-$MNO_2$ PHASE, Solid State Ionics, 57 (1992) 113-120.
[17] K. West, B. Zachauchristiansen, T. Jacobsen, S. Skaarup, SODIUM INSERTION IN VANADIUM-OXIDES, Solid State Ionics, 28 (1988) 1128-1131.
[18] K. West, B. Zachauchristiansen, T. Jacobsen, S. Skaarup, SOLID-STATE SODIUM CELLS - AN ALTERNATIVE TO LITHIUM CELLS, Journal of Power Sources, 26 (1989) 341-345.
[19] S. Tepavcevic, H. Xiong, V.R. Stamenkovic, X. Zuo, M. Balasubramanian, V.B. Prakapenka, C.S. Johnson, T. Rajh, Nanostructured Bilayered Vanadium Oxide Electrodes for Rechargeable Sodium-Ion Batteries, ACS Nano, 6 (2012) 530-538.
[20] H. Liu, H. Zhou, L. Chen, Z. Tang, W. Yang, Electrochemical insertion/deinsertion of sodium on $NaV_6O_{15}$ nanorods as cathode material of rechargeable sodium-based batteries, Journal of Power Sources, 196 (2011) 814-819.
[21] D. Hamani, M. Ati, J.-M. Tarascon, P. Rozier, $Na_xVO_2$ as possible electrode for Na-ion batteries, Electrochemistry Communications, 13 (2011) 938-941.
[22] C. Didier, M. Guignard, C. Denage, O. Szajwaj, S. Ito, I. Saadoune, J. Darriet, C. Delmas, Electrochemical Na-Deintercalation from $NaVO_2$, Electrochemical and Solid State Letters, 14 (2011) A75-A78.
[23] S. Komaba, T. Mikumo, A. Ogata, Electrochemical activity of nanocrystalline $Fe_3O_4$ in aprotic Li and Na salt electrolytes, Electrochemistry Communications, 10 (2008) 1276-1279.
[24] S. Komaba, T. Mikumo, N. Yabuuchi, A. Ogata, H. Yoshida, Y. Yamada, Electrochemical Insertion of Li and Na Ions into Nanocrystalline $Fe_3O_4$ and alpha-$Fe_2O_3$ for Rechargeable Batteries, Journal of the Electrochemical Society, 157 (2010) A60-A65.
[25] C. Delmas, C. Fouassier, P. Hagenmuller, STRUCTURAL CLASSIFICATION AND PROPERTIES OF THE LAYERED OXIDES, Physica B & C, 99 (1980) 81-85.

[26] J. Molenda, A. Stoklosa, D. Than, RELATION BETWEEN IONIC AND ELECTRONIC DEFECTS OF NA0.7MNO2 BRONZE AND ITS ELECTROCHEMICAL PROPERTIES, Solid State Ionics, 24 (1987) 33-38.
[27] F.C. Choua, E.T. Abel, J.H. Cho, Y.S. Lee, Electrochemical de-intercalation, oxygen non-stoichiometry, and crystal growth of NaxCoO2-delta, Journal of Physics and Chemistry of Solids, 66 (2005) 155-160.
[28] C. Fouassier, G. Matejka, J.M. Reau, P. Hagenmuller, NEW OXYGENATED BRONZES OF FORMULA NA(X)COO2 (X LESS-THAN-OR-EQUAL-TO ONE) - COBALT-OXYGEN-SODIUM SYSTEM, Journal of Solid State Chemistry, 6 (1973) 532-537.
[29] J.J. Braconnier, C. Delmas, C. Fouassier, P. Hagenmuller, ELECTROCHEMICAL BEHAVIOR OF THE PHASES NAXCOO2, Materials Research Bulletin, 15 (1980) 1797-1804.
[30] M. Roger, D.J.P. Morris, D.A. Tennant, M.J. Gutmann, J.P. Goff, J.U. Hoffmann, R. Feyerherm, E. Dudzik, D. Prabhakaran, A.T. Boothroyd, N. Shannon, B. Lake, P.P. Deen, Patterning of sodium ions and the control of electrons in sodium cobaltate, Nature, 445 (2007) 631-634.
[31] M. Pollet, M. Blangero, J.-P. Doumerc, R. Decourt, D. Carlier, C. Denage, C. Delmas, Structure and Properties of Alkali Cobalt Double Oxides A(0.6)CoO(2) (A = Li, Na, and K), Inorganic Chemistry, 48 (2009) 9671-9683.
[32] C. Delmas, J.J. Braconnier, C. Fouassier, P. Hagenmuller, ELECTROCHEMICAL INTERCALATION OF SODIUM IN NAXCOO2 BRONZES, Solid State Ionics, 3-4 (1981) 165-169.
[33] R. Berthelot, D. Carlier, C. Delmas, Electrochemical investigation of the P2-NaxCoO2 phase diagram, Nature Materials, 10 (2011) 74-U73.
[34] L.W. Shacklette, T.R. Jow, L. Townsend, RECHARGEABLE ELECTRODES FROM SODIUM COBALT BRONZES, Journal of the Electrochemical Society, 135 (1988) 2669-2674.
[35] Y.P. Ma, M.M. Doeff, S.J. Visco, L.C. Dejonghe, RECHARGEABLE NA NAXCOO2 AND NA15PB4 NAXCOO2 POLYMER ELECTROLYTE CELLS, Journal of the Electrochemical Society, 140 (1993) 2726-2733.
[36] M. Sawicki, L.L. Shaw, Advances and challenges of sodium ion batteries as post lithium ion batteries, RSC Advances, 5 (2015) 53129-53154.
[37] D. Kim, S.-H. Kang, M. Slater, S. Rood, J.T. Vaughey, N. Karan, M. Balasubramanian, C.S. Johnson, Enabling Sodium Batteries Using Lithium-Substituted Sodium Layered Transition Metal Oxide Cathodes, Advanced Energy Materials, 1 (2011) 333-336.
[38] X. Ma, H. Chen, G. Ceder, Electrochemical Properties of Monoclinic NaMnO2, Journal of the Electrochemical Society, 158 (2011) A1307-A1312.
[39] H. Kim, D.J. Kim, D.-H. Seo, M.S. Yeom, K. Kang, D.K. Kim, Y. Jung, Ab Initio Study of the Sodium Intercalation and Intermediate Phases in Na0.44MnO2 for Sodium-Ion Battery, Chemistry of Materials, 24 (2012) 1205-1211.
[40] S.-y. Yang, X.-y. Wang, Y. Wang, Q.-q. Chen, J.-j. Li, X.-k. Yang, Effects of Na content on structure and electrochemical performances of NaxMnO2+delta cathode material, Transactions of Nonferrous Metals Society of China, 20 (2010) 1892-1898.
[41] A. Mendiboure, C. Delmas, P. Hagenmuller, ELECTROCHEMICAL INTERCALATION AND DEINTERCALATION OF NAXMNO2 BRONZES, Journal of Solid State Chemistry, 57 (1985) 323-331.
[42] J. Zhao, H. Chen, J. Shi, J. Gu, X. Dong, J. Gao, M. Ruan, L. Yu, Electrochemical and oxygen desorption properties of nanostructured ternary compound NaxMnO2 directly templated from mesoporous SBA-15, Microporous and Mesoporous Materials, 116 (2008) 432-438.
[43] O.A. Vargas, A. Caballero, L. Hernan, J. Morales, Improved capacitive properties of layered manganese dioxide grown as nanowires, Journal of Power Sources, 196 (2011) 3350-3354.
[44] N. Bucher, S. Hartung, A. Nagasubramanian, Y.L. Cheah, H.E. Hoster, S. Madhavi, Layered NaxMnO2+z in Sodium Ion Batteries-Influence of Morphology on Cycle Performance, Acs Applied Materials & Interfaces, 6 (2014) 8059-8065.
[45] A. Caballero, L. Hernan, J. Morales, L. Sanchez, J.S. Pena, M.A.G. Aranda, Synthesis and characterization of high-temperature hexagonal P2-Na0.6MnO2 and its electrochemical behaviour as cathode in sodium cells, Journal of Materials Chemistry, 12 (2002) 1142-1147.
[46] J. Billaud, R.J. Clement, A.R. Armstrong, J. Canales-Vazquez, P. Rozier, C.P. Grey, P.G. Bruce, beta-NaMnO2: A High-Performance Cathode for Sodium-Ion Batteries, Journal of the American Chemical Society, 136 (2014) 17243-17248.
[47] I.-H. Jo, H.-S. Ryu, D.-G. Gu, J.-S. Park, I.-S. Ahn, H.-J. Ahn, T.-H. Nam, K.-W. Kim, The effect of electrolyte on the electrochemical properties of Na/α-NaMnO2 batteries, Materials Research Bulletin, 58 (2014) 74-77.
[48] J.J. Braconnier, C. Delmas, P. Hagenmuller, STUDY OF THE NAXCRO2 AND NAXNIO2 SYSTEMS BY ELECTROCHEMICAL DESINTERCALATION, Materials Research Bulletin, 17 (1982) 993-1000.
[49] S. Komaba, C. Takei, T. Nakayama, A. Ogata, N. Yabuuchi, Electrochemical intercalation activity of layered NaCrO2 vs. LiCrO2, Electrochemistry Communications, 12 (2010) 355-358.

[50] X. Xia, J.R. Dahn, NaCrO2 is a Fundamentally Safe Positive Electrode Material for Sodium-Ion Batteries with Liquid Electrolytes, Electrochemical and Solid State Letters, 15 (2012) A1-A4.
[51] Y. Takeda, K. Nakahara, M. Nishijima, N. Imanishi, O. Yamamoto, M. Takano, R. Kanno, SODIUM DEINTERCALATION FROM SODIUM IRON-OXIDE, Materials Research Bulletin, 29 (1994) 659-666.
[52] M. Tabuchi, C. Masquelier, T. Takeuchi, K. Ado, I. Matsubara, T. Shirane, R. Kanno, S. Tsutsui, S. Nasu, H. Sakaebe, O. Nakamura, Li+/Na+ exchange from alpha-NaFeO2 using hydrothermal reaction, Solid State Ionics, 90 (1996) 129-132.
[53] M.C. Blesa, E. Moran, C. Leon, J. Santamaria, J.D. Tornero, N. Menendez, alpha-NaFeO2: ionic conductivity and sodium extraction, Solid State Ionics, 126 (1999) 81-87.
[54] J. Zhao, L. Zhao, N. Dimov, S. Okada, T. Nishida, Electrochemical and Thermal Properties of alpha-NaFeO2 Cathode for Na-Ion Batteries, Journal of the Electrochemical Society, 160 (2013) A3077-A3081.
[55] D. Buchholz, A. Moretti, R. Kloepsch, S. Nowak, V. Siozios, M. Winter, S. Passerini, Toward Na-ion Batteries-Synthesis and Characterization of a Novel High Capacity Na Ion Intercalation Material, Chemistry of Materials, 25 (2013) 142-148.
[56] L.G. Chagas, D. Buchholz, L. Wu, B. Vortmann, S. Passerini, Unexpected performance of layered sodium-ion cathode material in ionic liquid-based electrolyte, Journal of Power Sources, 247 (2014) 377-383.
[57] H. Yu, S. Guo, Y. Zhu, M. Ishida, H. Zhou, Novel titanium-based O-3-type NaTi0.5Ni0.5O2 as a cathode material for sodium ion batteries, Chemical Communications, 50 (2014) 457-459.
[58] J. Wang, J. Yang, Y. Nuli, R. Holze, Room temperature Na/S batteries with sulfur composite cathode materials, Electrochemistry Communications, 9 (2007) 31-34.
[59] P. Barpanda, G. Oyama, S.-i. Nishimura, S.-C. Chung, A. Yamada, A 3.8-V earth-abundant sodium battery electrode, Nature Communications, 5 (2014) 1.
[60] K. Saravanan, C.W. Mason, A. Rudola, K.H. Wong, P. Balaya, The First Report on Excellent Cycling Stability and Superior Rate Capability of Na3V2(PO4)3 for Sodium Ion Batteries, Advanced Energy Materials, 3 (2013) 444-450.
[61] C. Zhu, K. Song, P.A. van Aken, J. Maier, Y. Yu, Carbon-Coated Na3V2(PO4)(3) Embedded in Porous Carbon Matrix: An Ultrafast Na-Storage Cathode with the Potential of Outperforming Li Cathodes, Nano Letters, 14 (2014) 2175-2180.
[62] Y.-U. Park, D.-H. Seo, H.-S. Kwon, B. Kim, J. Kim, H. Kim, I. Kim, H.-I. Yoo, K. Kang, A New High-Energy Cathode for a Na-Ion Battery with Ultrahigh Stability, Journal of the American Chemical Society, 135 (2013) 13870-13878.
[63] D.A. Stevens, J.R. Dahn, The mechanisms of lithium and sodium insertion in carbon materials, Journal of the Electrochemical Society, 148 (2001) A803-A811.
[64] S. Komaba, W. Murata, T. Ishikawa, N. Yabuuchi, T. Ozeki, T. Nakayama, A. Ogata, K. Gotoh, K. Fujiwara, Electrochemical Na Insertion and Solid Electrolyte Interphase for Hard-Carbon Electrodes and Application to Na-Ion Batteries, Advanced Functional Materials, 21 (2011) 3859-3867.
[65] Y. Liu, Y. Xu, Y. Zhu, J.N. Culver, C.A. Lundgren, K. Xu, C. Wang, Tin-Coated Viral Nanoforests as Sodium-Ion Battery Anodes, Acs Nano, 7 (2013) 3627-3634.
[66] J. Qian, X. Wu, Y. Cao, X. Ai, H. Yang, High Capacity and Rate Capability of Amorphous Phosphorus for Sodium Ion Batteries, Angewandte Chemie-International Edition, 52 (2013) 4633-4636.
[67] K. Dai, H. Zhao, Z. Wang, X. Song, V. Battaglia, G. Liu, Toward high specific capacity and high cycling stability of pure tin nanoparticles with conductive polymer binder for sodium ion batteries, Journal of Power Sources, 263 (2014) 276-279.
[68] Y. Zhu, X. Han, Y. Xu, Y. Liu, S. Zheng, K. Xu, L. Hu, C. Wang, Electrospun Sb/C Fibers for a Stable and Fast Sodium-Ion Battery Anode, Acs Nano, 7 (2013) 6378-6386.
[69] H. Zhu, Z. Jia, Y. Chen, N. Weadock, J. Wan, O. Vaaland, X. Han, T. Li, L. Hu, Tin Anode for Sodium-Ion Batteries Using Natural Wood Fiber as a Mechanical Buffer and Electrolyte Reservoir, Nano Letters, 13 (2013) 3093-3100.
[70] H. Xiong, M.D. Slater, M. Balasubramanian, C.S. Johnson, T. Rajh, Amorphous TiO2 Nanotube Anode for Rechargeable Sodium Ion Batteries, Journal of Physical Chemistry Letters, 2 (2011) 2560-2565.
[71] Y. Xu, E.M. Lotfabad, H. Wang, B. Farbod, Z. Xu, A. Kohandehghan, D. Mitlin, Nanocrystalline anatase TiO2: a new anode material for rechargeable sodium ion batteries, Chemical Communications, 49 (2013) 8973-8975.
[72] F. Klein, B. Jache, A. Bhide, P. Adelhelm, Conversion reactions for sodium-ion batteries, Physical Chemistry Chemical Physics, 15 (2013) 15876-15887.
[73] S.-M. Oh, J.-Y. Hwang, C.S. Yoon, J. Lu, K. Amine, I. Belharouak, Y.-K. Sun, High Electrochemical Performances of Microsphere C-TiO2 Anode for Sodium-Ion Battery, Acs Applied Materials & Interfaces, 6 (2014) 11295-11301.
[74] M. Mortazavi, J. Deng, V.B. Shenoy, N.V. Medhekar, Elastic softening of alloy negative electrodes for Na-ion batteries, Journal of Power Sources, 225 (2013) 207-214.

[75] Y. Xu, F.M. Mulder, TiF3 catalyzed MgH2 as a Li/Na ion battery anode, International Journal of Hydrogen Energy, 43 (2018) 20033-20040.
[76] J. Zhao, L. Zhao, K. Chihara, S. Okada, J.-i. Yamaki, S. Matsumoto, S. Kuze, K. Nakane, Electrochemical and thermal properties of hard carbon-type anodes for Na-ion batteries, Journal of Power Sources, 244 (2013) 752-757.
[77] A. Ponrouch, E. Marchante, M. Courty, J.-M. Tarascon, M. Rosa Palacin, In search of an optimized electrolyte for Na-ion batteries, Energy & Environmental Science, 5 (2012) 8572-8583.
[78] J. Ding, H. Wang, Z. Li, A. Kohandehghan, K. Cui, Z. Xu, B. Zahiri, X. Tan, E.M. Lotfabad, B.C. Olsen, D. Mitlin, Carbon Nanosheet Frameworks Derived from Peat Moss as High Performance Sodium Ion Battery Anodes, Acs Nano, 7 (2013) 11004-11015.
[79] Y. Cao, L. Xiao, M.L. Sushko, W. Wang, B. Schwenzer, J. Xiao, Z. Nie, L.V. Saraf, Z. Yang, J. Liu, Sodium Ion Insertion in Hollow Carbon Nanowires for Battery Applications, Nano Letters, 12 (2012) 3783-3787.
[80] J.Y. Jang, H. Kim, Y. Lee, K.T. Lee, K. Kang, N.-S. Choi, Cyclic carbonate based-electrolytes enhancing the electrochemical performance of Na4Fe3(PO4)(2)(P2O7) cathodes for sodium-ion batteries, Electrochemistry Communications, 44 (2014) 74-77.
[81] A. Ponrouch, R. Dedryvere, D. Monti, A.E. Demet, J.M.A. Mba, L. Croguennec, C. Masquelier, P. Johansson, M.R. Palacin, Towards high energy density sodium ion batteries through electrolyte optimization, Energy & Environmental Science, 6 (2013) 2361-2369.
[82] D. Monti, E. Jonsson, M. Rosa Palacin, P. Johansson, Ionic liquid based electrolytes for sodium-ion batteries: Na+ solvation and ionic conductivity, Journal of Power Sources, 245 (2014) 630-636.
[83] J.S. Moreno, G. Maresca, S. Panero, B. Scrosati, G.B. Appetecchi, Sodium-conducting ionic liquid-based electrolytes, Electrochemistry Communications, 43 (2014) 1-4.
[84] Z. Li, D. Young, K. Xiang, W.C. Carter, Y.-M. Chiang, Towards High Power High Energy Aqueous Sodium-Ion Batteries: The NaTi2(PO4)3/Na0.44MnO2 System, Advanced Energy Materials, 3 (2013) 290-294.
[85] D.J. Kim, R. Ponraj, A.G. Kannan, H.-W. Lee, R. Fathi, R. Ruffo, C.M. Mari, D.K. Kim, Diffusion behavior of sodium ions in Na0.44MnO2 in aqueous and non-aqueous electrolytes, Journal of Power Sources, 244 (2013) 758-763.
[86] W. Song, X. Ji, Y. Zhu, H. Zhu, F. Li, J. Chen, F. Lu, Y. Yao, C.E. Banks, Aqueous Sodium-Ion Battery using a Na3V2(PO4)(3) Electrode, Chemelectrochem, 1 (2014) 871-876.
[87] 5 - Beryllium, Magnesium, Calcium, Strontium, Barium and Radium, in: N.N. Greenwood, A. Earnshaw (Eds.) Chemistry of the Elements (Second Edition), Butterworth-Heinemann, Oxford, 1997, pp. 107-138.
[88] 4 - Lithium, Sodium, Potassium, Rubidium, Caesium and Francium, in: N.N. Greenwood, A. Earnshaw (Eds.) Chemistry of the Elements (Second Edition), Butterworth-Heinemann, Oxford, 1997, pp. 68-106.
[89] A. Brenner, Note on the Electrodeposition of Magnesium from an Organic Solution of a Magnesium-Boron Complex, Journal of The Electrochemical Society, 118 (1971) 99-100.
[90] T.D. Gregory, R.J. Hoffman, R.C. Winterton, Nonaqueous Electrochemistry of Magnesium: Applications to Energy Storage, Journal of The Electrochemical Society, 137 (1990) 775-780.
[91] D. Aurbach, Z. Lu, A. Schechter, Y. Gofer, H. Gizbar, R. Turgeman, Y. Cohen, M. Moshkovich, E. Levi, Prototype systems for rechargeable magnesium batteries, Nature, 407 (2000) 724-727.
[92] Y. Orikasa, T. Masese, Y. Koyama, T. Mori, M. Hattori, K. Yamamoto, T. Okado, Z.D. Huang, T. Minato, C. Tassel, J. Kim, Y. Kobayashi, T. Abe, H. Kageyama, Y. Uchimoto, High energy density rechargeable magnesium battery using earth-abundant and non-toxic elements, Scientific Reports, 4 (2014).
[93] H.D. Yoo, I. Shterenberg, Y. Gofer, G. Gershinsky, N. Pour, D. Aurbach, Mg rechargeable batteries: an on-going challenge, Energy & Environmental Science, 6 (2013) 2265-2279.
[94] M.Q. Zhao, C.E. Ren, M. Alhabeb, B. Anasori, M.W. Barsoum, Y. Gogotsi, Magnesium-Ion Storage Capability of MXenes, ACS Appl. Energ. Mater., 2 (2019) 1572-1578.
[95] M. Xu, S.L. Lei, J. Qi, Q.Y. Dou, L.Y. Liu, Y.L. Lu, Q. Huang, S.Q. Shi, X.B. Yan, Opening Magnesium Storage Capability of Two-Dimensional MXene by Intercalation of Cationic Surfactant, Acs Nano, 12 (2018) 3733-3740.
[96] R.G. Zhang, F. Mizuno, C. Ling, Fullerenes: non-transition metal clusters as rechargeable magnesium battery cathodes, Chemical Communications, 51 (2015) 1108-1111.
[97] H.S. Kim, T.S. Arthur, G.D. Allred, J. Zajicek, J.G. Newman, A.E. Rodnyansky, A.G. Oliver, W.C. Boggess, J. Muldoon, Structure and compatibility of a magnesium electrolyte with a sulphur cathode, Nature Communications, 2 (2011) 427.
[98] H.J. Tian, T. Gao, X.G. Li, X.W. Wang, C. Luo, X.L. Fan, C.Y. Yang, L.M. Suo, Z.H. Ma, W.Q. Han, C.S. Wang, High power rechargeable magnesium/iodine battery chemistry, Nature Communications, 8 (2017).
[99] D. Aurbach, Y. Gofer, Z. Lu, A. Schechter, O. Chusid, H. Gizbar, Y. Cohen, V. Ashkenazi, M. Moshkovich, R. Turgeman, E. Levi, A short review on the comparison between Li battery systems and rechargeable magnesium battery technology, Journal of Power Sources, 97-8 (2001) 28-32.

[100] D. Aurbach, G.S. Suresh, E. Levi, A. Mitelman, O. Mizrahi, O. Chusid, M. Brunelli, Progress in rechargeable magnesium battery technology, Advanced Materials, 19 (2007) 4260-+.
[101] R.E. Doe, R. Han, J. Hwang, A.J. Gmitter, I. Shterenberg, H.D. Yoo, N. Pour, D. Aurbach, Novel, electrolyte solutions comprising fully inorganic salts with high anodic stability for rechargeable magnesium batteries, Chemical Communications, 50 (2014) 243-245.
[102] P. Canepa, S.H. Bo, G.S. Gautam, B. Key, W.D. Richards, T. Shi, Y.S. Tian, Y. Wang, J.C. Li, G. Ceder, High magnesium mobility in ternary spinel chalcogenides, Nature Communications, 8 (2017).
[103] O. Tutusaus, R. Mohtadi, T.S. Arthur, F. Mizuno, E.G. Nelson, Y.V. Sevryugina, An Efficient Halogen-Free Electrolyte for Use in Rechargeable Magnesium Batteries, Angewandte Chemie-International Edition, 54 (2015) 7900-7904.
[104] Z. Zhao-Karger, M.E. Gil Bardaji, O. Fuhr, M. Fichtner, A new class of non-corrosive, highly efficient electrolytes for rechargeable magnesium batteries, Journal of Materials Chemistry A, 5 (2017) 10815-10820.
[105] T.S. Arthur, N. Singh, M. Matsui, Electrodeposited Bi, Sb and Bi1-xSbx alloys as anodes for Mg-ion batteries, Electrochemistry Communications, 16 (2012) 103-106.
[106] S.J. Su, Z.G. Huang, Y. Nuli, F. Tuerxun, J. Yang, J.L. Wang, A novel rechargeable battery with a magnesium anode, a titanium dioxide cathode, and a magnesium borohydride/tetraglyme electrolyte, Chemical Communications, 51 (2015) 2641-2644.
[107] R. Mohtadi, M. Matsui, T.S. Arthur, S.J. Hwang, Magnesium Borohydride: From Hydrogen Storage to Magnesium Battery, Angewandte Chemie-International Edition, 51 (2012) 9780-9783.
[108] X.H. Yao, J.R. Luo, Q. Dong, D.W. Wang, A rechargeable non-aqueous Mg-Br-2 battery, Nano Energy, 28 (2016) 440-446.
[109] R.G. Zhang, C. Ling, F. Mizuno, A conceptual magnesium battery with ultrahigh rate capability, Chemical Communications, 51 (2015) 1487-1490.
[110] X.P. Gao, A. Mariani, S. Jeong, X. Liu, X.W. Dou, M. Ding, A. Moretti, S. Passerini, Prototype rechargeable magnesium batteries using ionic liquid electrolytes, Journal of Power Sources, 423 (2019) 52-59.
[111] Z.Q. Rong, R. Malik, P. Canepa, G.S. Gautam, M. Liu, A. Jain, K. Persson, G. Ceder, Materials Design Rules for Multivalent Ion Mobility in Intercalation Structures, Chemistry of Materials, 27 (2015) 6016-6021.
[112] M.L. Mao, T. Gao, S.Y. Hou, C.S. Wang, A critical review of cathodes for rechargeable Mg batteries, Chemical Society Reviews, 47 (2018) 8804-8841.
[113] O. Peña, Chevrel phases: Past, present and future, Physica C: Superconductivity and its Applications, 514 (2015) 95-112.
[114] M.D. Levi, E. Lancri, E. Levi, H. Gizbar, Y. Gofer, D. Aurbach, The effect of the anionic framework of Mo6X8 Chevrel phase (X=S, Se) on the thermodynamics and the kinetics of the electrochemical insertion of Mg2+ ions, Solid State Ionics, 176 (2005) 1695-1699.
[115] L.F. Wan, B.R. Perdue, C.A. Apblett, D. Prendergast, Mg Desolvation and Intercalation Mechanism at the Mo6S8 Chevrel Phase Surface, Chemistry of Materials, 27 (2015) 5932-5940.
[116] M. Liu, Z.Q. Rong, R. Malik, P. Canepa, A. Jain, G. Ceder, K.A. Persson, Spinel compounds as multivalent battery cathodes: a systematic evaluation based on ab initio calculations, Energy & Environmental Science, 8 (2015) 964-974.
[117] M. Cabello, R. Alcántara, F. Nacimiento, G. Ortiz, P. Lavela, J.L. Tirado, Electrochemical and chemical insertion/deinsertion of magnesium in spinel-type MgMn2O4 and lambda-MnO2 for both aqueous and non-aqueous magnesium-ion batteries, CrystEngComm, 17 (2015) 8728-8735.
[118] Z. Feng, X. Chen, L. Qiao, A.L. Lipson, T.T. Fister, L. Zeng, C. Kim, T. Yi, N. Sa, D.L. Proffit, A.K. Burrell, J. Cabana, B.J. Ingram, M.D. Biegalski, M.J. Bedzyk, P. Fenter, Phase-Controlled Electrochemical Activity of Epitaxial Mg-Spinel Thin Films, ACS Applied Materials & Interfaces, 7 (2015) 28438-28443.
[119] M. Liu, A. Jain, Z.Q. Rong, X.H. Qu, P. Canepa, R. Malik, G. Ceder, K.A. Persson, Evaluation of sulfur spinel compounds for multivalent battery cathode applications, Energy & Environmental Science, 9 (2016) 3201-3209.
[120] X. Sun, P. Bonnick, V. Duffort, M. Liu, Z. Rong, K.A. Persson, G. Ceder, L.F. Nazar, A high capacity thiospinel cathode for Mg batteries, Energy & Environmental Science, 9 (2016) 2273-2277.
[121] A. Wustrow, B. Key, P.J. Phillips, N.Y. Sa, A.S. Lipton, R.F. Klie, J.T. Vaughey, K.R. Poeppelmeier, Synthesis and Characterization of MgCr2S4 Thiospinel as a Potential Magnesium Cathode, Inorganic Chemistry, 57 (2018) 8634-8638.
[122] X. Sun, P. Bonnick, L.F. Nazar, Layered TiS2 Positive Electrode for Mg Batteries, ACS Energy Letters, 1 (2016) 297-301.
[123] Y. Gu, Y. Katsura, T. Yoshino, H. Takagi, K. Taniguchi, Rechargeable magnesium-ion battery based on a TiSe2-cathode with d-p orbital hybridized electronic structure, Scientific Reports, 5 (2015) 12486.
[124] M. Arsentev, A. Missyul, A.V. Petrov, M. Hammouri, TiS3 Magnesium Battery Material: Atomic-Scale Study of Maximum Capacity and Structural Behavior, Journal of Physical Chemistry C, 121 (2017) 15509-15515.

[125] P. Novak, R. Imhof, O. Haas, Magnesium insertion electrodes for rechargeable nonaqueous batteries - a competitive alternative to lithium?, Electrochimica Acta, 45 (1999) 351-367.
[126] G. Gershinsky, H.D. Yoo, Y. Gofer, D. Aurbach, Electrochemical and Spectroscopic Analysis of Mg2+ Intercalation into Thin Film Electrodes of Layered Oxides: V2O5 and MoO3, Langmuir, 29 (2013) 10964-10972.
[127] Q.Y. An, Y.F. Li, H.D. Yoo, S. Chen, Q. Ru, L.Q. Mai, Y. Yao, Graphene decorated vanadium oxide nanowire aerogel for long-cycle-life magnesium battery cathodes, Nano Energy, 18 (2015) 265-272.
[128] P. Saha, P.H. Jampani, D. Hong, B. Gattu, J.A. Poston, A. Manivannan, M.K. Datta, P.N. Kumta, Synthesis and electrochemical study of Mg1.5MnO3: A defect spinel cathode for rechargeable magnesium battery, Materials Science and Engineering B-Advanced Functional Solid-State Materials, 202 (2015) 8-14.
[129] Y.R. Wang, X.L. Xue, P.Y. Liu, C.X. Wang, X. Yi, Y. Hu, L.B. Ma, G.Y. Zhu, R.P. Chen, T. Chen, J. Ma, J. Liu, Z. Jin, Atomic Substitution Enabled Synthesis of Vacancy-Rich Two-Dimensional Black TiO2-x Nanoflakes for High-Performance Rechargeable Magnesium Batteries, Acs Nano, 12 (2018) 12492-12502.
[130] N.H. Zainol, D. Hambali, Z. Osman, N. Kamarulzaman, R. Rusdi, Synthesis and characterization of Ti-doped MgMn2O4 cathode material for magnesium ion batteries, Ionics, 25 (2019) 133-139.
[131] M.J. Aragon, P. Lavela, P. Recio, R. Alcantara, J.L. Tirado, On the influence of particle morphology to provide high performing chemically desodiated C@NaV2(PO4)(3) as cathode for rechargeable magnesium batteries, Journal of Electroanalytical Chemistry, 827 (2018) 128-136.
[132] H. Tang, F.Y. Xiong, Y.L. Jiang, C.Y. Pei, S.S. Tan, W. Yang, M.S. Li, Q.Y. An, L.Q. Mai, Alkali ions pre-intercalated layered vanadium oxide nanowires for stable magnesium ions storage, Nano Energy, 58 (2019) 347-354.
[133] N. Wu, Z.Z. Yang, H.R. Yao, Y.X. Yin, L. Gu, Y.G. Guo, Improving the Electrochemical Performance of the Li4Ti5O12 Electrode in a Rechargeable Magnesium Battery by Lithium-Magnesium Co-Intercalation, Angewandte Chemie-International Edition, 54 (2015) 5757-5761.
[134] Z.Y. Li, X.K. Mu, Z. Zhao-Karger, T. Diemant, R.J. Behm, C. Kubel, M. Fichtner, Fast kinetics of multivalent intercalation chemistry enabled by solvated magnesium-ions into self-established metallic layered materials, Nature Communications, 9 (2018).
[135] L.M. Zhou, Q. Liu, Z.H. Zhang, K. Zhang, F.Y. Xiong, S.S. Tan, Q.Y. An, Y.M. Kang, Z. Zhou, L.Q. Mai, Interlayer-Spacing-Regulated VOPO4 Nanosheets with Fast Kinetics for High-Capacity and Durable Rechargeable Magnesium Batteries, Advanced Materials, 30 (2018).
[136] B. Anasori, M.R. Lukatskaya, Y. Gogotsi, 2D metal carbides and nitrides (MXenes) for energy storage, Nature Reviews Materials, 2 (2017) 16098.
[137] G.S. Gautam, P. Canepa, R. Malik, M. Liu, K. Perssonb, G. Ceder, First-principles evaluation of multi-valent cation insertion into orthorhombic V2O5, Chemical Communications, 51 (2015) 13619-13622.
[138] D. Muthuraj, S. Mitra, Reversible Mg insertion into chevrel phase Mo6S8 cathode: Preparation, electrochemistry and X-ray photoelectron spectroscopy study, Materials Research Bulletin, 101 (2018) 167-174.
[139] A. Djire, A. Bos, J. Liu, H. Zhang, E.M. Miller, N.R. Neale, Pseudocapacitive Storage in Nanolayered Ti2NTx MXene Using Mg-Ion Electrolyte, ACS Applied Nano Materials, 2 (2019) 2785-2795.
[140] J. Heath, H.R. Chen, M.S. Islam, MgFeSiO4 as a potential cathode material for magnesium batteries: ion diffusion rates and voltage trends, Journal of Materials Chemistry A, 5 (2017) 13161-13167.
[141] R.G. Zhang, C. Ling, Unveil the Chemistry of Olivine FePO4 as Magnesium Battery Cathode, Acs Applied Materials & Interfaces, 8 (2016) 18018-18026.
[142] Y.N. Nuli, J. Yang, Y.S. Li, J.L. Wang, Mesoporous magnesium manganese silicate as cathode materials for rechargeable magnesium batteries, Chemical Communications, 46 (2010) 3794-3796.
[143] Z. Zhao-Karger, M. Fichtner, Beyond Intercalation Chemistry for Rechargeable Mg Batteries: A Short Review and Perspective, Frontiers in Chemistry, 6 (2019).
[144] X. Zhou, J. Tian, J. Hu, C. Li, High Rate Magnesium–Sulfur Battery with Improved Cyclability Based on Metal–Organic Framework Derivative Carbon Host, Advanced Materials, 30 (2018) 1704166.
[145] J.G. Connell, B. Genorio, P.P. Lopes, D. Strmcnik, V.R. Stamenkovic, N.M. Markovic, Tuning the Reversibility of Mg Anodes via Controlled Surface Passivation by H2O/Cl- in Organic Electrolytes, Chemistry of Materials, 28 (2016) 8268-8277.
[146] B. Peng, J. Liang, Z.L. Tao, J. Chen, Magnesium nanostructures for energy storage and conversion, Journal of Materials Chemistry, 19 (2009) 2877-2883.
[147] D. Er, E. Detsi, H. Kumar, V.B. Shenoy, Defective Graphene and Graphene Allotropes as High-Capacity Anode Materials for Mg Ion Batteries, Acs Energy Letters, 1 (2016) 638-645.
[148] R. Attias, M. Salama, B. Hirsch, Y. Goffer, D. Aurbach, Anode-Electrolyte Interfaces in Secondary Magnesium Batteries, Joule, 3 (2019) 27-52.
[149] M. Matsui, H. Kuwata, D. Mori, N. Imanishi, M. Mizuhata, Destabilized Passivation Layer on Magnesium-Based Intermetallics as Potential Anode Active Materials for Magnesium Ion Batteries, Frontiers in Chemistry, 7 (2019).

[150] R.A. DiLeo, Q. Zhang, A.C. Marschilok, K.J. Takeuchi, E.S. Takeuchi, Composite Anodes for Secondary Magnesium Ion Batteries Prepared via Electrodeposition of Nanostructured Bismuth on Carbon Nanotube Substrates, Ecs Electrochemistry Letters, 4 (2015) A10-A14.
[151] Y.Y. Shao, M. Gu, X.L. Li, Z.M. Nie, P.J. Zuo, G.S. Li, T.B. Liu, J. Xiao, Y.W. Cheng, C.M. Wang, J.G. Zhang, J. Liu, Highly Reversible Mg Insertion in Nanostructured Bi for Mg Ion Batteries, Nano Letters, 14 (2014) 255-260.
[152] C.C. Chen, J.B. Wang, Q. Zhao, Y.J. Wang, J. Chen, Layered $Na_2Ti_3O_7/MgNaTi_3O_7/Mg_{0.5}NaTi_3O_7$ Nanoribbons as High-Performance Anode of Rechargeable Mg-Ion Batteries, Acs Energy Letters, 1 (2016) 1165-1172.
[153] X. Liu, S.Z. Liu, J.L. Xue, Discharge performance of the magnesium anodes with different phase constitutions for Mg-air batteries, Journal of Power Sources, 396 (2018) 667-674.
[154] H. Benzidi, M. Lakhal, M. Garara, M. Abdellaoui, A. Benyoussef, A. El kenz, O. Mounkachi, Arsenene monolayer as an outstanding anode material for (Li/Na/Mg)-ion batteries: density functional theory, Physical Chemistry Chemical Physics, 21 (2019) 19951-19962.
[155] J. Muldoon, C.B. Bucur, T. Gregory, Fervent Hype behind Magnesium Batteries: An Open Call to Synthetic Chemists-Electrolytes and Cathodes Needed, Angewandte Chemie-International Edition, 56 (2017) 12064-12084.
[156] D. Aurbach, I. Weissman, Y. Gofer, E. Levi, Nonaqueous magnesium electrochemistry and its application in secondary batteries, Chemical Record, 3 (2003) 61-73.
[157] C.J. Barile, R. Spatney, K.R. Zavadil, A.A. Gewirth, Investigating the Reversibility of in Situ Generated Magnesium Organohaloaluminates for Magnesium Deposition and Dissolution, The Journal of Physical Chemistry C, 118 (2014) 10694-10699.
[158] E.G. Nelson, S.I. Brody, J.W. Kampf, B.M. Bartlett, A magnesium tetraphenylaluminate battery electrolyte exhibits a wide electrochemical potential window and reduces stainless steel corrosion, Journal of Materials Chemistry A, 2 (2014) 18194-18198.
[159] A.J. Crowe, K.K. Stringham, B.M. Bartlett, Fluorinated Alkoxide-Based Magnesium-Ion Battery Electrolytes that Demonstrate Li-Ion-Battery-Like High Anodic Stability and Solution Conductivity, Acs Applied Materials & Interfaces, 8 (2016) 23060-23065.
[160] J.T. Herb, C.A. Nist-Lund, C.B. Arnold, A Fluorinated Alkoxyaluminate Electrolyte for Magnesium-Ion Batteries, ACS Energy Letters, 1 (2016) 1227-1232.
[161] T. Liu, Y. Shao, G. Li, M. Gu, J. Hu, S. Xu, Z. Nie, X. Chen, C. Wang, J. Liu, A facile approach using $MgCl_2$ to formulate high performance $Mg^{2+}$ electrolytes for rechargeable Mg batteries, Journal of Materials Chemistry A, 2 (2014) 3430-3438.
[162] J. Muldoon, C.B. Bucur, A.G. Oliver, T. Sugimoto, M. Matsui, H.S. Kim, G.D. Allred, J. Zajicek, Y. Kotani, Electrolyte roadblocks to a magnesium rechargeable battery, Energy & Environmental Science, 5 (2012) 5941-5950.
[163] F. Bertasi, K. Vezzu, G. Nawn, G. Pagot, V. Di Noto, Interplay Between Structure and Conductivity in 1-Ethyl-3-methylimidazolium tetrafluoroborate/(delta-$MgCl_2$)(f) Electrolytes for Magnesium Batteries, Electrochimica Acta, 219 (2016) 152-162.
[164] Z. Zhao-Karger, X. Zhao, O. Fuhr, M. Fichtner, Bisamide based non-nucleophilic electrolytes for rechargeable magnesium batteries, RSC Advances, 3 (2013) 16330-16335.
[165] S.Y. Ha, Y.W. Lee, S.W. Woo, B. Koo, J.S. Kim, J. Cho, K.T. Lee, N.S. Choi, Magnesium(II) Bis(trifluoromethane sulfonyl) Imide-Based Electrolytes with Wide Electrochemical Windows for Rechargeable Magnesium Batteries, Acs Applied Materials & Interfaces, 6 (2014) 4063-4073.
[166] N. Amir, Y. Vestfrid, O. Chusid, Y. Gofer, D. Aurbach, Progress in nonaqueous magnesium electrochemistry, Journal of Power Sources, 174 (2007) 1234-1240.
[167] Y.Y. Shao, T.B. Liu, G.S. Li, M. Gu, Z.M. Nie, M. Engelhard, J. Xiao, D.P. Lv, C.M. Wang, J.G. Zhang, J. Liu, Coordination Chemistry in magnesium battery electrolytes: how ligands affect their performance, Scientific Reports, 3 (2013).
[168] J. Chang, R.T. Haasch, J. Kim, T. Spila, P.V. Braun, A.A. Gewirth, R.G. Nuzzo, Synergetic Role of $Li^+$ during Mg Electrodeposition/Dissolution in Borohydride Diglyme Electrolyte Solution: Voltammetric Stripping Behaviors on a Pt Microelectrode Indicative of Mg-Li Alloying and Facilitated Dissolution, Acs Applied Materials & Interfaces, 7 (2015) 2494-2502.
[169] O. Zavorotynska, A. El-Kharbachi, S. Deledda, B.C. Hauback, Recent progress in magnesium borohydride $Mg(BH_4)_2$: Fundamentals and applications for energy storage, International Journal of Hydrogen Energy, 41 (2016) 14387-14403.
[170] T. Watkins, A. Kumar, D.A. Buttry, Designer Ionic Liquids for Reversible Electrochemical Deposition/Dissolution of Magnesium, Journal of the American Chemical Society, 138 (2016) 641-650.
[171] R. Mohtadi, F. Mizuno, Magnesium batteries: Current state of the art, issues and future perspectives, Beilstein Journal of Nanotechnology, 5 (2014) 1291-1311.

[172] M.N. Guzik, R. Mohtadi, S. Sartori, Lightweight complex metal hydrides for Li-, Na-, and Mg-based batteries, Journal of Materials Research, 34 (2019) 877-904.
[173] S. Su, Z. Huang, Y. NuLi, F. Tuerxun, J. Yang, J. Wang, A novel rechargeable battery with a magnesium anode, a titanium dioxide cathode, and a magnesium borohydride/tetraglyme electrolyte, Chemical Communications, 51 (2015) 2641-2644.
[174] X. Yao, J. Luo, Q. Dong, D. Wang, A rechargeable non-aqueous Mg-Br2 battery, Nano Energy, 28 (2016) 440-446.
[175] J. Tian, D. Cao, X. Zhou, J. Hu, M. Huang, C. Li, High-Capacity Mg–Organic Batteries Based on Nanostructured Rhodizonate Salts Activated by Mg–Li Dual-Salt Electrolyte, ACS Nano, 12 (2018) 3424-3435.
[176] D. Wang, X.W. Gao, Y.H. Chen, L.Y. Jin, C. Kuss, P.G. Bruce, Plating and stripping calcium in an organic electrolyte, Nature Materials, 17 (2018) 16-20.
[177] R. Cohen, Y. Lavi, E. Peled, CALORIMETRIC STUDY OF THE CALCIUM/SR(ALCL4)2-SOCL2 BATTERY, Journal of the Electrochemical Society, 137 (1990) 2648-2653.
[178] R.J. Gummow, G. Vamvounis, M.B. Kannan, Y.H. He, Calcium-Ion Batteries: Current State-of-the-Art and Future Perspectives, Advanced Materials, 30 (2018).
[179] D. Aurbach, R. Skaletsky, Y. Gofer, The Electrochemical Behavior of Calcium Electrodes in a Few Organic Electrolytes, Journal of The Electrochemical Society, 138 (1991) 3536-3545.
[180] A. Ponrouch, C. Frontera, F. Bardé, M.R. Palacín, Towards a calcium-based rechargeable battery, Nature Materials, 15 (2015) 169.
[181] C.P. Grey, J.M. Tarascon, Sustainability and in situ monitoring in battery development, Nature Materials, 16 (2016) 45.
[182] K.A. See, J.A. Gerbec, Y.S. Jun, F. Wudl, G.D. Stucky, R. Seshadri, A High Capacity Calcium Primary Cell Based on the Ca-S System, Advanced Energy Materials, 3 (2013) 1056-1061.
[183] M. Wang, C. Jiang, S. Zhang, X. Song, Y. Tang, H.-M. Cheng, Reversible calcium alloying enables a practical room-temperature rechargeable calcium-ion battery with a high discharge voltage, Nature Chemistry, 10 (2018) 667-672.
[184] S. Wu, F. Zhang, Y.B. Tang, A Novel Calcium-Ion Battery Based on Dual-Carbon Configuration with High Working Voltage and Long Cycling Life, Advanced Science, 5 (2018).
[185] M. Hayashi, H. Arai, H. Ohtsuka, Y. Sakurai, Electrochemical insertion/extraction of calcium ions using crystalline vanadium oxide, Electrochemical and Solid State Letters, 7 (2004) A119-A121.
[186] M. Bervas, L.C. Klein, G.G. Amatucci, Vanadium oxide–propylene carbonate composite as a host for the intercalation of polyvalent cations, Solid State Ionics, 176 (2005) 2735-2747.
[187] M.E. Arroyo-de Dompablo, C. Krich, J. Nava-Avendaño, M.R. Palacín, F. Bardé, In quest of cathode materials for Ca ion batteries: the CaMO3 perovskites (M = Mo, Cr, Mn, Fe, Co, and Ni), Physical Chemistry Chemical Physics, 18 (2016) 19966-19972.
[188] A.L. Lipson, S. Kim, B.F. Pan, C. Liao, T.T. Fister, B.J. Ingram, Calcium intercalation into layered fluorinated sodium iron phosphate, Journal of Power Sources, 369 (2017) 133-137.
[189] C.P. Grey, J.M. Tarascon, Sustainability and in situ monitoring in battery development, Nature Materials, 16 (2017) 45-56.
[190] Z. Li, O. Fuhr, M. Fichtner, Z. Zhao-Karger, Towards stable and efficient electrolytes for room-temperature rechargeable calcium batteries, Energy & Environmental Science, (2019).
[191] A.H.F. Niaei, T. Hussain, M. Hankel, D.J. Searles, Hydrogenated defective graphene as an anode material for sodium and calcium ion batteries: A density functional theory study, Carbon, 136 (2018) 73-84.
[192] Z.P. Yao, V.I. Hegde, A. Aspuru-Guzik, C. Wolverton, Discovery of Calcium-Metal Alloy Anodes for Reversible Ca-Ion Batteries, Advanced Energy Materials, 9 (2019).
[193] J.C. Pramudita, D. Sehrawat, D. Goonetilleke, N. Sharma, An Initial Review of the Status of Electrode Materials for Potassium-Ion Batteries, Advanced Energy Materials, 7 (2017) 1602911.
[194] C. Xu, B. Li, H. Du, F. Kang, Energetic Zinc Ion Chemistry: The Rechargeable Zinc Ion Battery, Angewandte Chemie International Edition, 51 (2012) 933-935.
[195] G. Fang, J. Zhou, A. Pan, S. Liang, Recent Advances in Aqueous Zinc-Ion Batteries, ACS Energy Letters, 3 (2018) 2480-2501.
[196] Y. Li, H. Dai, Recent advances in zinc–air batteries, Chemical Society Reviews, 43 (2014) 5257-5275.
[197] J.F. Parker, C.N. Chervin, I.R. Pala, M. Machler, M.F. Burz, J.W. Long, D.R. Rolison, Rechargeable nickel–3D zinc batteries: An energy-dense, safer alternative to lithium-ion, Science, 356 (2017) 415-418.
[198] F.Y. Cheng, J. Chen, X.L. Gou, P.W. Shen, High-Power Alkaline Zn–MnO2 Batteries Using γ-MnO2 Nanowires/Nanotubes and Electrolytic Zinc Powder, Advanced Materials, 17 (2005) 2753-2756.
[199] X. Wang, F. Wang, L. Wang, M. Li, Y. Wang, B. Chen, Y. Zhu, L. Fu, L. Zha, L. Zhang, Y. Wu, W. Huang, An Aqueous Rechargeable Zn//Co3O4 Battery with High Energy Density and Good Cycling Behavior, Advanced Materials, 28 (2016) 4904-4911.

[200] T. Shoji, M. Hishinuma, T. Yamamoto, Zinc-manganese dioxide galvanic cell using zinc sulphate as electrolyte. Rechargeability of the cell, Journal of Applied Electrochemistry, 18 (1988) 521-526.
[201] F. Wang, O. Borodin, T. Gao, X. Fan, W. Sun, F. Han, A. Faraone, J.A. Dura, K. Xu, C. Wang, Highly reversible zinc metal anode for aqueous batteries, Nature Materials, 17 (2018) 543-549.
[202] N. Zhang, F. Cheng, Y. Liu, Q. Zhao, K. Lei, C. Chen, X. Liu, J. Chen, Cation-Deficient Spinel $ZnMn_2O_4$ Cathode in $Zn(CF_3SO_3)_2$ Electrolyte for Rechargeable Aqueous Zn-Ion Battery, Journal of the American Chemical Society, 138 (2016) 12894-12901.
[203] L. Zhang, L. Chen, X. Zhou, Z. Liu, Towards High-Voltage Aqueous Metal-Ion Batteries Beyond 1.5 V: The Zinc/Zinc Hexacyanoferrate System, Advanced Energy Materials, 5 (2015) 1400930.
[204] G. Li, Z. Yang, Y. Jiang, C. Jin, W. Huang, X. Ding, Y. Huang, Towards polyvalent ion batteries: A zinc-ion battery based on NASICON structured $Na_3V_2(PO_4)_3$, Nano Energy, 25 (2016) 211-217.
[205] W. Li, K. Wang, S. Cheng, K. Jiang, A long-life aqueous Zn-ion battery based on $Na_3V_2(PO_4)_2F_3$ cathode, Energy Storage Materials, 15 (2018) 14-21.
[206] M.H. Alfaruqi, V. Mathew, J. Gim, S. Kim, J. Song, J.P. Baboo, S.H. Choi, J. Kim, Electrochemically Induced Structural Transformation in a γ-$MnO_2$ Cathode of a High Capacity Zinc-Ion Battery System, Chemistry of Materials, 27 (2015) 3609-3620.
[207] C. Wei, C. Xu, B. Li, H. Du, F. Kang, Preparation and characterization of manganese dioxides with nano-sized tunnel structures for zinc ion storage, Journal of Physics and Chemistry of Solids, 73 (2012) 1487-1491.
[208] D. Kundu, B.D. Adams, V. Duffort, S.H. Vajargah, L.F. Nazar, A high-capacity and long-life aqueous rechargeable zinc battery using a metal oxide intercalation cathode, Nature Energy, 1 (2016) 16119.
[209] M. Song, H. Tan, D. Chao, H.J. Fan, Recent Advances in Zn-Ion Batteries, Advanced Functional Materials, 28 (2018) 1802564.
[210] J. Ming, J. Guo, C. Xia, W. Wang, H.N. Alshareef, Zinc-ion batteries: Materials, mechanisms, and applications, Materials Science and Engineering: R: Reports, 135 (2019) 58-84.
[211] N. Zhang, Y. Dong, M. Jia, X. Bian, Y. Wang, M. Qiu, J. Xu, Y. Liu, L. Jiao, F. Cheng, Rechargeable Aqueous Zn–$V_2O_5$ Battery with High Energy Density and Long Cycle Life, ACS Energy Letters, 3 (2018) 1366-1372.
[212] L. Wang, K.-W. Huang, J. Chen, J. Zheng, Ultralong cycle stability of aqueous zinc-ion batteries with zinc vanadium oxide cathodes, Science Advances, 5 (2019) eaax4279.
[213] M.-C. Lin, M. Gong, B. Lu, Y. Wu, D.-Y. Wang, M. Guan, M. Angell, C. Chen, J. Yang, B.-J. Hwang, H. Dai, An ultrafast rechargeable aluminium-ion battery, Nature, 520 (2015) 324.
[214] N. Jayaprakash, S.K. Das, L.A. Archer, The rechargeable aluminum-ion battery, Chemical Communications, 47 (2011) 12610-12612.
[215] J.V. Rani, V. Kanakaiah, T. Dadmal, M.S. Rao, S. Bhavanarushi, Fluorinated Natural Graphite Cathode for Rechargeable Ionic Liquid Based Aluminum–Ion Battery, Journal of The Electrochemical Society, 160 (2013) A1781-A1784.
[216] F. Wang, F. Yu, X. Wang, Z. Chang, L. Fu, Y. Zhu, Z. Wen, Y. Wu, W. Huang, Aqueous Rechargeable Zinc/Aluminum Ion Battery with Good Cycling Performance, ACS Applied Materials & Interfaces, 8 (2016) 9022-9029.
[217] M.A. Reddy, M. Fichtner, Batteries based on fluoride shuttle, Journal of Materials Chemistry, 21 (2011) 17059-17062.
[218] C. Rongeat, M.A. Reddy, R. Witter, M. Fichtner, Solid Electrolytes for Fluoride Ion Batteries: Ionic Conductivity in Polycrystalline Tysonite-Type Fluorides, Acs Applied Materials & Interfaces, 6 (2014) 2103-2110.
[219] C. Rongeat, M.A. Reddy, T. Diemant, R.J. Behm, M. Fichtner, Development of new anode composite materials for fluoride ion batteries, Journal of Materials Chemistry A, 2 (2014) 20861-20872.
[220] I. Mohammad, R. Witter, M. Fichtner, M.A. Reddy, Introducing Interlayer Electrolytes: Toward Room-Temperature High-Potential Solid-State Rechargeable Fluoride Ion Batteries, ACS Applied Energy Materials, 2 (2019) 1553-1562.
[221] I. Mohammad, R. Witter, M. Fichtner, M. Anji Reddy, Room-Temperature, Rechargeable Solid-State Fluoride-Ion Batteries, ACS Applied Energy Materials, 1 (2018) 4766-4775.
[222] X. Zhao, S. Ren, M. Bruns, M. Fichtner, Chloride ion battery: A new member in the rechargeable battery family, Journal of Power Sources, 245 (2014) 706-711.
[223] P. Gao, M.A. Reddy, X. Mu, T. Diemant, L. Zhang, Z. Zhao-Karger, V.S.K. Chakravadhanula, O. Clemens, R.J. Behm, M. Fichtner, VOCl as a Cathode for Rechargeable Chloride Ion Batteries, Angewandte Chemie, 128 (2016) 4357-4362.
[224] I.V. Murin, O.V. Glumov, N.A. Mel'nikova, Solid electrolytes with predominant chloride conductivity, Russian Journal of Electrochemistry, 45 (2009) 411-416.
[225] N. Imanaka, K. Okamoto, G.Y. Adachi, Water-insoluble lanthanum oxychloride-based solid electrolytes with ultra-high chloride ion conductivity, Angewandte Chemie-International Edition, 41 (2002) 3890-3892.

[226] K. Yamada, Y. Kuranaga, K. Ueda, S. Goto, T. Okuda, Y. Furukawa, Phase transition and electric conductivity of ASnCl(3) (A = Cs and CH3NH3), Bulletin of the Chemical Society of Japan, 71 (1998) 127-134.
[227] C. Wan, R.V. Denys, V.A. Yartys, In situ neutron powder diffraction study of phase-structural transformations in the La–Mg–Ni battery anode alloy, Journal of Alloys and Compounds, 670 (2016) 210-216.
[228] F. Cuevas, J.M. Joubert, M. Latroche, A. Percheron-Guégan, Intermetallic compounds as negative electrodes of Ni/MH batteries, Applied Physics A, 72 (2001) 225-238.
[229] P. Notten, Rechargeable nickel-metalhydride batteries: a successful new concept, in: Interstitial intermetallic alloys, Springer, 1995, pp. 151-195.
[230] P. Notten, M. Latroche, Secondary batteries-nickel systems: nickel–metal hydride: metal hydrides, in: Encyclopedia of electrochemical power sources, Elsevier, 2009, pp. 502-521.
[231] V. Yartys, D. Noréus, M. Latroche, Metal hydrides as negative electrode materials for Ni–MH batteries, Applied Physics A, 122 (2016) 43.
[232] Kawasaki, Battery Power System (BPS) for Railways, in, 2019.
[233] Y. Khan, Intermetallic compounds in the cobalt-rich part of the R-cobalt systems (R= Ce, La, Ce・La), Journal of the Less Common Metals, 34 (1974) 191-200.
[234] R.V. Denys, A.B. Riabov, V.A. Yartys, M. Sato, R.G. Delaplane, Mg substitution effect on the hydrogenation behaviour, thermodynamic and structural properties of the La2Ni7–H(D)2 system, Journal of Solid State Chemistry, 181 (2008) 812-821.
[235] J.C. Crivello, J. Zhang, M. Latroche, Structural Stability of ABy Phases in the (La,Mg)–Ni System Obtained by Density Functional Theory Calculations, The Journal of Physical Chemistry C, 115 (2011) 25470-25478.
[236] T. Kohno, H. Yoshida, F. Kawashima, T. Inaba, I. Sakai, M. Yamamoto, M. Kanda, Hydrogen storage properties of new ternary system alloys: La2MgNi9, La5Mg2Ni23, La3MgNi14, Journal of Alloys and Compounds, 311 (2000) L5-L7.
[237] A. Férey, F. Cuevas, M. Latroche, B. Knosp, P. Bernard, Elaboration and characterization of magnesium-substituted La5Ni19 hydride forming alloys as active materials for negative electrode in Ni-MH battery, Electrochimica Acta, 54 (2009) 1710-1714.
[238] L. Lemort, M. Latroche, B. Knosp, P. Bernard, Elaboration and characterization of new pseudo-binary hydride-forming phases Pr1. 5Mg0. 5Ni7 and Pr3. 75Mg1. 25Ni19: a comparison to the binary Pr2Ni7 and Pr5Ni19 ones, The Journal of Physical Chemistry C, 115 (2011) 19437-19444.
[239] J.C. Crivello, R.V. Denys, M. Dornheim, M. Felderhoff, D.M. Grant, J. Huot, T.R. Jensen, P. de Jongh, M. Latroche, G.S. Walker, C.J. Webb, V.A. Yartys, Mg-based compounds for hydrogen and energy storage, Applied Physics A, 122 (2016) 85.
[240] C.C. Nwakwuo, T. Holm, R.V. Denys, W. Hu, J.P. Maehlen, J.K. Solberg, V.A. Yartys, Effect of magnesium content and quenching rate on the phase structure and composition of rapidly solidified La2MgNi9 metal hydride battery electrode alloy, Journal of Alloys and Compounds, 555 (2013) 201-208.
[241] W.-K. Hu, R.V. Denys, C.C. Nwakwuo, T. Holm, J.P. Maehlen, J.K. Solberg, V.A. Yartys, Annealing effect on phase composition and electrochemical properties of the Co-free La2MgNi9 anode for Ni-metal hydride batteries, Electrochimica Acta, 96 (2013) 27-33.
[242] M. Latroche, F. Cuevas, W.-K. Hu, D. Sheptyakov, R.V. Denys, V.A. Yartys, Mechanistic and Kinetic Study of the Electrochemical Charge and Discharge of La2MgNi9 by in Situ Powder Neutron Diffraction, The Journal of Physical Chemistry C, 118 (2014) 12162-12169.
[243] A.A. Volodin, R.V. Denys, G.A. Tsirlina, B.P. Tarasov, M. Fichtner, V.A. Yartys, Hydrogen diffusion in La1.5Nd0.5MgNi9 alloy electrodes of the Ni/MH battery, Journal of Alloys and Compounds, 645 (2015) S288-S291.
[244] V. Yartys, R. Denys, Structure–properties relationship in RE3−xMgxNi9H10–13 (RE=La,Pr,Nd) hydrides for energy storage, Journal of Alloys and Compounds, 645 (2015) S412-S418.
[245] A.A. Volodin, C. Wan, R.V. Denys, G.A. Tsirlina, B.P. Tarasov, M. Fichtner, U. Ulmer, Y. Yu, C.C. Nwakwuo, V.A. Yartys, Phase-structural transformations in a metal hydride battery anode La1.5Nd0.5MgNi9 alloy and its electrochemical performance, International Journal of Hydrogen Energy, 41 (2016) 9954-9967.
[246] N.S. Nazer, R.V. Denys, V.A. Yartys, W.-K. Hu, M. Latroche, F. Cuevas, B.C. Hauback, P.F. Henry, L. Arnberg, In operando neutron diffraction study of LaNdMgNi9H13 as a metal hydride battery anode, Journal of Power Sources, 343 (2017) 502-512.
[247] I.E. Gabis, E.A. Evard, A.P. Voyt, V.G. Kuznetsov, B.P. Tarasov, J.C. Crivello, M. Latroche, R.V. Denys, W. Hu, V.A. Yartys, Modeling of metal hydride battery anodes at high discharge current densities and constant discharge currents, Electrochimica Acta, 147 (2014) 73-81.
[248] R.V. Denys, V.A. Yartys, Effect of magnesium on the crystal structure and thermodynamics of the La3−xMgxNi9 hydrides, Journal of Alloys and Compounds, 509 (2011) S540-S548.

[249] H. Wang, H. Lin, W. Cai, L. Ouyang, M. Zhu, Tuning kinetics and thermodynamics of hydrogen storage in light metal element based systems–a review of recent progress, Journal of Alloys and Compounds, 658 (2016) 280-300.
[250] C. Eyövge, T. Öztürk, Nafion Coated Mg50Ni50 and (La, Mg) 2Ni7 Negative Electrodes for NiMH Batteries, Journal of The Electrochemical Society, 165 (2018) A2203-A2208.
[251] L. Pasquini, The effects of nanostructure on the hydrogen sorption properties of magnesium-based metallic compounds: a review, Crystals, 8 (2018) 106.
[252] T. Yang, Q. Li, C. Liang, X. Wang, C. Xia, H. Wang, F. Yin, Y. Zhang, Microstructure and hydrogen absorption/desorption properties of Mg24Y3M (M= Ni, Co, Cu, Al) alloys, International Journal of Hydrogen Energy, 43 (2018) 8877-8887.
[253] L. Ouyang, J. Huang, H. Wang, J. Liu, M. Zhu, Progress of hydrogen storage alloys for Ni-MH rechargeable power batteries in electric vehicles: A review, Materials Chemistry and Physics, 200 (2017) 164-178.
[254] K.-H. Young, J. Nei, C. Wan, V.R. Denys, A.V. Yartys, Comparison of C14- and C15-Predomiated AB2 Metal Hydride Alloys for Electrochemical Applications, Batteries, 3 (2017).
[255] K.-H. Young, M.J. Koch, C. Wan, V.R. Denys, A.V. Yartys, Cell Performance Comparison between C14- and C15-Predomiated AB2 Metal Hydride Alloys, Batteries, 3 (2017).
[256] C. Wan, R.V. Denys, M. Lelis, D. Milčius, V.A. Yartys, Electrochemical studies and phase-structural characterization of a high-capacity La-doped AB2 Laves type alloy and its hydride, Journal of Power Sources, 418 (2019) 193-201.
[257] A.A. Volodin, R.V. Denys, C. Wan, I.D. Wijayanti, Suwarno, B.P. Tarasov, V.E. Antonov, V.A. Yartys, Study of hydrogen storage and electrochemical properties of AB2-type Ti0.15Zr0.85La0.03Ni1.2Mn0.7V0.12Fe0.12 alloy, Journal of Alloys and Compounds, 793 (2019) 564-575.
[258] I.D. Wijayanti, L. Mølmen, R.V. Denys, J. Nei, S. Gorsse, K. Young, M.N. Guzik, V. Yartys, The electrochemical performance of melt-spun C14-Laves type TiZr-based alloy, International Journal of Hydrogen Energy, (2019).
[259] I.D. Wijayanti, L. Mølmen, R.V. Denys, J. Nei, S. Gorsse, M.N. Guzik, K. Young, V. Yartys, Studies of Zr-based C15 type metal hydride battery anode alloys prepared by rapid solidification, Journal of Alloys and Compounds, 804 (2019) 527-537.
[260] M. Gutjahr, A new type of reversible negative electrode for alkaline storage batteries based on metal alloy hydrides, Power sources, 4 (1973) 79-91.
[261] F. Cuevas, M. Latroche, P. Ochin, A. Dezellus, A. Percheron-Guégan, Influence of polymorphism on the electrochemical properties of (Ti0.64Zr0.36)Ni alloys, Journal of Alloys and Compounds, 356-357 (2003) 730-733.
[262] A. Szajek, M. Jurczyk, E. Jankowska, The electronic and electrochemical properties of the TiFe-based alloys, Journal of alloys and compounds, 348 (2003) 285-292.
[263] B. Guiose, F. Cuevas, B. Décamps, E. Leroy, A. Percheron-Guégan, Microstructural analysis of the ageing of pseudo-binary (Ti, Zr) Ni intermetallic compounds as negative electrodes of Ni-MH batteries, Electrochimica Acta, 54 (2009) 2781-2789.
[264] K.-h. Young, J. Nei, The current status of hydrogen storage alloy development for electrochemical applications, Materials, 6 (2013) 4574-4608.
[265] H. Miyamura, M. Takada, K. Hirose, S. Kikuchi, Metal hydride electrodes using titanium–iron-based alloys, Journal of alloys and compounds, 356 (2003) 755-758.
[266] M.A. Gutjahr, Etude du comportement électrochimique de certains hydrures de métaux de transition en vue de leur application comme masse active de l'électrode négative dans des batteries secondaires, in, Université de Genève, 1974.
[267] E.W. Justi, H.H. Ewe, A.W. Kalberlah, N.M. Saridakis, M.H. Schaefer, Electrocatalysis in the nickel·titanium system, Energy Conversion, 10 (1970) 183-187.
[268] J. Kleperis, G. Wójcik, A. Czerwinski, J. Skowronski, M. Kopczyk, M. Beltowska-Brzezinska, Electrochemical behavior of metal hydrides, Journal of Solid State Electrochemistry, 5 (2001) 229-249.
[269] H. Emami, F. Cuevas, Cobalt induced multi-plateau behavior in TiNi-based Ni-MH electrodes, Energy Storage Materials, 8 (2017) 189-193.
[270] H. Emami, F. Cuevas, M. Latroche, Ti(Ni,Cu) pseudobinary compounds as efficient negative electrodes for Ni–MH batteries, Journal of Power Sources, 265 (2014) 182-191.
[271] L. Croguennec, M.R. Palacin, Recent Achievements on Inorganic Electrode Materials for Lithium-Ion Batteries, Journal of the American Chemical Society, 137 (2015) 3140-3156.
[272] L. Aymard, Y. Oumellal, J.-P. Bonnet, Metal hydrides: an innovative and challenging conversion reaction anode for lithium-ion batteries, Beilstein Journal of Nanotechnology, 6 (2015) 1821-1839.
[273] S. Sartori, F. Cuevas, M. Latroche, Metal hydrides used as negative electrode materials for Li-ion batteries, Applied Physics A, 122 (2016) 135.

[274] M. Latroche, D. Blanchard, F. Cuevas, A. El Kharbachi, B.C. Hauback, T.R. Jensen, P.E. de Jongh, S. Kim, N.S. Nazer, P. Ngene, S.-i. Orimo, D.B. Ravnsbæk, V.A. Yartys, Full-cell hydride-based solid-state Li batteries for energy storage, International Journal of Hydrogen Energy, 44 (2019) 7875-7887.
[275] A. El kharbachi, Y. Hu, M.H. Sørby, P.E. Vullum, J.P. Mæhlen, H. Fjellvåg, B.C. Hauback, Understanding Capacity Fading of MgH2 Conversion-Type Anodes via Structural Morphology Changes and Electrochemical Impedance, The Journal of Physical Chemistry C, 122 (2018) 8750-8759.
[276] Z. Qian, H. Zhang, G. Jiang, Y. Bai, Y. Ren, W. Du, R. Ahuja, Ab Initio Screening of Doped Mg(AlH4)2 Systems for Conversion-Type Lithium Storage, Materials, 12 (2019) 2599.
[277] Y. Oumellal, A. Rougier, G. Nazri, J. Tarascon, L. Aymard, Metal hydrides for lithium-ion batteries, Nature materials, 7 (2008) 916.
[278] Y. Oumellal, C. Zlotea, S. Bastide, C. Cachet-Vivier, E. Léonel, S. Sengmany, E. Leroy, L. Aymard, J.-P. Bonnet, M. Latroche, Bottom-up preparation of MgH 2 nanoparticles with enhanced cycle life stability during electrochemical conversion in Li-ion batteries, Nanoscale, 6 (2014) 14459-14466.
[279] W. Zaïdi, Y. Oumellal, J.-P. Bonnet, J. Zhang, F. Cuevas, M. Latroche, J.-L. Bobet, L. Aymard, Carboxymethylcellulose and carboxymethylcellulose-formate as binders in MgH2–carbon composites negative electrode for lithium-ion batteries, Journal of Power Sources, 196 (2011) 2854-2857.
[280] A. El kharbachi, H.F. Andersen, M.H. Sørby, P.E. Vullum, J.P. Mæhlen, B.C. Hauback, Morphology effects in MgH2 anode for lithium ion batteries, International Journal of Hydrogen Energy, 42 (2017) 22551-22556.
[281] N. Berti, E. Hadjixenophontos, F. Cuevas, J. Zhang, A. Lacoste, P. Dubot, G. Schmitz, M. Latroche, Thin films as model system for understanding the electrochemical reaction mechanisms in conversion reaction of MgH2 with lithium, Journal of Power Sources, 402 (2018) 99-106.
[282] L. Huang, L. Aymard, J.-P. Bonnet, MgH 2–TiH 2 mixture as an anode for lithium-ion batteries: synergic enhancement of the conversion electrode electrochemical performance, Journal of Materials Chemistry A, 3 (2015) 15091-15096.
[283] N. Berti, F. Cuevas, J. Zhang, M. Latroche, Enhanced reversibility of the electrochemical Li conversion reaction with MgH2–TiH2 nanocomposites, International Journal of Hydrogen Energy, 42 (2017) 22615-22621.
[284] A.H. Dao, N. Berti, P. López-Aranguren, J. Zhang, F. Cuevas, C. Jordy, M. Latroche, Electrochemical properties of MgH2 – TiH2 nanocomposite as active materials for all-solid-state lithium batteries, Journal of Power Sources, 397 (2018) 143-149.
[285] W. Zaidi, J.-P. Bonnet, J. Zhang, F. Cuevas, M. Latroche, S. Couillaud, J.-L. Bobet, M.T. Sougrati, J.-C. Jumas, L. Aymard, Reactivity of complex hydrides Mg2FeH6, Mg2CoH5 and Mg2NiH4 with lithium ion: Far from equilibrium electrochemically driven conversion reactions, International Journal of Hydrogen Energy, 38 (2013) 4798-4808.
[286] J. Zhang, W. Zaïdi, V. Paul-Boncour, K. Provost, A. Michalowicz, F. Cuevas, M. Latroche, S. Belin, J.-P. Bonnet, L. Aymard, XAS investigations on nanocrystalline Mg 2 FeH 6 used as a negative electrode of Li-ion batteries, Journal of Materials Chemistry A, 1 (2013) 4706-4717.
[287] K. Provost, J. Zhang, W. Zaïdi, V. Paul-Boncour, J.-P. Bonnet, F. Cuevas, S. Belin, L. Aymard, M. Latroche, X-ray Absorption Spectroscopy and X-ray Diffraction Studies of the Thermal and Li-Driven Electrochemical Dehydrogenation of Nanocrystalline Complex Hydrides Mg2MH x (M= Co, Ni), The Journal of Physical Chemistry C, 118 (2014) 29554-29567.
[288] P. Huen, D.B. Ravnsbæk, Insight into poor cycling stability of MgH2 anodes, Journal of The Electrochemical Society, 164 (2017) A3138-A3143.
[289] P. Huen, D.B. Ravnsbæk, All-solid-state lithium batteries–The Mg2FeH6-electrode LiBH4-electrolyte system, Electrochemistry Communications, 87 (2018) 81-85.
[290] L. Zeng, K. Kawahito, S. Ikeda, T. Ichikawa, H. Miyaoka, Y. Kojima, Metal hydride-based materials towards high performance negative electrodes for all-solid-state lithium-ion batteries, Chemical Communications, 51 (2015) 9773-9776.
[291] P. López-Aranguren, N. Berti, A.H. Dao, J. Zhang, F. Cuevas, M. Latroche, C. Jordy, An all-solid-state metal hydride–Sulfur lithium-ion battery, Journal of Power Sources, 357 (2017) 56-60.
[292] L. Zeng, T. Ichikawa, K. Kawahito, H. Miyaoka, Y. Kojima, Bulk-Type All-Solid-State Lithium-Ion Batteries: Remarkable Performances of a Carbon Nanofiber-Supported MgH2 Composite Electrode, ACS Applied Materials & Interfaces, 9 (2017) 2261-2266.
[293] Z. Qian, A.D. Sarkar, T.A. Maark, X. Jiang, M.D. Deshpande, M. Bououdina, R. Ahuja, Pure and Li-doped NiTiH: Potential anode materials for Li-ion rechargeable batteries, Applied Physics Letters, 103 (2013) 033902.
[294] A. El kharbachi, Y. Hu, M.H. Sørby, J.P. Mæhlen, P.E. Vullum, H. Fjellvåg, B.C. Hauback, Reversibility of metal-hydride anodes in all-solid-state lithium secondary battery operating at room temperature, Solid State Ionics, 317 (2018) 263-267.

[295] P. Huen, D.B. Ravnsbæk, All-solid-state lithium batteries – The Mg2FeH6-electrode LiBH4-electrolyte system, Electrochemistry Communications, 87 (2018) 81-85.
[296] A. El Kharbachi, H. Uesato, H. Kawai, S. Wenner, H. Miyaoka, M.H. Sørby, H. Fjellvåg, T. Ichikawa, B.C. Hauback, MgH2–CoO: a conversion-type composite electrode for LiBH4-based all-solid-state lithium ion batteries, RSC Advances, 8 (2018) 23468-23474.
[297] C. Wang, M. Sawicki, S. Emani, C. Liu, L.L. Shaw, Na3MnCO3PO4 - A High Capacity, Multi-Electron Transfer Redox Cathode Material for Sodium Ion Batteries, Electrochimica Acta, 161 (2015) 322-328.
[298] P. Senguttuvan, G. Rousse, V. Seznec, J.-M. Tarascon, M. Rosa Palacin, Na2Ti3O7: Lowest Voltage Ever Reported Oxide Insertion Electrode for Sodium Ion Batteries, Chemistry of Materials, 23 (2011) 4109-4111.
[299] D. Aurbach, Y. Cohen, M. Moshkovich, The study of reversible magnesium deposition by in situ scanning tunneling microscopy, Electrochemical and Solid State Letters, 4 (2001) A113-A116.
[300] T.D. Gregory, R.J. Hoffman, R.C. Winterton, DEVELOPMENT OF AN AMBIENT TEMPERATURE-SECONDARY MAGNESIUM BATTERY, Journal of the Electrochemical Society, 135 (1988) C119-C119.
[301] M. Matsui, Study on electrochemically deposited Mg metal, Journal of Power Sources, 196 (2011) 7048-7055.
[302] M. Jackle, A. Gross, Microscopic properties of lithium, sodium, and magnesium battery anode materials related to possible dendrite growth, Journal of Chemical Physics, 141 (2014).
[303] R. Davidson, A. Verma, D. Santos, F. Hao, C. Fincher, S. Xiang, J. Van Buskirk, K. Xie, M. Pharr, P.P. Mukherjee, S. Banerjee, Formation of Magnesium Dendrites during Electrodeposition, ACS Energy Letters, 4 (2019) 375-376.
[304] C.M. MacLaughlin, Status and Outlook for Magnesium Battery Technologies: A Conversation with Stan Whittingham and Sarbajit Banerjee, Acs Energy Letters, 4 (2019) 572-575.
[305] H.R. Yao, Y. You, Y.X. Yin, L.J. Wan, Y.G. Guo, Rechargeable dual-metal-ion batteries for advanced energy storage, Physical Chemistry Chemical Physics, 18 (2016) 9326-9333.
[306] D.S. Tchitchekova, D. Monti, P. Johansson, F. Bardé, A. Randon-Vitanova, M.R. Palacín, A. Ponrouch, On the Reliability of Half-Cell Tests for Monovalent (Li+, Na+) and Divalent (Mg2+, Ca2+) Cation Based Batteries, Journal of The Electrochemical Society, 164 (2017) A1384-A1392.
[307] G.G. Amatucci, F. Badway, A. Singhal, B. Beaudoin, G. Skandan, T. Bowmer, I. Plitz, N. Pereira, T. Chapman, R. Jaworski, Investigation of Yttrium and Polyvalent Ion Intercalation into Nanocrystalline Vanadium Oxide, Journal of The Electrochemical Society, 148 (2001) A940-A950.
[308] T. Ouchi, H. Kim, B.L. Spatocco, D.R. Sadoway, Calcium-based multi-element chemistry for grid-scale electrochemical energy storage, Nature Communications, 7 (2016) 10999.
[309] D. Chao, C. Zhu, M. Song, P. Liang, X. Zhang, N.H. Tiep, H. Zhao, J. Wang, R. Wang, H. Zhang, H.J. Fan, A High-Rate and Stable Quasi-Solid-State Zinc-Ion Battery with Novel 2D Layered Zinc Orthovanadate Array, Advanced Materials, 30 (2018) 1803181.
[310] J. Zhao, K.K. Sonigara, J. Li, J. Zhang, B. Chen, J. Zhang, S.S. Soni, X. Zhou, G. Cui, L. Chen, A Smart Flexible Zinc Battery with Cooling Recovery Ability, Angewandte Chemie International Edition, 56 (2017) 7871-7875.
[311] S. Wang, S. Jiao, W.-L. Song, H.-S. Chen, J. Tu, D. Tian, H. Jiao, C. Fu, D.-N. Fang, A novel dual-graphite aluminum-ion battery, Energy Storage Materials, 12 (2018) 119-127.
[312] X. Zhao, Q. Li, Z. Zhao-Karger, P. Gao, K. Fink, X. Shen, M. Fichtner, Magnesium Anode for Chloride Ion Batteries, ACS Applied Materials & Interfaces, 6 (2014) 10997-11000.
[313] C. Chen, T. Yu, M. Yang, X. Zhao, X. Shen, An All-Solid-State Rechargeable Chloride Ion Battery, Advanced Science, 6 (2019) 1802130.
[314] W.H. Zhu, Y. Zhu, B.J. Tatarchuk, Self-discharge characteristics and performance degradation of Ni-MH batteries for storage applications, International Journal of Hydrogen Energy, 39 (2014) 19789-19798.
[315] W.H. Zhu, Y. Zhu, Z. Davis, B.J. Tatarchuk, Energy efficiency and capacity retention of Ni–MH batteries for storage applications, Applied Energy, 106 (2013) 307-313.
[316] A. Taniguchi, N. Fujioka, M. Ikoma, A. Ohta, Development of nickel/metal-hydride batteries for EVs and HEVs, Journal of Power Sources, 100 (2001) 117-124.
[317] K. Kubota, M. Dahbi, T. Hosaka, S. Kumakura, S. Komaba, Towards K-Ion and Na-Ion Batteries as "Beyond Li-Ion", The Chemical Record, 18 (2018) 459-479.
[318] R.A.H. Niessen, P.H.L. Notten, Electrochemical Hydrogen Storage Characteristics of Thin Film MgX (X = Sc，Ti，V，Cr) Compounds, Electrochemical and Solid-State Letters, 8 (2005) A534-A538.
[319] P.H.L. Notten, M. Ouwerkerk, H. van Hal, D. Beelen, W. Keur, J. Zhou, H. Feil, High energy density strategies: from hydride-forming materials research to battery integration, Journal of Power Sources, 129 (2004) 45-54.
[320] T. Meng, K.-H. Young, D.F. Wong, J. Nei, Ionic Liquid-Based Non-Aqueous Electrolytes for Nickel/Metal Hydride Batteries, Batteries, 3 (2017) 4.

[321] Y. Oumellal, A. Rougier, G.A. Nazri, J.M. Tarascon, L. Aymard, Metal hydrides for lithium-ion batteries, Nature Materials, 7 (2008) 916-921.
[322] R. Demir-Cakan, M.R. Palacin, L. Croguennec, Rechargeable aqueous electrolyte batteries: from univalent to multivalent cation chemistry, Journal of Materials Chemistry A, 7 (2019) 20519-20539.
[323] Battery University™ www.batteryuniversity.com (accessed Oct. 2019).
[324] S. Wu, F. Zhang, Y. Tang, A Novel Calcium-Ion Battery Based on Dual-Carbon Configuration with High Working Voltage and Long Cycling Life, Advanced Science, 5 (2018) 1701082.
[325] A.L. Lipson, S. Kim, B. Pan, C. Liao, T.T. Fister, B.J. Ingram, Calcium intercalation into layered fluorinated sodium iron phosphate, Journal of Power Sources, 369 (2017) 133-137.
[326] A. Unemoto, T. Ikeshoji, S. Yasaku, M. Matsuo, V. Stavila, T.J. Udovic, S.-i. Orimo, Stable Interface Formation between TiS2 and LiBH4 in Bulk-Type All-Solid-State Lithium Batteries, Chemistry of Materials, 27 (2015) 5407-5416.
[327] A. El Kharbachi, I. Nuta, F. Hodaj, M. Baricco, Above room temperature heat capacity and phase transition of lithium tetrahydroborate, Thermochimica Acta, 520 (2011) 75-79.
[328] A. El kharbachi, Y. Hu, K. Yoshida, P. Vajeeston, S. Kim, M.H. Sørby, S.-i. Orimo, H. Fjellvåg, B.C. Hauback, Lithium ionic conduction in composites of Li(BH4)0.75I0.25 and amorphous 0.75Li2S·0.25P2S5 for battery applications, Electrochimica Acta, 278 (2018) 332-339.
[329] M. Anji Reddy, M. Fichtner, Batteries based on fluoride shuttle, Journal of Materials Chemistry, 21 (2011) 17059-17062.

**Graphical Abstract**

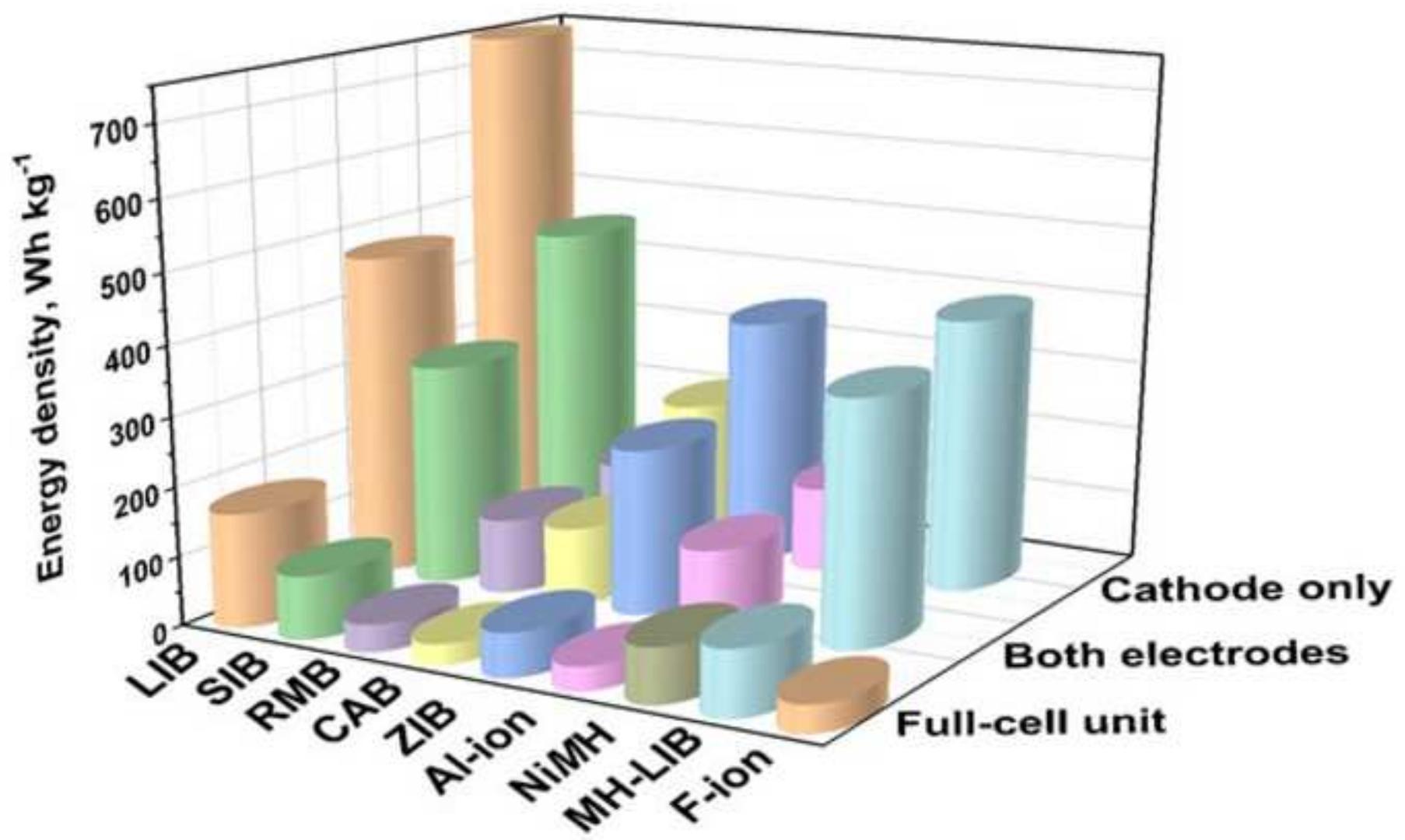

# Highlights

- Battery market expansion needs to follow the energy transition

- Beyond Li-ion batteries are of high importance for specific applications

- Comparison between different battery prototypes and configurations

- The multitude applications may allow the emergence of novel batteries

Table_1

| Table 1. | | Properties | | | | | |
|---|---|---|---|---|---|---|---|
| Year | Composition anode / electrolyte / cathode (I or C) [§] | Operating $T$, $^{o}C$ | Cathode capacity (1st – last cycle)/ mAh g$^{-1}$ | Operating voltage / V | Stability (cycles and/or CE[*]) | Comments | Ref. |
| 1990 | Mg / Mg(BBu$_2$Ph$_2$) in THF-DME / Co$_3$O$_4$ (I) | R$T$ | ca. 185 | 1.5 | 4 | low potential, high polarisation, low oxidative stability of the electrolyte | [90] |
| 2000 | Mg / THF/Mg(AlCl$_2$BuEt)$_2$ /Mo$_6$S$_8$ (I) | -20 to 80 °C | ca.90 – ca. 75 | 1-1.3 | 580 | low capacity, but long durability (up to 2000 cycles) | [91] |
| 2015 | Mg / Mg(BH$_4$)$_2$+LiBH$_4$ in tetraglyme / TiO$_2$ (I) | R$T$ | 168-148 | 0.9-1.1 | 100 | good stability and rate capability | [173] |
| 2016 | Mg / Mg(TFSI)$_2$ in DME/ Diglyme(1:1vol)+ Mg(TFSI)$_2$ -PYR$_{14}$TFSI(IL)[#]-MgBr$_2$ / Br (C) | R$T$ | ca. 275 | 2.4-3.2 | 20, 95% | dual-electrolyte, few cycles only demonstrated | [174] |
| 2017 | Mg / Mg-HMDS / I$_2$ (C) | R$T$ | 180 | 2.2 | 120 | Absence of solid-state diffusion, suitable for semi-flow batteries | [98] |
| 2018 | Mg / Mg-Li dual-salt / Na$_2$C$_6$O$_6$ (co-I) | R$T$ | 450-125 (at various rates) | 1.1 | 600 | Multi-process intercalation, dominated by Li-ions | [175] |
| 2019 | Mg / [Mg(BH$_4$)$_2$]$_{0.3}$[N$_{07}$TFSI]$_{0.7}$-PYR$_{14}$TFSI(IL)[#]/ V$_2$O$_5$ aerogel (I) | R$T$ | 100 - 80 | 1.4 – 1.8 | 40 | halide-free non-corrosive electrolyte; large capacity loss with cycling | [110] |

[*] CE: Coulombic efficiency (%)
[§] I: Intercalation cathode; C: Conversion cathode
[#] IL: Ionic liquid

Table_2

| Battery technology | Cathode | Anode | Electrolyte composition | Assessed performance full-cells | | Major advantages | Drawbacks | Ref. |
| --- | --- | --- | --- | --- | --- | --- | --- | --- |
| | | | | Energy density [a] (Wh kg$^{-1}$) | Durability (cycles) | | | |
| **LIB** | LiFePO$_4$ | Graphitic carbon | 1M LiClO$_4$/ EC-DMC | 90-120 | 2,000 | High energy density | Cost and safety | [321] |
| | LiCoO$_2$ | | 1M LiPF$_6$/ EC-DMC | 150-240 | 500-1000 | High voltage and energy density | | |
| **SIB** | Na$_{1.5}$VPO$_{4.8}$F$_{0.7}$ | Hard or nanostructured carbon | 1M NaClO$_4$/ EC$_{0.45}$·PC$_{0.45}$·DMC$_{0.1}$ | 80-100 [b] | 120-210 | Abundance, low cost at large-scale and better safety, wide temperature range | Moderate energy density | [36,62,64,78,81] |
| | Na$_{0.45}$Ni$_{0.22}$Co$_{0.11}$Mn$_{0.66}$O$_2$ | | 10 mol.% NaTFSI / PYR$_{14}$FSI | 70-114 | 100 | | *idem*, cost of ILs | [56] |
| **RMB** | Mo$_6$S$_8$ | Mg (2200) | Mg(AlCl$_2$BuEt)$_2$/THF | 38-42 | 2,000 | Abundance, low cost | Low capacity | [91] |
| | TiO$_2$ | | Mg(BH$_4$)$_2$-LiBH$_4$ /tetraglyme | 40-45 | 120 | | | [173] |
| **CAB** | Carbon-based layered | Carbon-based layered | 0.7M Ca(PF$_6$)$_2$/EC-DMC-EMC | 40-45 | 300 | Abundance, low cost, low temperature operation | Low energy density | [322] |
| | Na$_2$FePO$_4$F | BP2000 carbon | Ca(PF$_6$)$_2$/EC-PC | 16-20 | 50 | | | [323] |
| **ZIB** | V$_2$O$_5$ | Zn | 3M Zn(CF$_3$SO$_3$)$_2$ aq. electrolyte | 70-75 | 4,000 | Safety, low cost, facile manufacturing, wide temperature | Low voltage | [211] |
| | Zn$_{0.3}$V$_2$O$_5$.1.5H$_2$O | | | 60-85 | 20,000 | | | [195,210] |
| **Al-ion** | 3D graphitic foam | Al | AlCl$_3$/1-ethyl-3-methylimidazolium chloride | 35-40 | 7500 | Cost, safety, high power | High Lewis acidity issues | [213] |
| | Ultrathin graphite nanosheets | Zn | Al$_2$(SO$_4$)$_3$/Zn(CHCOO)$_2$ aq. electrolyte | 15-25 | 200 | Cost, large-scale stationary storage | Low energy density | [216] |
| **Ni*M*H** | Ni(OH)2 | LaNi$_5$-type | Conc. KOH alkaline electrolyte | 40-120 | 300-500 | High specific power | Low voltage (aq. medium) | [245,321] |
| **MH-LIB** | TiS$_2$ | Li | LiBH$_4$ | 110-120 | 300 | High energy density | Operating at 120 °C | [274,324,325] |
| | TiS$_2$ | 75MgH$_2$-25CoO | Li(BH$_4$)$_{0.75}$I$_{0.25}$·(Li$_2$S)$_{0.75}$·(P$_2$S$_5$)$_{0.25}$ | 70-90 [b] | -- | High energy density, dentrite free, safety, low *T* | Low voltage | [274,292,294,324,326] |
| **F-ion** | BiF$_3$ | Ce metal | La$_{0.9}$Ba$_{0.1}$F$_{2.9}$ - BaSnF$_4$ (R*T*-150°C) | < 40 | 10-50 | Safety, energy density | Capacity fading at R*T* | [220,221,327] |

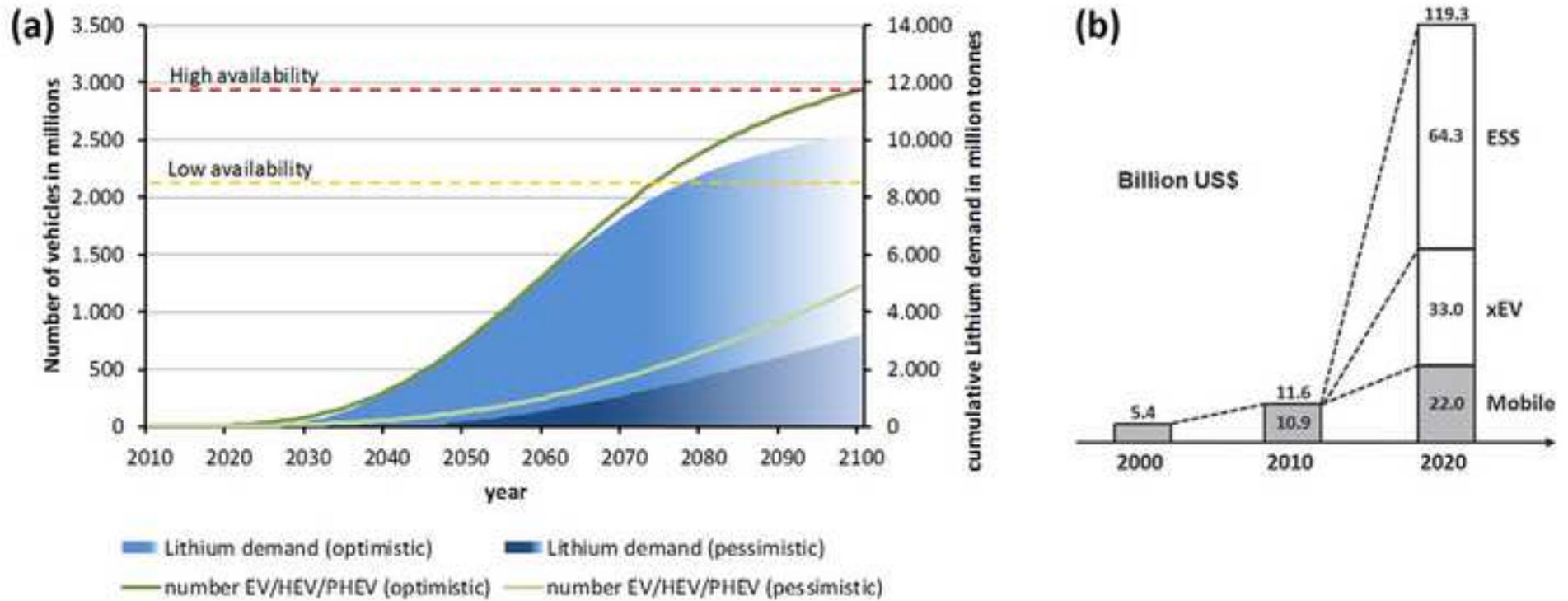

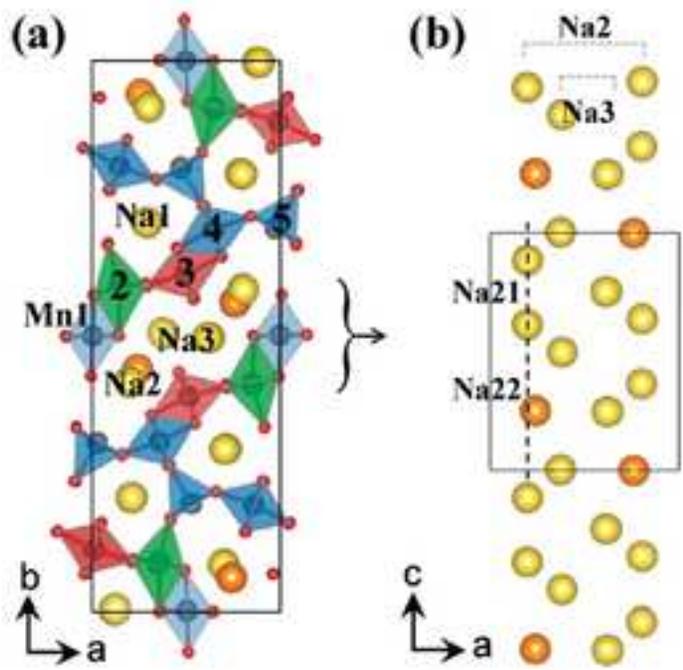
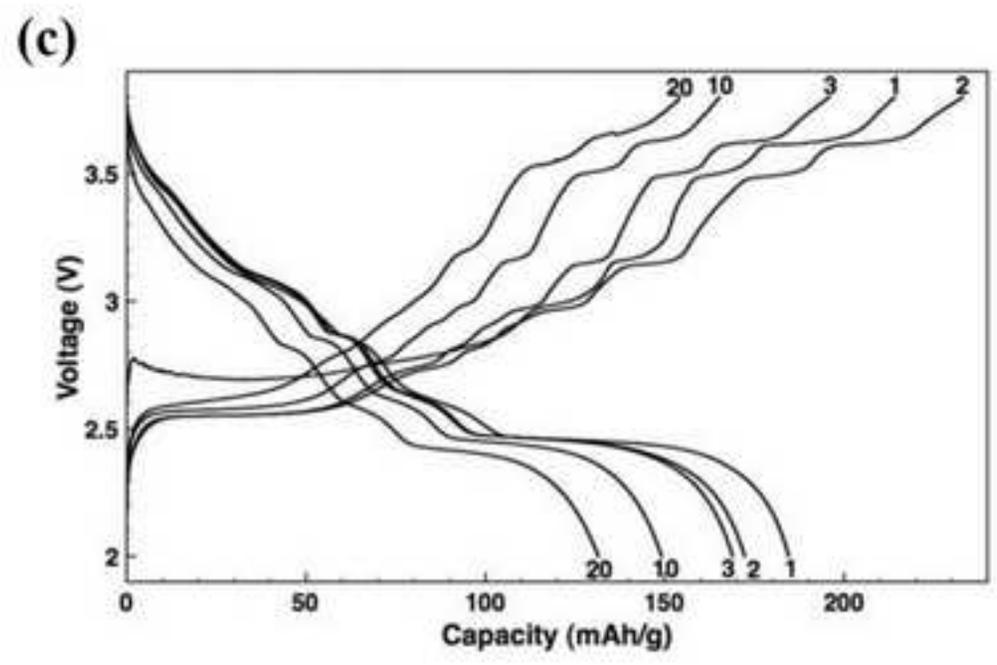

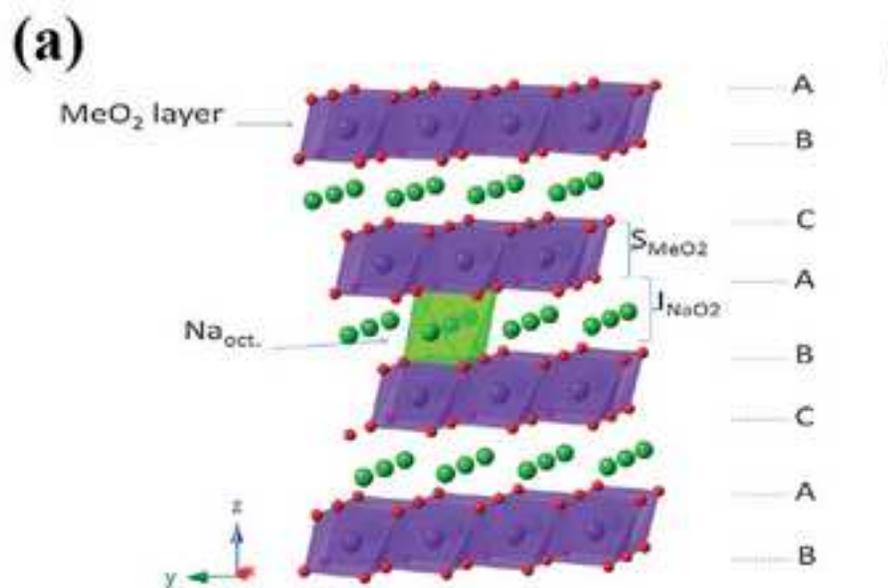
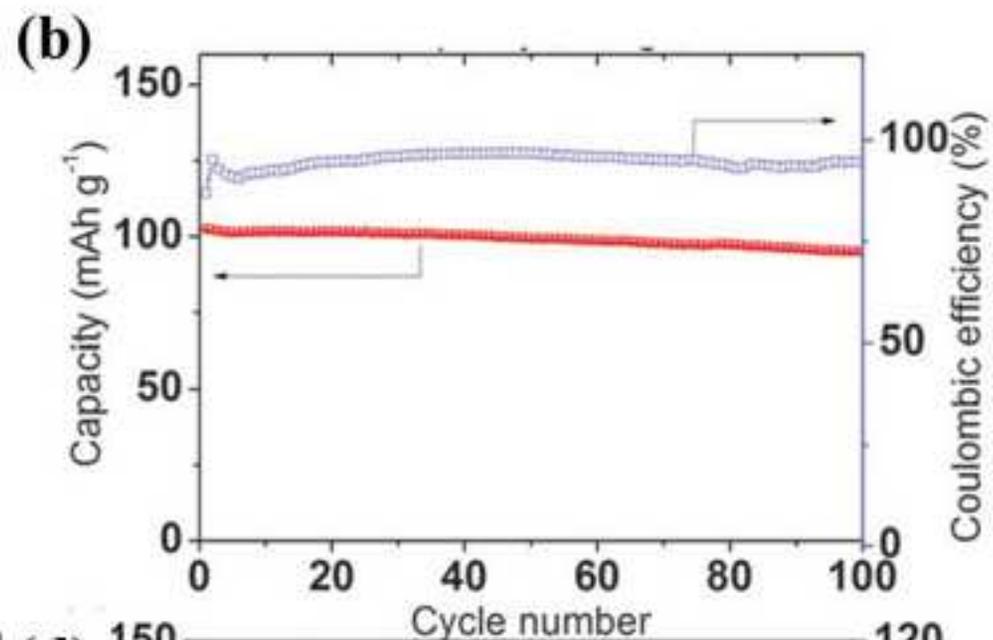
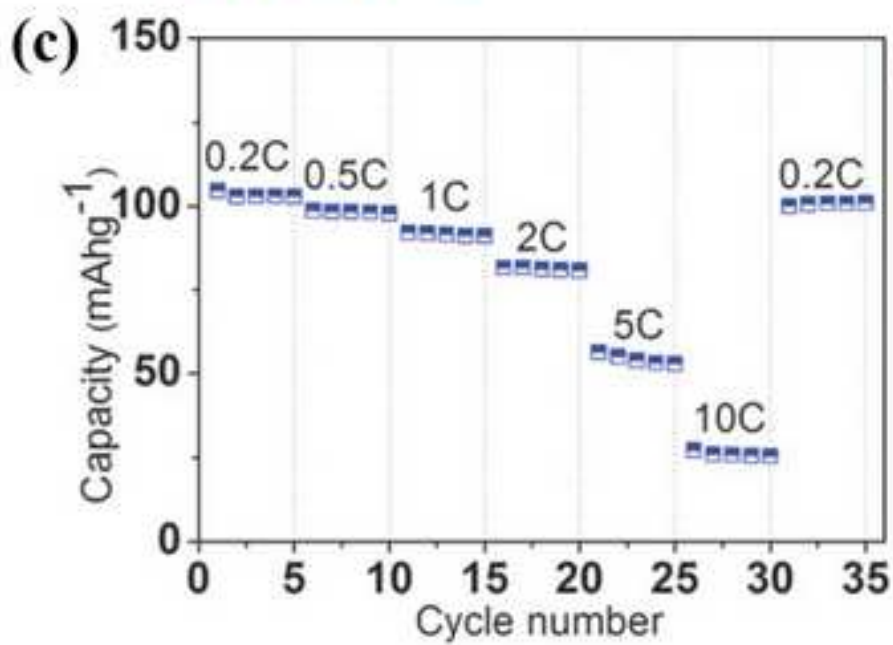
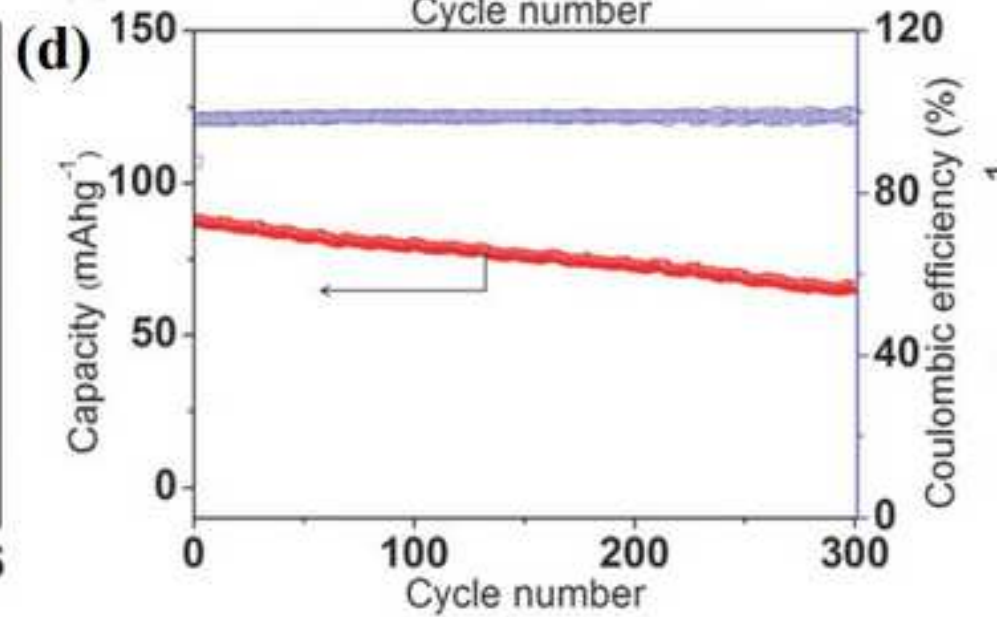

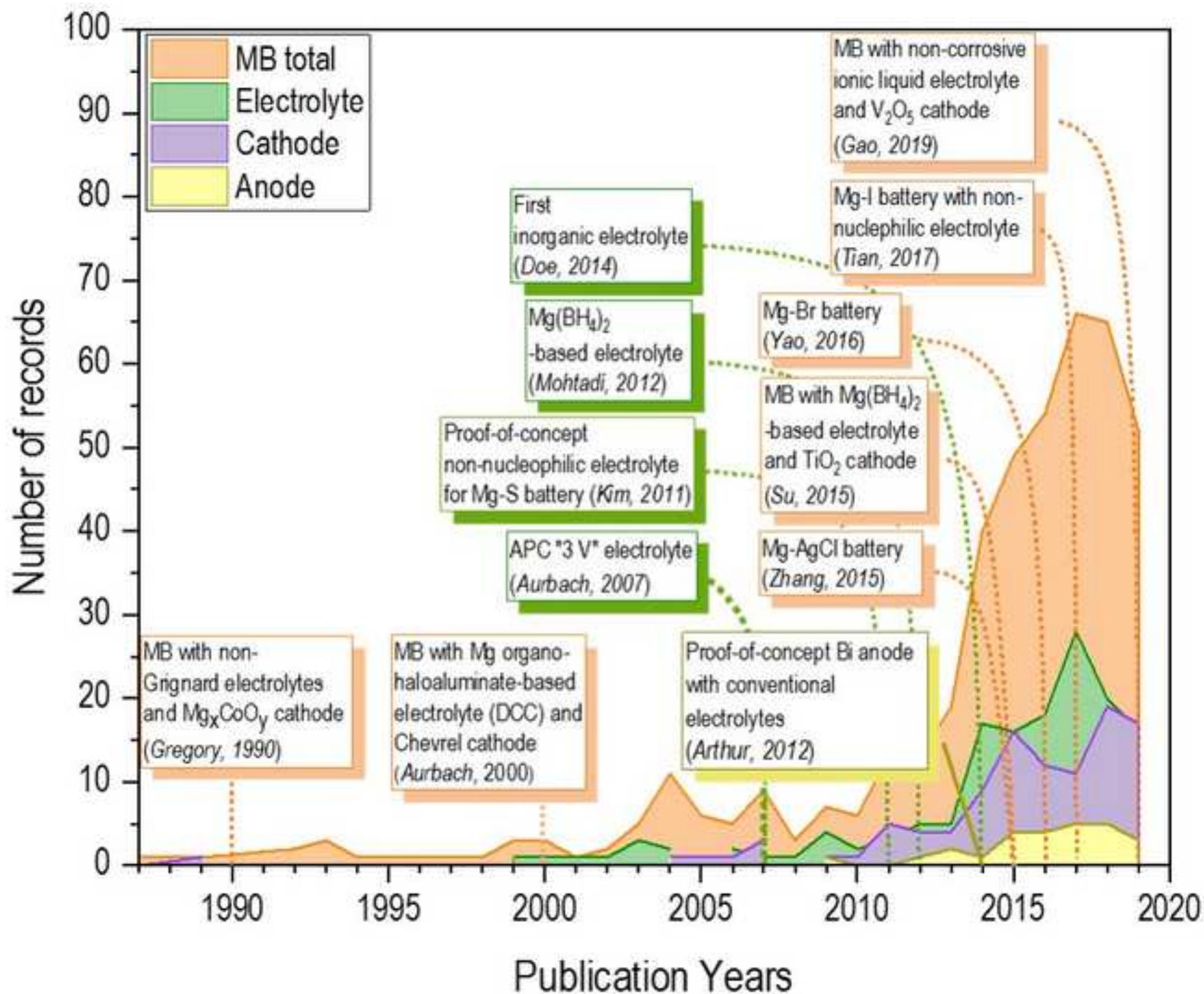

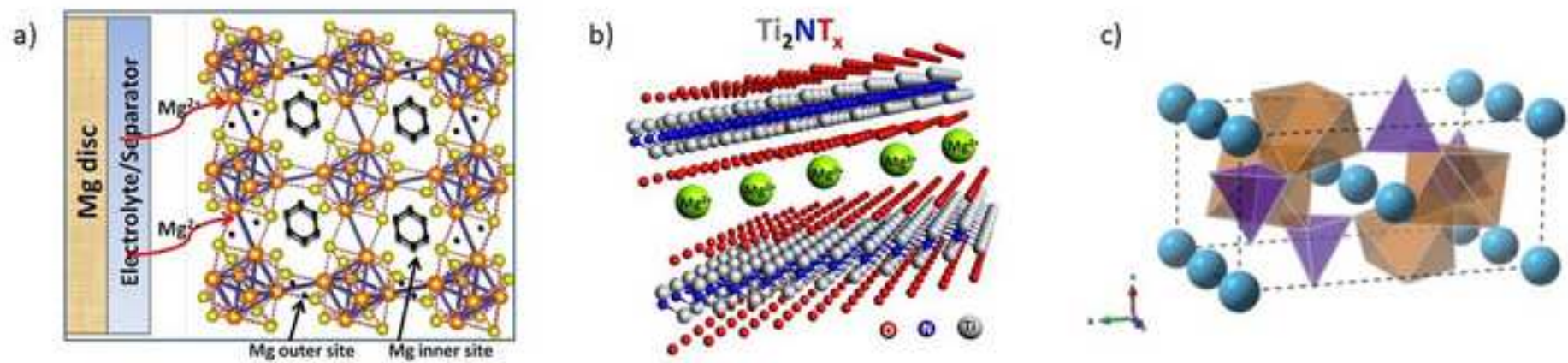

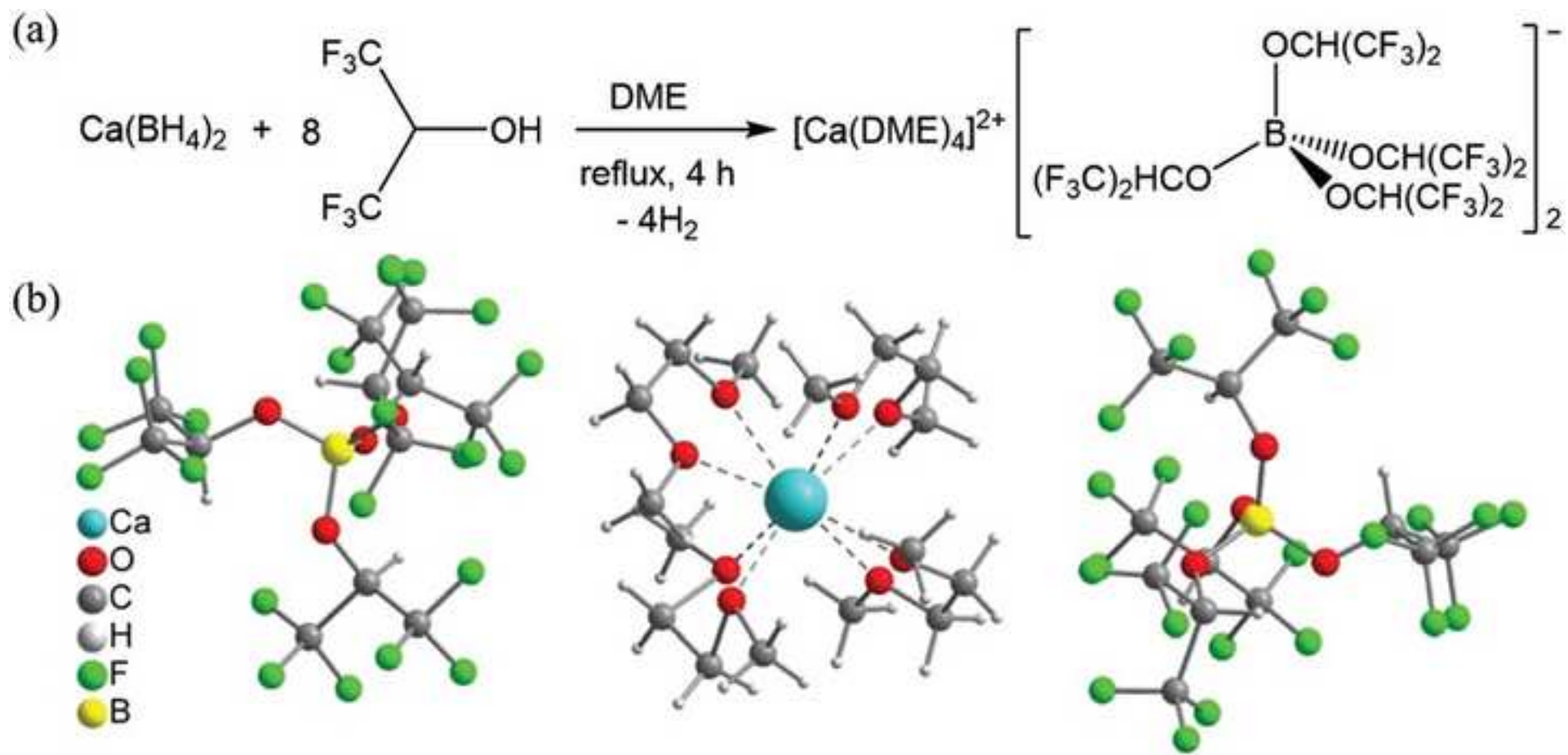

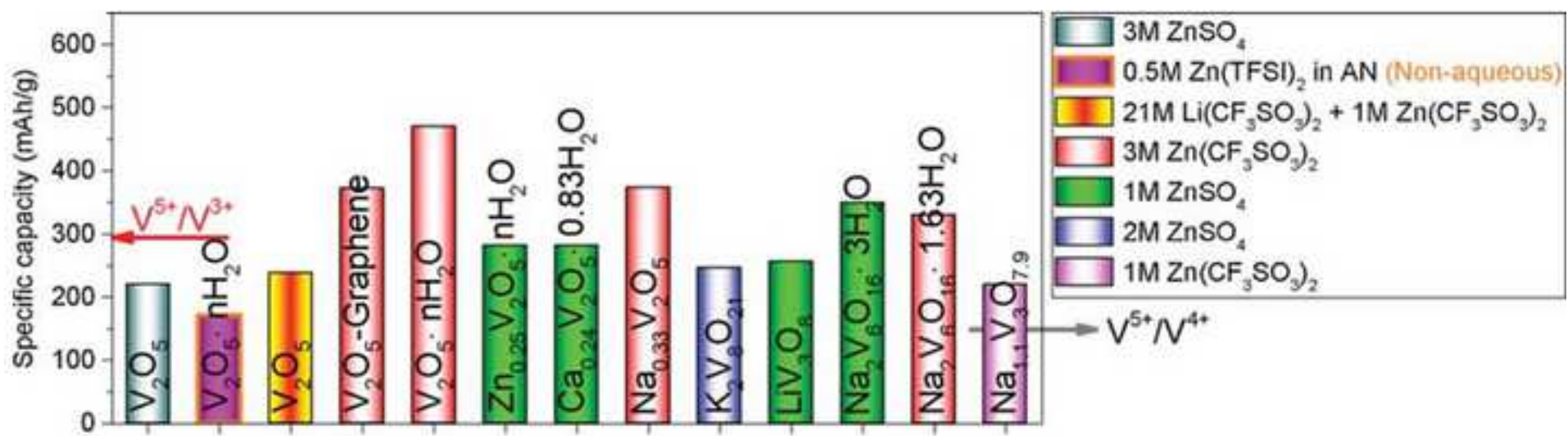

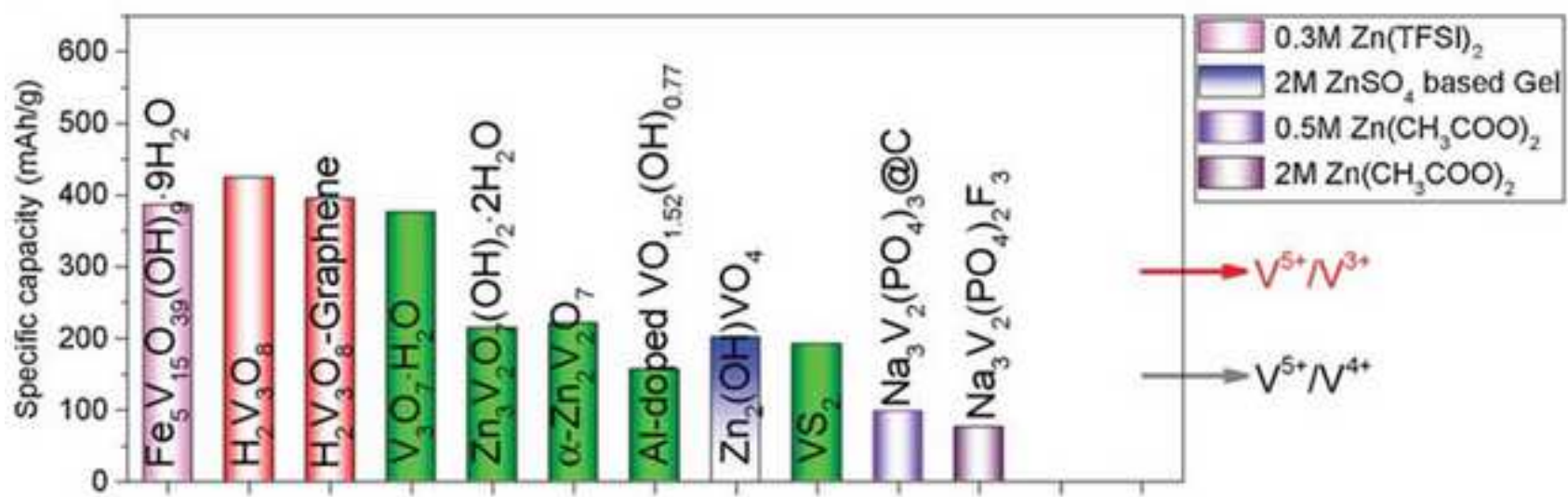

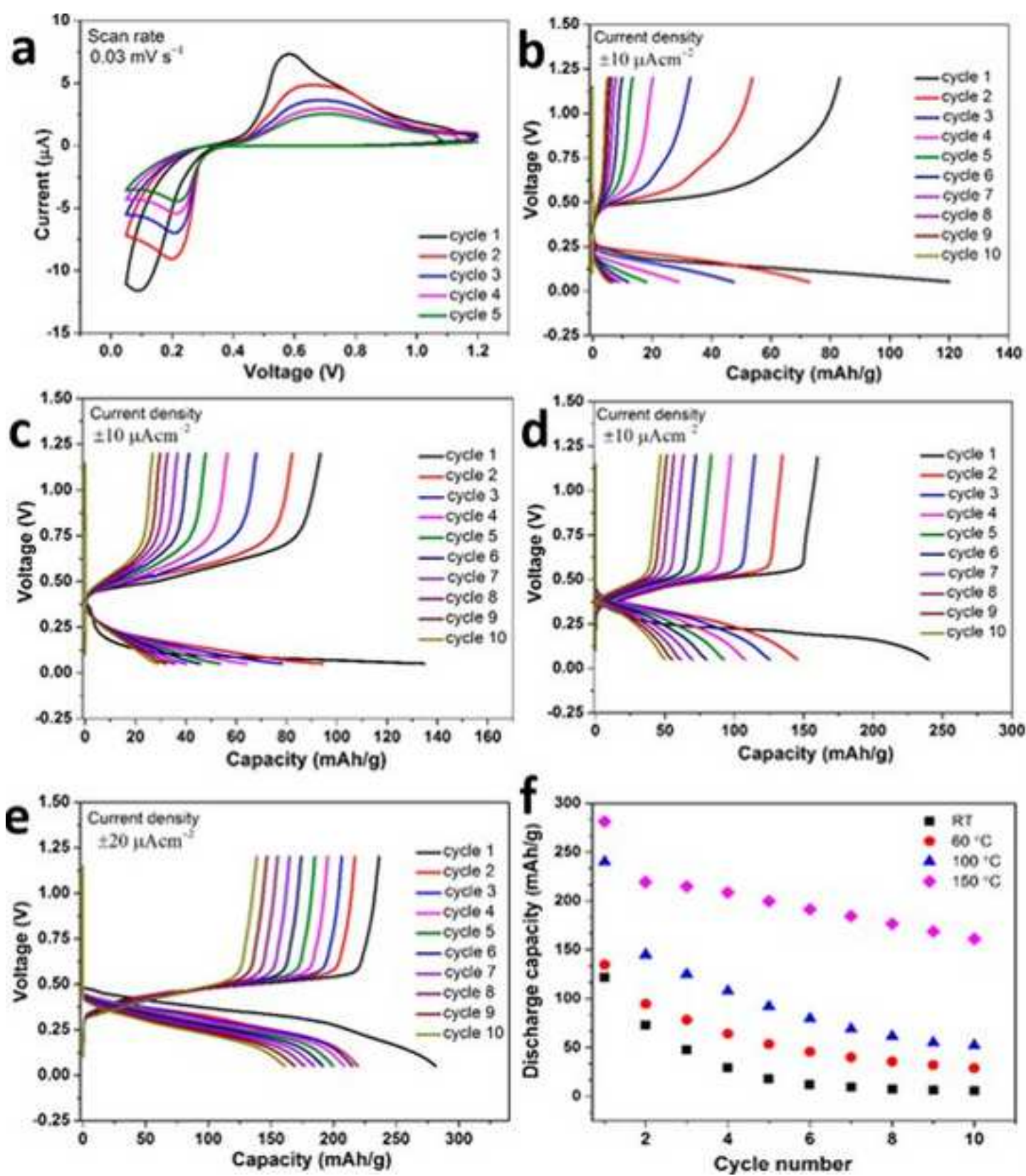

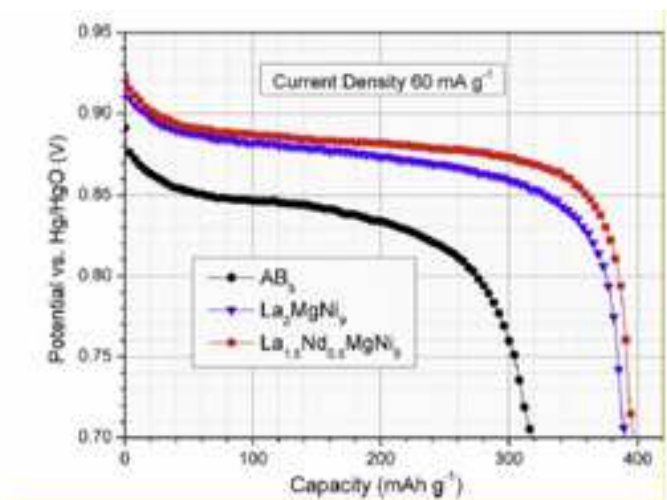 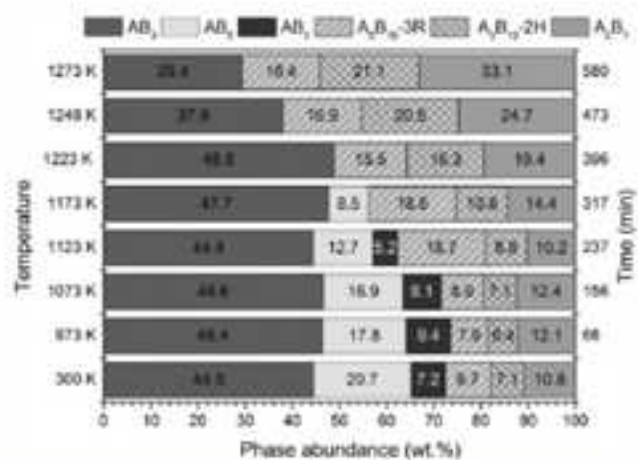 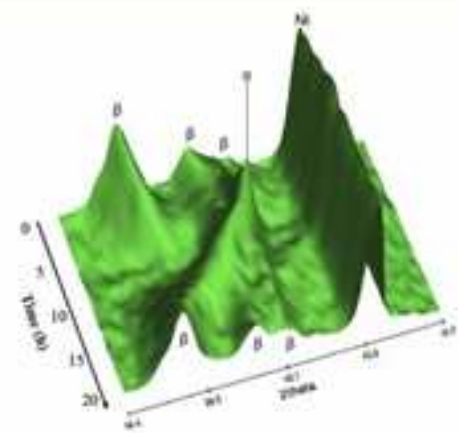

*a*  *b*  *c*

**Figure_10**
**Click here to download high resolution image**

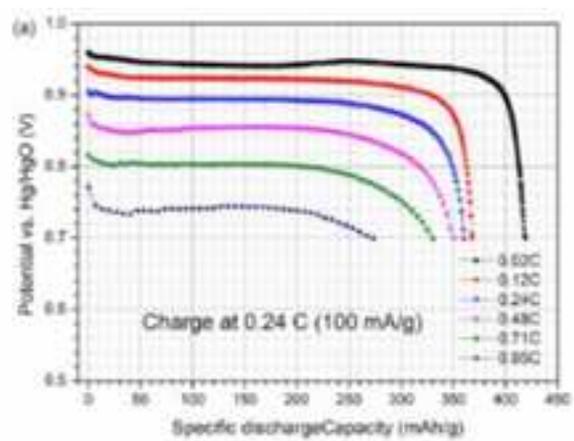 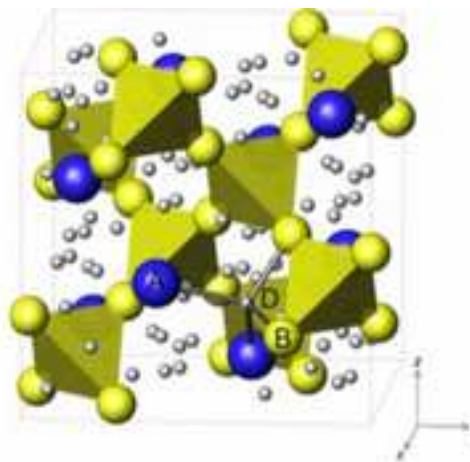 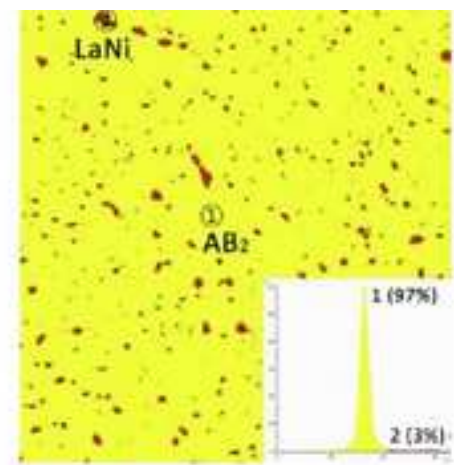

a  b  c

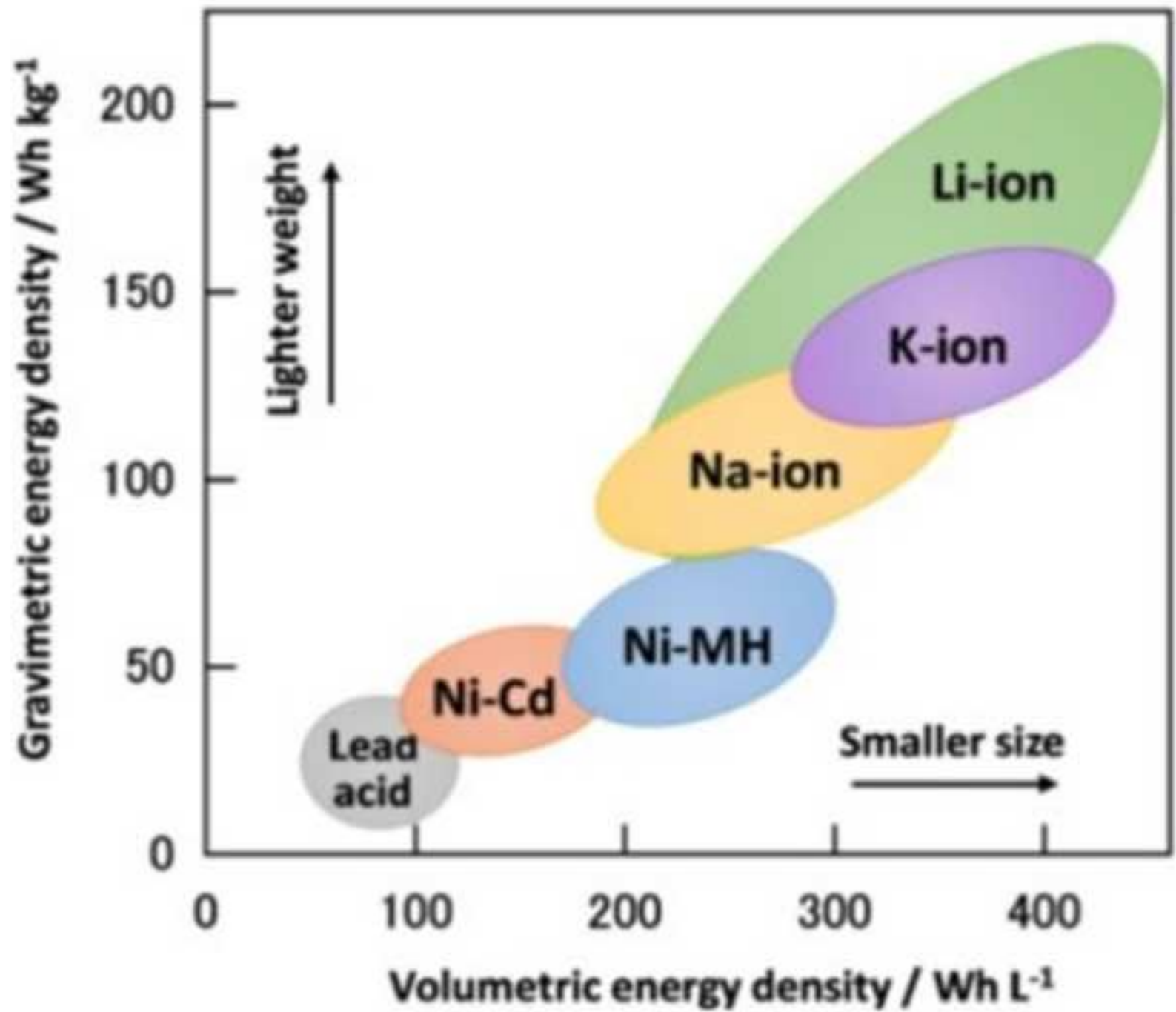

Figure_12
Click here to download high resolution image

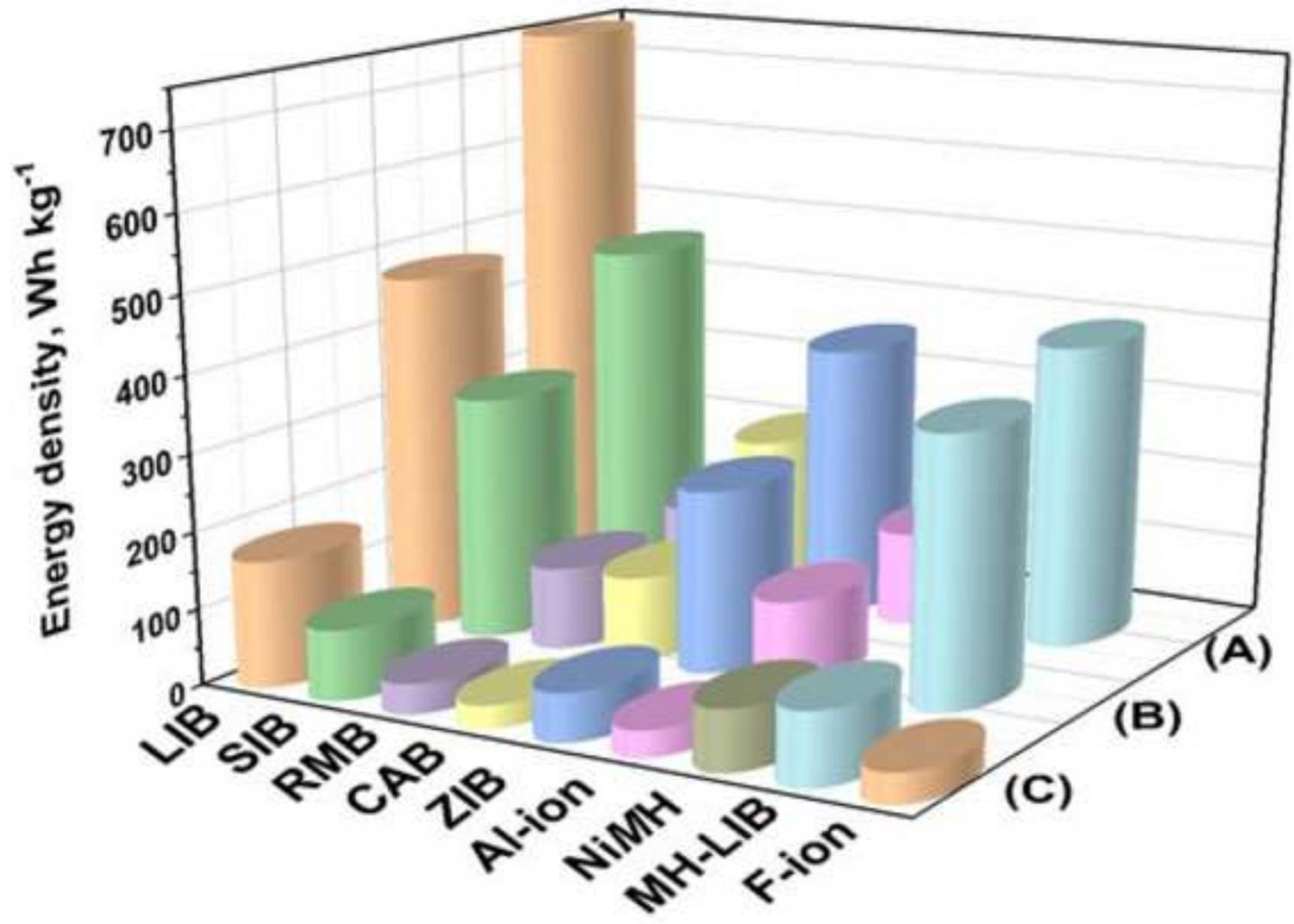

# Author Contributions Section

A.E.-K., V.Y. and MF planned and elaborate the first draft. O.Z. contributed with Mg and Ca-ion batteries sections. M.L. and F.C. contributed with the MH- based batteries part. All authors contributed to the discussion. A.E.-K. and V.Y. prepared and revised the final manuscript.